\documentclass[usenatbib]{mnras}
\ifx\pdftexversion\undefined \usepackage{epsf,epsfig}
\else \usepackage[pdftex]{graphicx} \fi

\DeclareGraphicsExtensions{.pdf,.png}

\usepackage{lscape}
\usepackage{color}
\usepackage{amsmath}
\usepackage{multirow}
\usepackage{longtable}
\usepackage[normalem]{ulem}
\usepackage{subfig}

\newcommand{\Herschel}{\textit{Herschel}}
\newcommand{\Spitzer}{\textit{Spitzer}}
\newcommand{\masterlist}{\textit{master list}}
\newcommand{\um}{\micron\  }

\title[HELP: The \Herschel\ Extragalactic Legacy Project]{HELP: The \Herschel\ Extragalactic Legacy Project\footnote{\Herschel\ is an ESA space observatory with science instruments provided by European-led Principal Investigator consortia and with important participation from NASA.}}
\author[R. Shirley et al.]{
  \parbox{\textwidth}{
    R.~Shirley,$^{1,2,3}$
  K.~Duncan,$^{4,5}$
  M.C.~Campos~Varillas,$^{1}$
  P.D.~Hurley,$^{1}$
  K.~Ma\l{}ek,$^{6,7}$
  Y.~Roehlly,$^{1,7}$
  M.W.L.~Smith,$^{8}$
  H.~Aussel,$^{9}$
  T.~Bakx,$^{8,10,11}$
  V.~Buat,$^{7,12}$
  D.~Burgarella,$^{7}$
  N.~Christopher,$^{13}$
  S.~Duivenvoorden,$^{1}$
  S.~Eales,$^{8}$
  A.~Efstathiou,$^{13}$
  E.A.~Gonz\'alez~Solares,$^{3}$
  M.~Griffin,$^{8}$
  M.~Jarvis,$^{14,15}$
  B.~Lo~Faro,$^{7}$
  L.~Marchetti,$^{15,16,17}$
  I.~McCheyne,$^{1}$
  A.~Papadopoulos,$^{13}$
  K.~Penner,$^{8}$
  E.~Pons,$^{3,7}$
  M.~Prescott,$^{15}$
  E.~Rigby,$^{4}$
  H.~Rottgering,$^{4}$
  A.~Saxena,$^{4,18}$
  J.~Scudder,$^{1,19}$
  M.~Vaccari,$^{15,17}$
  L.~Wang, $^{20,21}$
and S.J.~Oliver$^{1}$\thanks{S.Oliver@Sussex.ac.uk}
}
\\
\\
$^{1}$Astronomy Centre, Department of Physics \& Astronomy, University of Sussex, Brighton, BN1 9QH, UK\\
$^{2}$Astronomy Centre, Department of Physics \& Astronomy, University of Southampton, Southampton, SO17 1BJ, UK\\
$^{3}$Institute of Astronomy, University of Cambridge, Madingley Road, Cambridge, CB3 0HA, UK\\
$^{4}$Sterrewacht Leiden, Universiteit Leiden, Leiden, Netherlands\\
$^{5}$Royal Observatory, Edinburgh, Blackford Hill,Edinburgh, EH9 3HJ, UK\\
$^{6}$National Centre for Nuclear Research, ul. Ho$\dot{z}$a 69, 00-681 Warszawa, Poland\\
$^{7}$Aix Marseille Univ, CNRS, CNES, LAM, Marseille, France\\
$^{8}$School of Physics and Astronomy, Cardiff University, Queens Buildings, The Parade, Cardiff CF24 3AA, UK\\
$^{9}$Laboratoire AIM-Paris-Saclay, CEA/DSM/Irfu - CNRS - Universit\'e Paris, CE-Saclay, France\\
$^{10}$Division of Particle and Astrophysical Science, Graduate School of Science, Nagoya University, Aichi 464-8602, Japan.\\
$^{11}$National Astronomical Observatory of Japan, 2-21-1, Osawa, Mitaka, Tokyo 181-8588, Japan.\\
$^{12}$Institut Universitaire de France (IUF) France\\
$^{13}$School of Sciences, European University Cyprus, Engomi, 1516, Nicosia, Cyprus\\
$^{14}$Astrophysics, Department of Physics, University of Oxford, Keble Road, Oxford OX1 3RH, UK\\
$^{15}$Department of Physics and Astronomy, University of the Western Cape, 7535 Bellville, Cape Town, South Africa\\
$^{16}$Department of Astronomy, University of Cape Town, 7701 Rondebosch, Cape Town, South Africa\\
$^{17}$INAF - Istituto di Radioastronomia, via Gobetti 101, 40129 Bologna, Italy\\
$^{18}$Department of Physics and Astronomy, University College London, Gower Street, London, WC1E 6BT\\
$^{19}$Department of Physics \& Astronomy, Oberlin College, Oberlin, OH, 44074, USA\\
$^{20}$SRON Netherlands Institute for Space Research, Landleven 12, 9747 AD Groningen, The Netherlands\\
$^{21}$Kapteyn Astronomical Institute, University of Groningen, Postbus 800, 9700 AV Groningen, the Netherlands.
}

\pubyear{2021}
\begin{document}
\maketitle



\begin{abstract}
  We present the \Herschel\ Extragalactic Legacy Project (HELP). This project collates, curates, homogenises, and creates derived data products for most 
  of the premium multi-wavelength extragalactic data sets. The sky boundaries for the first data release cover 1270 ${\rm
  deg}^2$ defined by the \Herschel\ SPIRE extragalactic survey fields;
  notably the \Herschel\ Multi-tiered Extragalactic Survey (HerMES) and the
  \Herschel\ Atlas survey (H-ATLAS). Here, we describe the motivation and principal
  elements in the design of the project. Guiding principles are transparent or 
  ``open" methodologies with care for reproducibility and identification of provenance. 
  A key element of the design focuses
  around the homogenisation of calibration, meta data and the provision of
  information required to define the selection of the data for statistical analysis. We apply
  probabilistic methods that extract information directly from the images at long
  wavelengths, exploiting the prior information available at shorter wavelengths
  and providing full posterior distributions rather than maximum likelihood estimates and associated uncertainties as in traditional catalogues. 
With this project definition paper we provide full access to the first data release of HELP; Data Release 1 (DR1),
including a monolithic map of the largest SPIRE extragalactic
  field at 385 deg$^2$ and 18 million measurements of PACS and SPIRE fluxes. We also provide tools to access and analyse the full HELP database. 
 This new data set includes far-infrared photometry, photometric redshifts, and derived physical properties estimated from modelling the spectral energy distributions over the full HELP sky.
  All the software and data presented is publicly available.  
\end{abstract}

\begin{keywords}
  techniques: photometric -- catalogues -- surveys -- infrared: galaxies --
  submillimetre: galaxies -- galaxies: evolution
\end{keywords}
\section[Introduction]{Introduction}\label{sec:intro}

A fundamental requirement for rigorous testing of any theories of galaxy
formation and evolution is a complete statistical audit or census of the stellar
content and star-formation rates of galaxies in the Universe at different times
and as a function of the mass of the dark matter halos that host them.

This audit requires many elements. We need un-biased maps of large volumes of the Universe made with telescopes that probe different wavelengths at which different physical processes of interest manifest themselves. We need catalogues of the galaxies contained within these maps with photometry estimated
uniformly from field-to-field, from telescope-to-telescope, and from
wavelength-to-wavelength. We need to understand the probability of a galaxy of
given properties appearing in our data sets.  We need the machinery to bring
together these various data sets and calculate the ``value-added" physical data
of primary interest, e.g. the distances, stellar masses,
star-formation rates, active galactic nuclei (AGN) fractions and the intrinsic number densities of the different galaxy
populations.

For decades many  teams  have been undertaking ambitious coordinated
multi-wavelength programmes to study large volumes of the distant Universe.
These surveys are becoming sufficiently complete that we are now able to
undertake the necessary homogenising and adding value, and thus provide the first
representative and comprehensive census of the galaxy populations in the Universe.

ESA's \Herschel\ \citep{Pilbratt:2010lr} mission has a unique role in these studies, probing the
obscured star-formation activity, which at high redshifts forms about 80\% of
all star formation. The \Herschel\ extragalactic surveys were a major goal
of \Herschel\ and occupied around 10\% of the \Herschel\ mission.  

The \Herschel\ Spectral and Photometric Imaging Receiver (SPIRE) instrument is sufficiently sensitive that the images can
detect nearly all of the emission making up the Cosmic Infrared Background
Radiation (CIRB) \citep{Duivenvoorden:2020}, which itself makes up roughly half of the total background
radiation from galaxies.  However, the large beam size means that the objects
that can be clearly seen as individual sources only make up about 15\% of the
CIRB.  The \Herschel\ Photoconductor Array Camera and Spectrometer (PACS) instrument complements the SPIRE observations with bands centered 
at 100~$\mu$m and 160~$\mu$m but at lower depths than SPIRE.  

A particular focus of HELP is to employ new methods to learn from our
large statistically meaningful samples. This requires harnessing the ancillary
data and the \Herschel\ data to unlock the full information from the
\Herschel\ images and then make that available as a legacy to the community.

The science possible with the \Herschel\ data will be significantly enhanced 
with ongoing optical, NIR and radio surveys.  The VISTA near-infrared surveys detect the radiation from the old
stellar population in galaxies, which accounts for most of the stellar mass,
while the radio surveys being carried out over the next few years with LOFAR, MeerKAT
and ASKAP detect radiation associated with the young stellar population and with radio-loud AGN.

The challenge for astronomers wishing to exploit these rich data sets is to collate the data, understand the selection effects and derive physical parameters. 
Collation of multi-wavelength data has been undertaken for very deep surveys
over small areas (less than few deg$^2$) in particular COSMOS \citep{cosmos,cosmos_2,cosmos_3} and ASTRODEEP \citep{astrodeep_1,astrodeep_2} and for wide nearby surveys (over 200-1000 deg$^2$)
especially SDSS \citep{sdss} and GAMA \citep{gama_1,gama_2}.
However, due to size of the data and complexity arising from the variety of
observatories used, little concerted effort has been made to assemble the
deep surveys over 10-1000 deg$^2$. These surveys are particularly important as
they are large enough to probe representative  ranges of environments and to
provide large statistical samples to  fully explore the range of galaxy
phenomena in detail including rare and transitory phenomena.

Dealing with this complexity and volume of data is not trivial. It requires cross-matching hundreds of millions of objects observed at different bands, identifying spurious sources in a robust and reliable manner and this needs to be done consistently across all fields with varying depths and bands. Dealing with such volumes of data is also memory intensive and requires huge compute power to process the resulting far infrared (FIR) photometry, photometric redshifts, and SED fitting. 

This paper presents the \Herschel\ Extragalactic Legacy Project (HELP) Data Release 1 (DR1) and details the pipelines and methods used to tackle the fore-mentioned challenges of complexity and volume size inherent to collating large, deep heterogeneous survey data. This paper follows specific HELP papers detailing specific parts of the project \citep[e.g.][]{Hurley:2017lr,Duncan:2018a,Duncan:2018b,Malek:2018,Shirley:2019} and science results using data from DR1 \citep[e.g.][]{Scudder:2016,Duivenoorden:2016,LoFaro:2017,Pearson:2017,Pearson:2018,Buat:2018,Scudder:2018,Donevski:2020,Duivenvoorden:2020,Mountrichas2021}. In Section~\ref{sec:fields} we define the HELP fields.   In Section~\ref{sec:strategy} we describe the overall HELP strategy. In
Section~\ref{sec:workflow} we describe the specific work-flow for DR1.  In Section~\ref{sec:results} we present some statistics of the data release. In Section~\ref{sec:discussion} 
discuss the uses of this data set and conclude.

\section[The HELP fields]
{The HELP fields}\label{sec:fields}

Many extragalactic surveys from different observatories and at different
wavelengths have been coordinated in their planning and execution. Each survey 
had different motivations and factors constraining their choice of field locations, sizes, and thus their individual footprint on the sky. The superset of all survey footprints would be large and include many areas with only a few data sets. The primary motivation for HELP is the \Herschel\ coverage so DR1 is limited to the main wide area extragalactic \Herschel\ surveys. 

Given that there is no imminent successor to  \Herschel\  the data
from that mission provides a legacy benchmark.  Within the \Herschel\ observatory
the SPIRE instrument \citep{Griffin:2010lr} mapped larger areas than the PACS
instrument \citep{Poglitsch:2010lr}. We thus define the boundaries of the project on the basis of the extragalactic surveys carried out with SPIRE. The specific \Herschel\ OBSIDS chosen to define the project are listed in
Appendix~\ref{appendix:obsids}. The footprint of these observations is conveniently captured in HEALPix Multi Order Coverage maps, MOCs \citep{MOC} which are provided online\footnote{\url{http://hedam.lam.fr/HELP/dataproducts/dmu2/}}. 

\begin{figure*}
  \centering \includegraphics[width=16cm]{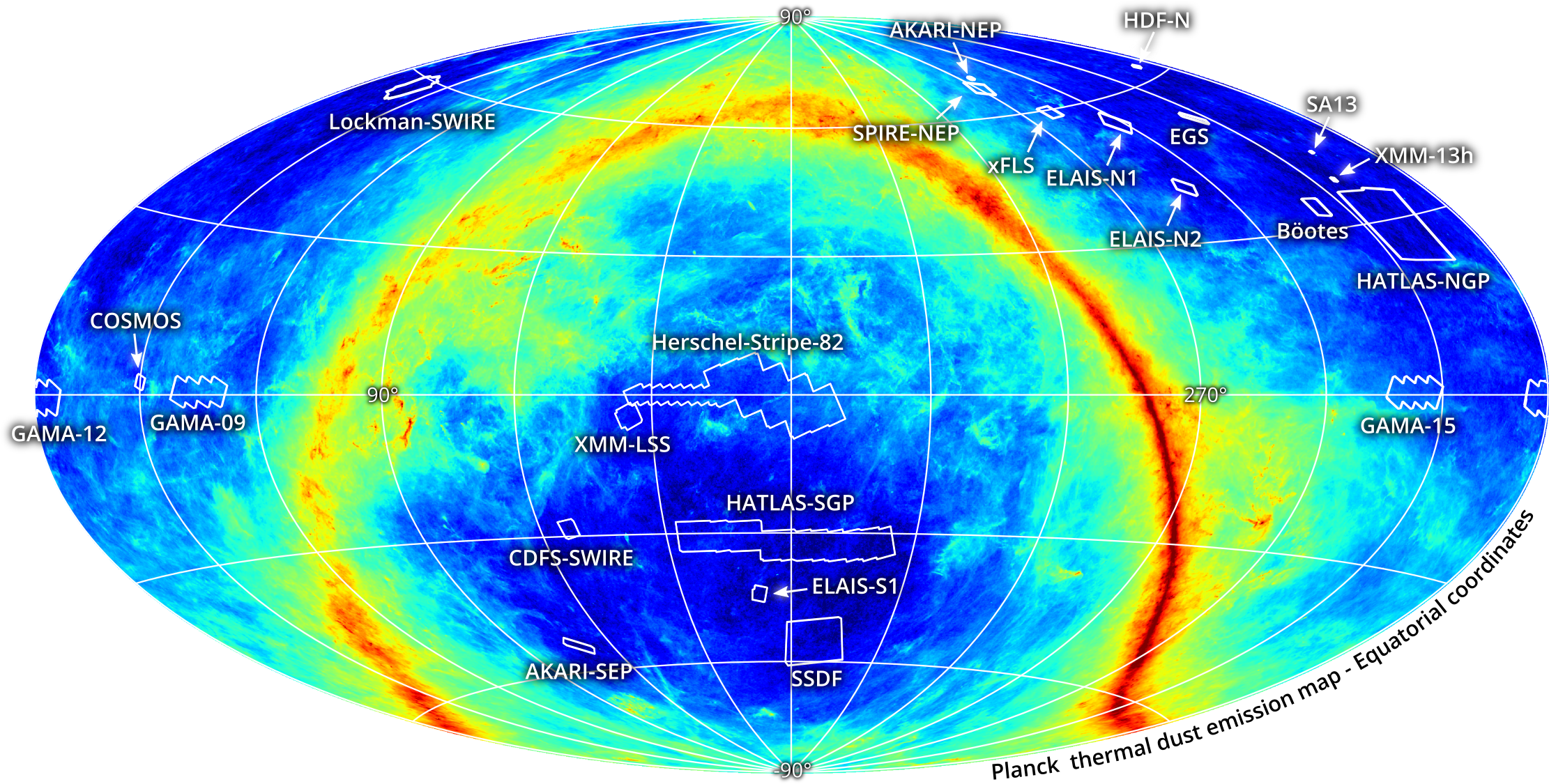}
  \caption[HELP Sky]{Projection of the HELP fields onto the dust emission from
    our own Galaxy. Reproduced from \citet{Shirley:2019}}.\label{fig:helpsky}
\end{figure*}

Some basic properties of the fields are tabulated in Table~\ref{tab:fields} and the  footprints are illustrated on a map of the Galactic dust from  Planck \citep{Planck:2014lr} in Figure~\ref{fig:helpsky}. As expected from the requirement of the infrared surveys and alignment with other multi-wavelength surveys,  we can see that the HELP fields: avoid the emission of dust from our Galaxy; are distributed in right ascension; have some concentration at the celestial equator; and include fields near both ecliptic poles.

 \begin{table*}
  \caption{Names, locations and areas of the individual HELP fields in alphabetical order.  The total area is 1269.1 square degrees.}\label{tab:fields}

\begin{tabular}{|l|r|r|r|r|r|r|r|}
\hline
  \multicolumn{1}{|c|}{Name} &
  \multicolumn{1}{c|}{RA} &
  \multicolumn{1}{c|}{Dec} &
  \multicolumn{1}{c|}{RA min} &
  \multicolumn{1}{c|}{RA max} &
  \multicolumn{1}{c|}{Dec min} &
  \multicolumn{1}{c|}{Dec max} &
  \multicolumn{1}{c|}{Area} \\
  \multicolumn{1}{|c|}{} &
  \multicolumn{1}{c|}{[deg]} &
  \multicolumn{1}{c|}{[deg]} &
  \multicolumn{1}{c|}{[deg]} &
  \multicolumn{1}{c|}{[deg]} &
  \multicolumn{1}{c|}{[deg]} &
  \multicolumn{1}{c|}{[deg]} &
  \multicolumn{1}{c|}{[deg.$^2$]} \\
\hline
  AKARI-NEP & 270.0 & 66.6 & 264.6 & 275.3 & 64.5 & 68.5 & 9.2\\
  AKARI-SEP & 70.8 & -53.9 & 66.2 & 75.4 & -55.9 & -51.7 & 8.7\\
  Bo\"otes & 218.1 & 34.2 & 215.7 & 220.6 & 32.2 & 36.1 & 11.4\\
  CDFS-SWIRE & 53.1 & -28.2 & 50.8 & 55.4 & -30.4 & -26.0 & 13.0\\
  COSMOS & 150.1 & 2.2 & 148.7 & 151.6 & 0.8 & 3.6 & 5.1\\
  EGS & 215.0 & 52.7 & 212.4 & 217.5 & 51.2 & 54.2 & 3.6\\
  ELAIS-N1 & 242.9 & 55.1 & 237.9 & 247.9 & 52.4 & 57.5 & 13.5\\
  ELAIS-N2 & 249.2 & 41.1 & 246.1 & 252.3 & 39.1 & 43.0 & 9.2\\
  ELAIS-S1 & 8.8 & -43.6 & 6.4 & 11.2 & -45.5 & -41.6 & 9.0\\
  GAMA-09 & 134.7 & 0.5 & 127.2 & 142.2 & -2.5 & 3.5 & 62.0\\
  GAMA-12 & 179.8 & -0.5 & 172.3 & 187.3 & -3.5 & 2.5 & 62.7\\
  GAMA-15 & 217.6 & 0.5 & 210.0 & 225.2 & -2.5 & 3.4 & 61.7\\
  HDF-N & 189.2 & 62.2 & 188.1 & 190.4 & 61.8 & 62.7 & 0.67\\
  Herschel-Stripe-82 & 14.3 & 0.0 & 348.4 & 36.2 & -9.1 & 8.9 & 363.4\\
  Lockman-SWIRE & 161.2 & 58.1 & 154.8 & 167.7 & 55.0 & 60.8 & 22.4\\
  HATLAS-NGP & 199.5 & 29.2 & 189.9 & 209.2 & 21.7 & 36.1 & 177.7\\
  SA13 & 198.0 & 42.7 & 197.6 & 198.5 & 42.4 & 43.0 & 0.27\\
  HATLAS-SGP & 1.5 & -32.7 & 337.2 & 26.9 & -35.6 & -24.5 & 294.6\\
  SPIRE-NEP & 265.0 & 69.0 & 263.7 & 266.4 & 68.6 & 69.4 & 0.6\\
  SSDF & -8.1 & -55.1 & -357.8 & -18.5 & -60.5 & -48.5 & 110.4\\
  xFLS & 259.0 & 59.4 & 255.6 & 262.5 & 57.9 & 60.8 & 7.4\\
  XMM-13hr & 203.6 & 37.9 & 202.9 & 204.4 & 37.4 & 38.5 & 0.76\\
  XMM-LSS & 35.1 & -4.5 & 32.2 & 38.1 & -7.5 & -1.6 & 21.8\\

\hline\end{tabular}
\end{table*}

The \Herschel\ Multi-tiered Extragalactic Survey (HerMES, \citealt{Oliver:2012})
is a major survey conducted by the \Herschel\ mission~\citep{Pilbratt:2010lr}
using the SPIRE \citep{Griffin:2010lr} and PACS \citep{Poglitsch:2010lr}
instruments covering 380 deg.$^2$. A number of important \Herschel\ surveys are contained within the footprint of the SPIRE data in HerMES, notably the PACS evolutionary Probe (PEP, \citealt{pep}). The largest and shallowest of the HerMES SPIRE tiers is the HerMES Large Mode Survey, HeLMS, which adjoins the 70 deg$^2$ HerS survey \citep{2014ApJS..210...22V} to form the largest contiguous extragalactic SPIRE field, which we refer to as the \Herschel\ Stripe 82 field. The largest SPIRE footprint comes from the \Herschel\ Astrophysical Terahertz Large Area Survey' (H-ATLAS, \citealt{Eales:2010lr}) which comprises 660 deg.$^2$ \citep{MSmith:2017}. Additional SPIRE coverage comes from: the \Herschel -AKARI NEP Deep Survey \citep{akari-nep}; and the SPIRE coverage of South Pole Telescope deep field (SSDF, \citealt{2013ApJ...771L..16H}); and the SPIRE calibration field in the North Ecliptical Pole.

The multi-wavelength data available in these fields is extremely rich.  This is important scientifically through providing the key for basic properties of the objects such as their redshift and probing different physical emission processes. The wealth of data is partly because the choice of these fields by the \Herschel\ teams was motivated by existing surveys. In addition, new surveys have been carried out through coordination between survey teams and an appreciation of the value of the accumulated data in these fields has encouraged independent surveys.  There are also many very large area surveys that overlap with these fields by accident. A primary goal of HELP is to collate these data sets together. The number of overlapping surveys is continually expanding, so the current collation can only be a snapshot.




\if

The first stage of the HELP pipeline is to build a catalogue of auxiliary optical to mid infrared catalogues that combine to make the \emph{prior list} for extraction of far-infrared fluxes. This positional cross match of over 50 public surveys is described in detail in \citep{Shirley:2019} and the process is summarised in section~\ref{sec:masterlist}. This catalogue forms the basis of all the forced photometry and will be referred to as the \masterlist\ throughout. Here we summarise the various input data sets and their overall depths.

\subsubsection[The depth of data]{The depth of data}

The depths of the input catalogues are of crucial importance in determining the depth of fluxes extracted from the far-infrared red images. They also determine the depths and completeness of photometric redshifts and must be properly accounted for in building statistical samples of galaxies.
\fi

\section[HELP Strategy]{HELP Strategy}
\label{sec:strategy}

The area, depth and wavelength coverage of the data in the surveys within the HELP fields have enormous potential for addressing important scientific questions, particularly addressing the questions of galaxy evolution. The volume of the Universe probed is phenomenal allowing studies of rare or transitory phenomena. This volume also provides large samples of galaxies that can be divided into meaningful sub-samples to test galaxy formation scenarios in more detail.  The variety of areas and depth allows probes of faint and distant galaxies and enables comparison between distant and nearby samples, i.e. to study galaxy evolution.  The volume also provides a complete sampling of the range of galaxy environments.
The wealth of multi-wavelength data allows for study of the different emission from  all the important physical processes e.g. stellar mass, star formation, active galactic nuclei and provides the basic information like positions and distances.

The challenge to realise this potential is that the information from different survey teams, from different wavelengths, from different facilities, and from different fields is not curated. This means that astronomers will tend to use a limited subset of the available data and also that basic analysis is unnecessarily repeated by many researchers.

The HELP strategy is to curate these data sets so that they can be used in their entirety by the whole astronomical community with the minimum of specialist knowledge and to add value to these data to enable more efficient and extensive scientific exploitation. 

HELP is designed to create a framework for wide-area multiwavelength studies that can be continuously updated with new observations. The scope for Data Release 1, DR1, is to curate object catalogues and photometry at near-IR and optical wavelengths that have been provided by the survey teams from images at mid to far-IR wavelengths alongside spectroscopic redshifts. HELP also provides tools to access the original imaging for manual inspection of interesting sources identified in the final catalogues based on their far infrared flux or physical properties.

The most fundamental element of the curation is by providing homogeneous data products with consistent measurements, units, and data formats.  We provide comprehensive meta data describing the data, using Virtual Observatory (VO) standards. In particular we provide the user with access to the original references and data from the survey teams, providing written descriptions of all the data in addition to machine readable files with links to papers, summaries of coverage and descriptions of instruments, including definitions of bands and links to transmission curves.

A key type of meta data for undertaking statistical studies of galaxy evolution is the selection function.  The selection function is the probability of an object being detected and included in a given sample as a function of the galaxy properties. Determining the form of these functions is a major challenge for collated surveys.  We therefore need to develop tools to reverse engineer the selection function and protocols to provide those to users.  We provide the following for capturing selections functions at increasing levels of sophistication:
\begin{description}
  \item{\bf Binary coverage maps:} These contain the basic information of where, on the sky, data exists.  We choose Multi-Order Coverage maps, MOCs \citep{MOC} to capture this.
  \item{\bf Depth maps:} These capture a simple, scalar, estimate of the depth of data at any sky position in a given band. We use HEALPix order 10 cells to provide a map of depths. This order can be changed and is chosen to be a compromise between attaining large enough samples to accurately measure depth and attaining a usefully high resolution. 
  \item{\bf Completeness maps:} These capture the probability that an object of a given intrinsic flux at a particular sky position would be included in a catalogue. These are derived from the depth maps and separately provided for photometric redshift availability. 
\end{description}

\subsection[Open Science]{Open Science}
The project has been implemented using open science frameworks with the following general principles:

\begin{itemize}
    \item All code is publicly available through a version controlled git repository. 
    \item Production code is embedded in  extensively annotated Jupyter Notebooks with integrated diagnostic plots.
    \item Every version of each data product is associated with the git commit code for any code used at the time of production
\end{itemize}

These key principles enable rerunning of any section of the pipeline in order to facilitate both verification and extension of the work by external researchers. By using Jupyter notebooks to document all the processing on GitHub, all the information about data quality is readily available and the code can be rerun with future additional survey data. 

\subsection{Tools}

The HELP philosophy is that astronomers can easily carry out their scientific investigations
without a high degree of instrument or survey specific expertise.  We have defined
some specific scientific use-cases which should be achievable at the end of the
project.  Our target is that these recipes could be used by a postgraduate
student to produce
meaningful scientific results.  Our intention is that all scientific results
from the team are easily reproduced using these tools.  Some of these tools are database operations.  Our database is VO enabled
with ADQL interfaces.  Some tools are traditional client/server interfaces.
Other tools are developed to provide containers (e.g. Docker) that the user
can download and run on their own CPU resources.  We supply extensive examples and documentation to aid the uptake of all the tools developed and presented here.


\section{The DR1 workflow}\label{sec:workflow}

\begin{figure*}
\centering \includegraphics[width=18cm]{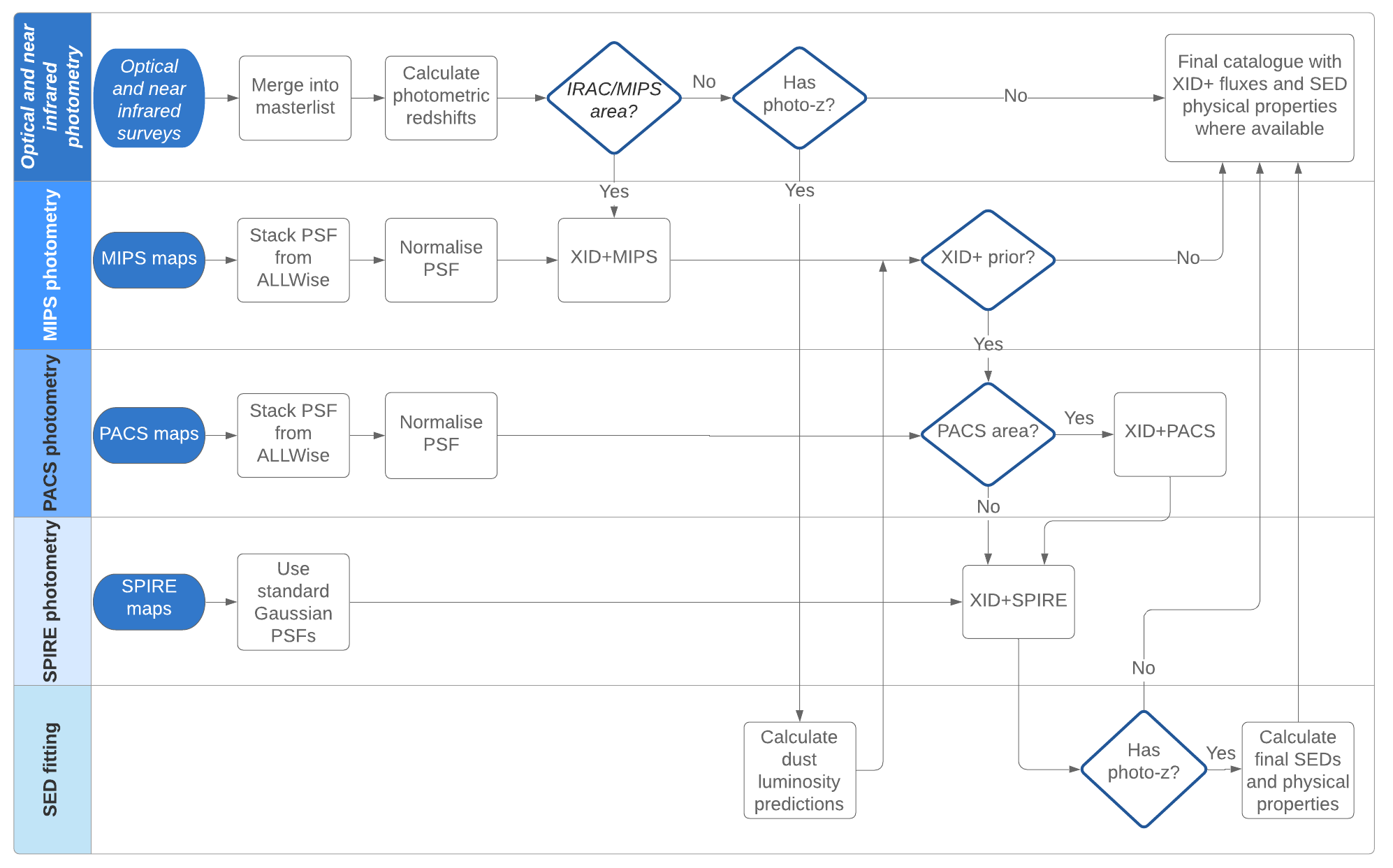}
\caption[HELP Sky]{Overview of the full workflow. All objects in the original \masterlist\ make it through to the final list regardless of which added value derived quantities are available. A given astronomical object from an input optical or near-infrared survey can be traced through the pipeline using this high level schematic. For each of the {\sc XID+} runs and final CIGALE SED run there are further criteria applied to each object that are not shown here for simplicity but are described in the relevant sections. }.\label{fig:pipeline}
\end{figure*}

In this section we describe the HELP workflow, outlining the key data analysis steps, the decisions taken and the outputs resulting from the workflow.
Figure~\ref{fig:pipeline} is a visual representation of the workflow which we summarise below, with additional details for each stage in the subsections that follow.

First, we create the \masterlist\ of astronomical sources and collate photometry measurements for these sources at all wavelengths between $0.36-4.5~\mu$m. Part of the photometry collation process involves determining the highest quality measurements available in a given field and wavelength region. 
In order for subsequent data processing to work effectively, there should be high quality photometry across a wide spread of wavelengths. 
This stage also allows us to investigate the depths available in a given area for a given band. 
Some of the fields in the HELP area have deeper surveys available and wider wavelength coverage than others. After the production of the \masterlist\ which includes all the compiled spectroscopic redshifts the catalogue is used to calculate photometric redshifts as described in section~\ref{sec:photoz}. These are required for spectral energy distribution (SED) modelling. 
The next stage is to produce the {\em prior list} which is required for {\sc XID+} forced FIR photometry. The forced photometry performed by XID+ takes the \emph{prior list} as a hard positional prior for objects that are most likely to be detectable in the FIR based on the optical to NIR photometry available. The exact selection of the \emph{prior list} is defined in section~\ref{sec:xid}. Objects with fitted FIR fluxes are then fed through to the final stage where SED modelling is used to calculate galaxy properties.

The final merged catalogue contains all the objects from the \masterlist\ and any subsequent quantities added by the HELP pipeline. The final catalogue can thus be broadly grouped into three hierarchical categories:
\begin{itemize}
    \item The \masterlist : Objects detected in an optical or NIR survey. 
    \item The \emph{prior list}: Objects included in the {\sc XID+} list of prior positions with FIR fluxes in any of the MIPS, PACS, and SPIRE bands available
    \item The \emph{A list}: Objects selected for SED modelling with {\sc XID+} detections, a photometric redshift, an SED model, and physical properties estimated.
\end{itemize}

We will now describe the details of each stage of the pipeline.

\subsection[Description of map workflow]{Mid and far-infrared images}\label{sec:maps}

This is the first time that all \Herschel\ extragalactic blank field survey images are presented together in an homogeneous form. We also provide the \emph{Spitzer} Multiband Image Photometer \citep[MIPS,][]{Rieke:2004} 24~$\mathrm{\mu}$m band images that are also used for computing forced photometry as part of the general pipeline. There are a total of severn mid or far-infrared (FIR) imaging bands presented here and used to compute forced photometry for far-infrared fluxes. These are the MIPS 24~$\mu$m band, the PACS 70~$\mu$m, 100~$\mu$m bands and 160~$\mu$m, and the SPIRE 250~$\mu$m, 350~$\mu$m, and 500~$\mu$m bands.

\subsubsection{Spitzer MIPS 24~$\mu$m images}
The MIPS images are from two different data sources depending on the field. The Spitzer Enhanced Imaging Products (SEIP) is a collection of Super Mosaics of Spitzer MIPS data. They are not presented in contiguous form but as individual sometimes overlapping images as originally provided in NASA/IPAC Infrared Science Archive\footnote{ \url{https://irsa.ipac.caltech.edu/Missions/spitzer.html}}. 
We do not mosaic them here because each image may reuse data and taking account of this requires decision which reduce the general applicability of the data sets.
The Spitzer Legacy Program was motivated by a desire to enable major science observing projects early in the Spitzer mission. Starting with 6 projects, the Legacy program was expanded to include 20 extragalactic projects over the cryogenic lifespan of Spitzer. The data products produced by the Legacy projects are monolithic images covering the full observed field and are used when available.  

As the data come from different Spitzer projects, the point spread function (PSF) for MIPS is highly variable across fields, therefore we compute the MIPS PSF for each field independently. The process for computing the PSFs is described in section \ref{sec:xid+pipeline}.

\subsubsection{PACS images}
PACS observations are available for a sub-set of the SPIRE area. These observations were sometimes taken in parallel with SPIRE observations 
(typically larger fields, due to the offset between detectors),
while some fields were taken with PACS alone. We have re-processed all the PACS data, to create an optimised set of images across all fields. The timeline data are initially processed using the \Herschel\ Data Processing System \citep[HIPE,][]{Ott:2010}, and a basic `PhotProject' image created per observation.
We correct the individual observations for any shift in astrometry by stacking the images on the position of WISE sources. The measured RA and Dec shift are applied to the timelines and the data exported using the UniHipe plugin. The final map for the field is created using {\sc Unimap} \citep{unimap_2,unimap_3,unimap_4,unimap_5,unimap_6,unimap_7} combining all available ovservations. The PSF for PACS is made using the same procedure as for MIPS.

\subsubsection{SPIRE images}

Each of these projects had different processing pipelines but had similar procedures for producing images from the instrument timelines described in \cite{Oliver:2012,Chapin:2011lr,Levenson:2010lr,Viero:2013rt,Viero:2014lr, MSmith:2017}. We have compared images produced by the H-ATLAS and HerMES pipelines using the same input data and found no significant differences.
The images presented and used here are also all homogenised to the same units and storage format. 
The SPIRE images presented here have a number of layers containing the homogenised image, the error image, and {\em nebulised} image. The {\em nebulisation} process removes large scale structure caused by cirrus with the method presented in \cite{MSmith:2017} using the {\sc Nebuliser} algorithm developed by the Cambridge Astronomical Survey Unit \footnote{\url{http://casu.ast.cam.ac.uk/surveys-projects/software-release/background-filtering}}.

\begin{figure*}
  \centering \includegraphics[width=16cm, angle=0]{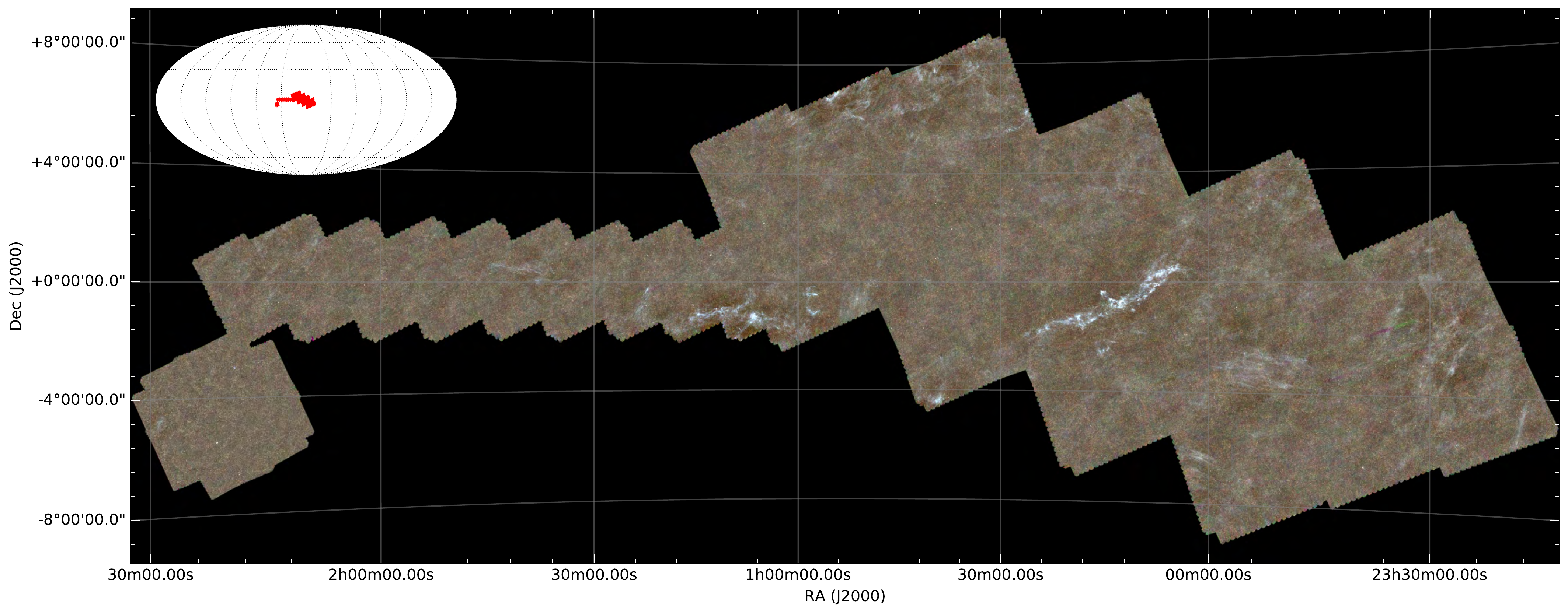}

  \caption[Three-colour image of \Herschel\ Stripe 82 region]{RGB representation of
    the \Herschel\ Stripe 82 and XMM-LSS field, with 250\um, 350\um and 500\um
    represented by blue, green and red respectively. This is the largest
    contiguous extragalactic region observed by \Herschel .  The maximum scale
    of the field from the East to West tips is  50\degr and the separation from
    edge-to-edge (following the zig-zag, roughly North-to-South) is 11\degr. The
    inset indicates the location of this region on an all-sky equatorial
    projection. The total area of  this field is 385 deg$^2$. Readily apparent
    is the strong cirrus structure throughout the map, including a ``seagull"
    like shape in the centre.  The data comes from three different observations
    (XMM-LSS, HELMS from HerMES \citealt{Oliver:2012} and HerS). This image
    was built for HELP from the processed SPIRE time-lines using the HerMES SMAP
    processing.}\label{fig:hs82}
\end{figure*}

\subsection[Masterlist]{The \masterlist }
\label{sec:masterlist}

The HELP \masterlist\ contains optical, near- and mid-infrared (e.g. \emph{Spitzer} IRAC) catalogues. It includes every source with a measurement in any band. 
A positional cross match is then used to combine the various wavelengths. Sources are flagged to indicate data coverage to discriminate lack of detection from lack of data. Full details of the cross-match criteria and mis-association fractions are given in \cite{Shirley:2019}. While cross match radii are determined for each input catalogue depending on positional errors it is typically around 0.4 arcseconds.
We also provide a table of the original catalogue IDs and the original catalogues. 
This means that where additional useful information is included in the input catalogue, it can be quickly recovered using the table of cross identifiers. 
All this data is provided in a simple and well documented structure to facilitate independent validation and external use.
Full details on the production of the \masterlist\ are presented in \cite{Shirley:2019} and through the code itself on Github.

The \masterlist\ is central to the HELP pipeline and data products. As the \masterlist\ progresses through the pipeline, some fraction of objects satisfy the criteria required for the next additional processing stage. 
For instance, if there are not sufficient optical photometry points then the photometric redshift (photo-$z$) calculation is likely to fail and the object will not have a photo-$z$, and so cannot be used for SED fitting. 
Likewise, when \textsc{XID+} is run, some objects will have low signal to noise so will not be `detected' and will not be used for SED fitting.
We aim to model each of these selection effects so that the full selection function can be understood as accurately as possible. \citet{Shirley:2019} provides depth maps in order to model the detection of objects in the original catalogues. Here we provide additional depth maps for both the photo=$z$ catalogues, which can significantly affect selection at the margins, and for the new photometry presented here.

\subsubsection{Star masks and artefacts}
Astronomical catalogues contains spurious artefacts resulting from instrument noise and dependent on the extraction methods. 
A key source of artefacts is bright stars, where the wings of the PSF or scattered light raise the background and spurious signals exceed the object detection threshold. 
We therefore developed a star masking pipeline which highlights regions likely to suffer from these artefacts. Our approach is to look at the excess number density of catalogued sources as a function of distance from bright stars. 

For our bright star list we select all Gaia \citep{lindegren2016gaia, brown2016gaia} stars with $g<16$~mag.
The reference band used on a given field is determined by which has most impact the \emph{prior list} for {\sc XID+}. 
In IRAC regions this is the IRAC 3.6~$\mu$m band, in other regions it is the deepest $K$ band. Within magnitude bins we then determine an effective exclusion radius, $r_{50}$, at which the excess number density (above the background level) drops to fifty percent of its peak. We choose fifty percent because the decline is steep and taking the mid point is a robust measure of its location. We then fit a linear relation: 
\begin{equation}
\label{eqn:starmask}
    \textrm{log}_{10}(r_{50}) = A + B M_{\textrm{star}}.
\end{equation}
\noindent This function defines the radius of a circle around each Gaia star, within which all objects are excluded from the \emph{prior list} and which should be excluded from all statistical analysis.  This function typically reduces to zero for all objects below 14 mag. We fit the parameters $A$ and $B$ based on the magnitude bins of size 0.5 and generate star masks for each field and target band independently. The final star masks are provided in the DS9 and MOC formats.


\subsection{{\sc XID+}: the probabilistic deblender for confusion dominated images}
\label{sec:xid}
For many \Herschel\ fields, in addition to the SPIRE images we also have \Spitzer\ MIPS 24 $\mathrm{\mu m}$ and \Herschel\ PACS 100 and 160 $\mathrm{\mu m}$ images that cover the mid to far-infrared part of the electromagnetic spectrum. However, due to the relatively large beam size of the these images compared to the galaxy density ($\approx 30$ per SPIRE beam for optical sources with $B < 28$), multiple galaxies can be located within the same instrument beam. This is referred to as the problem of source confusion.

To obtain accurate photometry from these infrared images, accounting for source confusion is essential. One way to solve the problem is to use prior information to accurately distribute the flux in the images to the underlying astronomical objects. For example, if we know the location of a galaxy to a reasonable tolerance (e.g. from an optical image where resolution is much smaller than the \Herschel\ beam), we may expect a galaxy to be found in the MIPS, PACS and SPIRE images at the same location. Typically the position of known objects have errors significantly less than the FIR point spread function such that we assume the positions are known precisely. 

As part of HELP, we have developed {\sc XID+} \citep{Hurley:2017lr} which uses a probabilistic Bayesian approach that provides a framework in which to include prior information and obtain the full posterior probability distribution on flux estimates. Obtaining the full posterior probability distribution is particularly important for getting accurate uncertainties on source flux. A given {\sc XID+} model is described as: 
\begin{equation}
d_j = \sum_{i=1}^{S}P_{ij} f_i + N (0, \Sigma_{\textup{inst}}) + N(B, \Sigma_{\textup{conf}}),
\end{equation}
where $d_j$ is the model of the map pixel $j$, $P_{ij}$ is the Point Response Function (PRF) for source $i$ in pixel $j$, $f_i$ is the flux density for source \textit{i} and two independent noise terms for instrumental and confusion noise: $N(0, \Sigma_{\textup{inst}})$ and $N(B, \Sigma_{\textup{conf}})$ respectively.

Rather than find just the flux values that maximises the likelihood, {\sc XID+} maps out the entire posterior, $p(f|d)$, which can be defined as:
\begin{equation}
    p(f|d)\quad\propto\quad p(d|f) \quad \times \quad p(f),
\end{equation}
where $p(d|f)$ is the likelihood, the probability of the data given the flux densities, and $p(f)$ is the prior probability distribution on the fluxes. The method is fully described in \cite{Hurley:2017lr}.

\subsubsection[HELP XID+ pipeline]{HELP {\sc XID+} pipeline}\label{sec:xid+pipeline}
HELP uses {\sc XID+} to carry out forced photometry on the \Spitzer\ MIPS and the \Herschel\ PACS and SPIRE images to produce catalogue fluxes for the HELP database. Our prior source list for these images are constructed using two different pipelines, which we describe in detail in the following paragraphs. For flux priors, we use uninformative flux priors (i.e. uniform flux prior bounded with reasonable limits derived locally from the image) to enable an unrestricted range of analysis with the HELP data products. More informative prior information would be preferable for specific science projects and is a powerful approach to extract more information out of the data \citep[e.g.][]{Pearson:2017, Pearson:2018}, however their use must be fully understood and taken into account, such that they are more suited for bespoke projects than for a data product. If not then apparent results might reflect the ancilliary data more than the far-infrared maps directly. In the next paragraphs we describe the steps followed to run {\sc XID+} across the HELP fields.   


Our list of prior sources is constructed from the \masterlist . Fitting all the sources in the \masterlist\ to the source confused infrared images results in fluxes that are degenerate without using more informative flux priors. We therefore have to limit the number of sources that go into our prior source list to those that are most likely to be detectable in the images. This approach fits the Bayesian philosophy of model building; build a simple model, fit to the data, evaluate, and finally improve the model where necessary. The prior source list is an integral part of our model for the infrared images. Limiting the number of sources to those that are most likely to be detected simplifies the model and the Bayesian P value maps described later, which provide a data product to carry out model evaluation. They identify where additional sources are needed to model the images. We depict the two ways we have constructed the prior source lists in Figure~\ref{fig:pipeline}, one for fields where there is \Spitzer\ and another for when there is no \Spitzer\ coverage.

For fields covered by \Spitzer , we use sources detected in any of the \Spitzer\ IRAC bands as they are known to be a good tracer for the \Spitzer\ MIPS images \citep{Rodighiero2006}. To remove any possible artefacts in the IRAC catalogues, we impose an additional constraint that sources must also have a detection in either the optical or NIR wavelength range. Sources that meet this criteria, constitute the {\sc XID+} \emph{prior list} for MIPS images. 

Once we have the output from {\sc XID+} on the \Spitzer\ MIPS images, we use our definition of detection level to select sources to be used for the {\sc XID+} \emph{prior list} for the \Herschel\ PACS and SPIRE images, which are fit independently (we do not use PACS {\sc XID+} detections as a prior for SPIRE given that the PACS data tends to be shallower than SPIRE). Detection is determined by the MIPS flux level where the Gaussian approximation to uncertainties is valid. Below a certain flux level, the map uncertainty is too large to be able to constrain the source flux and the flux posterior probability distribution for sources becomes skewed as the uniform flux prior dominates over the likelihood. This point is determined by manual inspection and given in Table~\ref{tab:xid+flux_cuts}.

For areas that have not been observed by \Spitzer\ IRAC, we compute a total dust luminosity ($L_{dust}=\int_{8-1000\mu m} B_{\lambda} d\lambda$) for each object using the CIGALE code, as described in Section~\ref{sec:physical_modelling}. We used the relationship between the ratio $\frac{L_{dust}}{f_{250}} $ and redshift, based on data from COSMOS field where we have both $L_{dust}$ and $f_{250}$. We apply this relationship to the sources for which we have dust luminosity and redshift predictions to estimate $f_{250}$.  Sources that have a predicted $f_{250}>5\mathrm{mJy}$ are added to the prior source list and {\sc XID+} is run on the \Herschel\ PACS and SPIRE images. This flux cut was chosen after running {\sc XID+} on a small region within \Herschel\ Stripe 82, using a range of flux cuts (e.g. using different \emph{prior lists}), and comparing the Bayesian $P$-value maps described in Section~\ref{sec:xid} to check whether the cuts applied to the prior source lists provided a good fit to the map. This was to manually check whether bright sources were being missed at a given predicted $L_{dust}$ cutoff value. 

For a number of fields, the dust luminosity relationship varies slightly due to an early bug in the prediction. Upon investigation, this has the affect of missing out 17 percent of the lowest flux sources that would otherwise have been included, whilst including 48 percent of sources that otherwise would have been removed from the source list. In this data release this will have the effect of introducing a selection effect and increasing the effective flux cut in predicted 250 $\mu$m flux. This will be propagated in to the effective forced photometry depth maps we provide and so will be automatically accounted for by modelling the selection using these maps.

As described in Section \ref{sec:masterlist}, star masks are used to define regions where bright stars cause large numbers of artefacts and spurious sources, resulting in these regions having no prior list sources. As {\sc XID+} is used for source de-blending rather than source detecting, it is not appropriate to apply {\sc XID+} to areas of the map where you have no prior knowledge of sources. We therefore exclude pixels from the area defined by the star masks from {\sc XID+} fitting using the Multi-Order Coverage maps (MOCs) built from the star masks. 

{\sc XID+} uses a Bayesian framework and so the flux parameters require a prior probability distribution. We use non-informative, uniform distributions with sensible limits. For \Spitzer\ MIPS images, the upper and lower 24 micron flux limits are based on the longest wavelength IRAC flux available. For a lower limit we take ${f_{\rm IRAC}}/{500}$ and for upper limit $f_{\rm IRAC} \times 500$. For \Herschel\ SPIRE and PACS, we set the flux prior lower limit to zero and source specific upper limit equal to the local (as defined by the PRF) maximum pixel value plus the absolute value of the prior mean for background plus two times the standard deviation of the background prior. This combination of maximum pixel value alongside value and width of the background prior gives a conservative but not extreme upper limit on the flux.

An important part of the model is the Point Response Function (PRF). The PRF is the convolution of the point spread function and the transfer through to a pixel response function via the detection and map building process. This fully maps the contribution a point source makes to each pixel. 

In \Herschel\ SPIRE images the PSF is assumed to be a Gaussian, with full width half-maximum (FWHM) of 18.15, 25.15, and 36.3 arcsec for 250, 350, and 500 $\mathrm{\mu m}$, respectively \citep{Griffin:2010lr}. This is convolved with the pixel space to produce the PRF. In the case of both \Spitzer\ MIPS and \Herschel\ PACS the PRF is calculated by stacking the flux of point-like sources from the astrometry corrected AllWISE catalogues, referenced into the GAIA reference frame. Morphological outliers (with a reduced $\chi^2>4$ in the AllWISE profile fit) were excluded before stacking.

The PRF obtained in the stacking will not be as high signal-to-noise as the instrumental PSF and will not track the extended wings of the PDF. To get the correct normalisation, we match the  curve-of-growth of the instrumental PSF to that determined from our PRF. 

Having defined the PRF, we use it to populate the pointing matrix, which describes how much each source contributes to each pixel in the map. It is calculated by taking the PRF for each band, centring it on the position for each source and carrying out a nearest neighbour interpolation to establish the contribution each source makes to each pixel in the map.

Running {\sc XID+} on the full images simultaneously is computationally unfeasible. We therefore divide the image into equal areas using the Hierarchical Equal Area isoLatitude Pixelization of a sphere (HEALPix). The resolution of the pixels are determined by the HEALPix level, with optimum order for \Spitzer\ MIPS and \Herschel\ PACS set at 11, and 9 for \Herschel\ SPIRE, which correspond to $\approx$ 1.718 arcmin, and $\approx$ 6.871 arcmin respectively.
When we fit each tile, the perimeter is extended by one HEALPix pixel with two levels higher resolution (i.e. level 13 for MIPS/PACS, level 11 for SPIRE with a resolution $\approx$ 25.77 arcsec and 1.718 arcmin respectively), so any sources that could contribute within the HEALPix pixel of interest are taken into account.


As with other MCMC fitting, we need to run chains long enough to ensure we converge locally and with multiple chains to ensure we have found a global minimum. We use the default number of chains, four, and discard the first half of each chain as `warm up' or `burn in'. In order to assess the convergence of each parameter we use the same diagnostics, $\hat{R}$ and $n_{\rm effective}$ used in \cite{Hurley:2017lr} and described in \cite{gelman2013bayesian}.

\subsubsection{HELP XID+ data products}


One of the key strengths of {\sc XID+} is that it maps out the posterior rather than just the maximum likelihood estimate. For individual objects of interest the full posterior can be used to verify the quality of the fit. This also allows using the joint posterior probability distribution of two correlated sources, getting the full uncertainty on the fluxes. The full posterior is stored in a \textit{`.pkl'} file for each HEALPix tile. This data can be provided on request.

    
The posterior distribution also allows us to perform a probabilistic check of the {\sc XID+} fit. When examining goodness of fits, the traditional method is to look at the residuals. i.e. $({\rm data} - {\rm  model})/{\sigma}$. Because we have many samples from the posterior, we can create a distribution of model images, that cover all the possible images {\sc XID+} generates from the posterior parameter values. Having a distribution of model images that we can compare to the original data provides a more robust check than if we were to use one, best fit model map coming from likelihood. Using a distribution of models drawn from the posterior, and comparing to the original data is called a posterior predictive check \citep{gelman1996posterior}.

\begin{figure}
	\centering
	\includegraphics[width=1 \columnwidth]{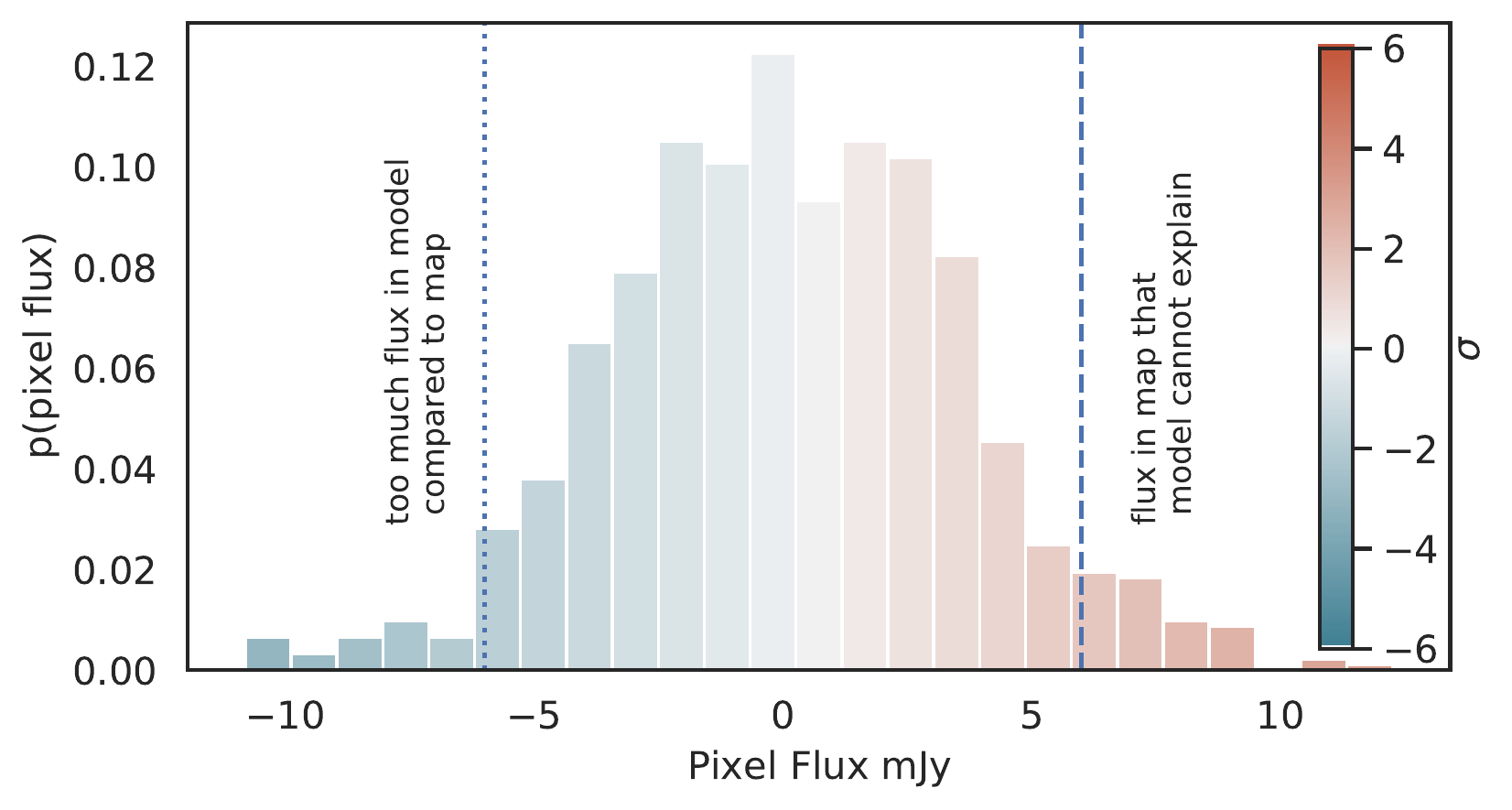}
\caption[pvalue]{An example distribution of posterior flux estimates for a single pixel. Comparison with the observation gives a measure of how unlikely the observed value is, given the posterior model, and is comparable to a residual. The resulting probability is converted to an equivalent Gaussian $\sigma$ level. A high value of $\sigma$ can be an indication of a missing source i.e. there is flux in the map that the model cannot explain. This can be a means to find interesting new  objects not seen in our prior catalogues.}\label{fig:pvalue}
\end{figure}

For {\sc XID+}, our approach to posterior predictive checks is to compare the observed flux of a pixel to the distribution realised in the model images. By calculating the fraction of model realisations that are above the observed value, we obtain the Bayesian $P$-value. 
A $P$-value of $\sim 0.5$ means the model is consistent with the data. Values close to $0$ shows that there is too much flux in the model compared to the map, whereas values close to $1$ indicates there is flux in the map that our model cannot explain. We convert these probabilities to a typical `$\sigma$' level. Figure~\ref{fig:pvalue} illustrates this process, which is repeated for every pixel in the map to produce Bayesian $P$-value maps for each band and field.

By using the full statistical power of the posterior probability distribution, these maps are more robust and less noisy than a traditional residual image and can be used to identify where the {\sc XID+} model assumptions break down. Examples of where {\sc XID+} might provide a bad fit are extended sources, which will appear in the Bayesian P value maps as a negative centre and positive rings. Another more interesting example would be missing sources: i.e sources in the map that are not in the \emph{prior list} which will appear in the Bayesian $P$-value maps as positive peaks. Obtaining a catalogue of sources from the Bayesian $P$-value maps can be added to the prior source list for {\sc XID+} to obtain updated catalogues or to find interesting objects which are drop outs in the lower wavelength images.

For the final catalogues we summarise the posterior flux distribution in the form of 16th, 50th, and 84th percentile This is equivalent to mean, and mean $\pm \sigma$ if the posterior distribution is Gaussian. These values are used for SED fitting and most science cases. We use the skewness level of the catalogue to determine a detection level, as described in \cite{Hurley:2017lr} and flag sources that are below this level. We also carry out model checking to identify whether the local (defined by the PRF) area of the map for each source is a good fit. To quantify this check, we define a Bayesian $P$-value residual statistic as follows; from the {\sc XID+} posterior probability samples, we use the same model images as our Bayesian $P$-value maps and calculate weighted residuals for the local pixels for every sample. We then calculate what percentage of our model images have a $\chi^2$ statistic greater than we would expect. This value is a probability, with zero indicating our inferred model will always provide a fit deemed good given the uncertainties, and 1 indicating our inferred model would always provide a fit deemed poor given the uncertainties. We provide this Bayesian P value residual statistic for each source and each band. Sources with a value $>0.5$ are flagged as unreliable. 

The final product consists of a catalogue with the flux percentiles, the median background, and the convergence statistics, Bayesian $P$-value residual statistic and a flag for sources that are below detection level or have a high Bayesian P value residual statistic. In addition to the final products, the {\sc XID+} example notebooks also provides some visualization tools to conduct a comprehensive analysis of the posterior. It is possible to create posterior replicated images and animations, create the marginalised posterior plots or reproduce the Bayesian $P$-value maps. There are examples available in the {\sc XID+} user guide.

\subsection[Blind catalogues, cross-matching and supplementary lists]{Blind catalogues }

An important additional step for providing a legacy data
set, suitable for community exploitation is to construct a catalogue of objects
detected in the SPIRE images without reference to any other data and with fluxes
extracted at the SPIRE wavelengths (a `blind' catalogue).  These catalogues give
a perspective of the sub-mm sky unaffected by any prior prejudice. Again, the most
significant challenge is the large SPIRE beam, leading to source confusion (e.g.
\citealt{Nguyen:2010lr}) which requires careful de-blending and the resultant
catalogues of sources do not necessarily correspond one-to-one to individual
galaxies. To enable statistical studies there are a number of key metrics required: positional and flux biases and accuracy; completeness and
reliability. Similar catalogues and metrics have been produced and made public
for the other HerMES fields \citep{Smith:2012lr,Wang:2013lr}, H-ATLAS fields \citep{Maddox2016}, and for all SPIRE fields in the \Herschel\ SPIRE Point source catalogue \citep{Herschel-spire-point2017}. We have produced new blind catalogues for all the HELP fields using a similar method to \cite{Chapin:2011lr}, and described below:

\subsubsection{Peak finding in Match Filtered images}
The blind sources are identified by searching for peaks in the matched filtered (MF) images, as they are optimised for identifying sources in source-confused images \citep{Chapin:2011lr}. We require peaks to have a flux density above the 85 per cent completeness level for each SPIRE band, where completeness is quantified as $1-\frac{N_{spurious}}{N_{real}}$ as a function of flux density. $N_{real}$ is the number of identified peaks in a given flux bin and  $N_{spurious}$ are the number of identified peaks in the negative version of a map (we assume the noise in the map is symmetrical about zero, thereby identifying peaks in the negative map will quantify the number of peaks that are from random fluctuations in the noise, as a function of flux). 

\subsubsection{Determining accurate positions}
Having found the peaks for each band independently, we determine accurate centres for each source by determining the best-fit flux density for the three SPIRE bands using inverse variance weighting and for sub-pixel steps around each peak. For each sub-pixel position, we calculate the Pearson correlation coefficient between best fit and map. The position with largest correlation is taken as the new position. We combine the three resulting SPIRE catalogues by removing duplicates at 350$\mathrm{\mu m}$ and 500$\mathrm{\mu m}$ using a nearest neighbour matching algorithm with 12 arcsec and 18 arcsec radius, respectively, adopting the position in the shortest wavelength available for each merged source.

\subsubsection{Determining fluxes with {\sc XID+}} 
Having obtained accurate positions and removed duplicates, {\sc XID+} is applied to the standard image map\footnote{we apply {\sc XID+} to the standard map rather than the nebulised map since it simultaneously fits for the background alongside the individual source}, using the merged, blind source matched filtered catalogue as the prior source list. The resulting {\sc XID+} catalogue, as described in section \ref{sec:xid} is merged with the blind source matched filtered catalogue (so as to preserve information about the sources used as priors for {\sc XID+}) to produce the HELP blind source catalogue.

\subsection{Spectroscopic redshifts}\label{sec:speczs}

As part of the HELP project we have collected spectroscopic redshifts from 101 different origins across all HELP fields, and created one merged catalogue for each field. Merging catalogues can be problematic due to the varying degrees of information and data quality available. Here we briefly describe the process used to make the matched catalogues.

\subsubsection{Spectroscopic redshift catalogue homogenisation}
The first step is to homogenise the individual redshift catalogues into a `standard format', where we extract the RA, Dec, redshift, and if available whether the spectra classify the object as a QSO or AGN. We also assign a `Quality Flag' ($Q$) to each spectra, using the information provided by the individual survey. For HELP we adopted the same $Q$ definition as used by the 2dF survey \citep{Colless2001}, and GAMA team \citep{Hopkins2013}, where $Q$ characterises confidence level on a five point scale.
In reality, assigning an exact $Q$ particularly for small surveys where reliability information is not given is difficult, therefore if a survey does not give an estimated reliability and claims the redshift is `good' or `reliable' it is normally assigned to $Q=3$. If available we also record the exact reliability given for the individual survey, as many large surveys have reliabilities higher than 90\% but not high enough to be assigned $Q=4$; this enables the user to set a individual reliability threshold (for example $>$95\%). For science studies we recommend using any redshifts with $Q \geq 3$. We also search the individual catalogue to ensure there are no duplicate entries by checking for no self-matches within 0.4\arcsec. Any duplicates were manually investigated and the best redshift kept following the procedure outlined below.

Before the catalogues for each field are merged, each catalogue is given a unique binary identifier (i.e., 1, 2, 4, 8, etc...). This means if the same object is observed by multiple surveys the source identifier numbers are added together, and the new source identifier can be used to see each catalogue that provided the corresponding redshift. For example a redshift with a source identifier of 11 would have been observed by surveys 1, 2 and 8 (but not 4). All individual redshift catalogues and their identifier are listed in Table~\ref{table:specZ}.

\subsubsection{Merging of spectroscopic redshift catalogues}
To merge the individual catalogues, we match each catalogue sequentially by using STILTS \citep{Taylor2006} to perform a sky match in a radius of 1--2\arcsec and checking up to 5\arcsec, where the radius is chosen to give the optimum matches. By performing a manual check of objects close to the matching radius the code can be modified so that any above/below the matching threshold should be merged/split. 

For any redshifts that have been found to be a match between the merged catalogue and the new individual catalogue the redshift with the higher $Q$ flag is kept. For the case where both $Q$ flags are $\geq 3$ a check is performed to see if the redshifts have a $\Delta z < 0.01$ and $<5$\% (for lower redshifts). If the two reliable redshifts disagree, a manual choice is made to decide the best redshift based on available information (i.e., if multiple sources, quality of data etc...). The fraction of conflicts between reliable redshifts is very small, for example in the SGP field we have 16 conflicts out of 47\,213 redshifts. For redshifts from PRIMUS \citep{Coil2011} due its lower resolution we increased the matching to $\Delta z < 0.03$ and used the higher-resolution spectra values.
For any merged galaxies the QSO/AGN flag is switched on if any of the QSO/AGN flags is set.

In total for HELP we collected 891\,317 spectroscopic redshifts, of which 713\,660 are considered reliable ($Q \geq 3$), and of these 621\,407 are unique sources. 
Table~\ref{table:dr1_overview} gives the number of reliable spectroscopic redshifts available for each field.

\subsection[Photometric redshifts]{Photometric redshifts}
\label{sec:photoz}
For the majority of HELP fields, where extensive multi-wavelength photometry is available, photo-$z$s have been calculated and are presented here for the first time. For 8 of the small HELP fields, where the best available optical photometry is provided only by all-sky photometric surveys and therefore is not improved by combining multiple surveys, we make use of photo-$z$s presented in the literature \cite{Zou:2019}. These account for under 1\% of HELP DR1 photo-$z$s.

\subsubsection{HELP Derived Photometric Redshifts}
Photo-$z$s for the prime HELP optical datasets are estimated based on the method presented by \citet{Duncan:2018a} and \citet{Duncan:2018b}. 
The approach combines multiple templates and machine-learning estimates to produce a hybrid consensus photo-$z$ estimate with accurately calibrated uncertainties. This method is only possible on fields with sufficient spectroscopic redshift samples as described below. 
We refer the reader to the original papers for a detailed discussion on the motivation and implementation. Below we summarise the implementation of this method for HELP. 

As in \citet{Duncan:2018a}, three different template-based estimations are calculated using the EAZY software \citep{Brammer:2008} with three different template sets: one set of stellar-only templates, the EAZY default library \citep{Brammer:2008}, and two sets including both stellar and AGN/QSO contributions, the XMM-COSMOS templates \citep{Salvato:2008,Salvato:2011} and the Atlas of Galaxy SEDs \citep{Brown:2014}. The individual template fitting results are optimized using zero-point offsets calculated from the spectroscopic redshift sample\footnote{The zero point offsets derived for each template set are stored and made available to the user for reference} and the posterior redshift predictions calibrated such that they accurately represent the uncertainties in the estimates. 
When sufficient spectroscopic training sets are available for a given field, additional Gaussian process photo-$z$ estimates \citep[\textsc{GPz};][]{Almosallam:2016a, Almosallam:2016b} are produced by training on one or more subsets of the \masterlist\ photometry. 
The multiple individual photo-$z$ estimates are then combined following the Hierarchical Bayesian combination method first presented in \citep{Dahlen:2013}, incorporating the additional improvements outlined in \citet{Duncan:2018a,Duncan:2018b}.

A key step in the Hierarchical Bayesian photo-$z$ method outlined in \citet{Duncan:2018b} is the separate treatment of priors and uncertainty calibration for known AGN. 
For HELP, we identified known AGN as follows:
\begin{itemize}
\item \emph{Optical AGN}: We identify known optical AGN through cross-matching of the \masterlist\ with the Million Quasar Catalogue compilation of optical AGN \citep{Flesch:2015}. 
Sources which have been spectroscopically classified as AGN are also flagged (see \ref{sec:speczs}).

\item \emph{X-ray AGN:} In HELP fields that have been targeted by deep pointed X-ray surveys, we make use of any publicly available X-ray catalogues and associated optical IDs to identify known X-ray AGN in the \masterlist . Outside of the publicly available deep X-ray surveys, we make use of the Second Rosat all-sky survey \citep[2RXS;][]{Boller:2016} and the XMM-Newton slew survey (XMMSL2)\footnote{\hyperlink{https://www.cosmos.esa.int/web/xmm-newton/xmmsl2-ug}{https://www.cosmos.esa.int/web/xmm-newton/xmmsl2-ug}} all-sky surveys.
X-ray sources were matched to their HELP \masterlist\ optical counterparts using the published AllWISE cross-matches of \citet{Salvato:2018}.
AGN and star-forming (or stellar) X-ray source populations are identified based on the colour criteria presented in \citet{Salvato:2018}:
\begin{equation}
 [W1] > -1.625 \times \log_{10}(F_{0.5-2\textup{keV}}) -8.8	,
\end{equation}
where $[W1]$ is the AllWISE W1 magnitude \emph{in Vega} magnitudes and $F_{0.5-2\textup{keV}}$ the 2RXS or XMMSL2 flux in units of erg$^{-1}$~s$^{-1}$~cm$^{-2}$.

\item \emph{Infrared AGN:} When \textit{Spitzer} photometry is available for a given field, IR AGN are also identified using the updated IR colour criteria presented in \citet{Donley:2012}.
\end{itemize}

Based on these criteria, \masterlist\ sources classified as AGN are flagged and processed following the AGN \citep{Duncan:2018a, Duncan:2018b}.
We note that by design these selections are not intended to provide complete samples of the AGN population within the HELP \masterlist.

For each HELP field we provide documentation outlining the specific \masterlist\ and spectroscopic redshift compilation versions that were used for the photo-$z$ estimation, calibration and validation.
We also document the precise set of optical filters included in the template fitting along with the lists of filter combinations used for \textup{GPz} machine-learning estimates (where available).

The photo-$z$ estimates produced by HELP are provided in two ways. Firstly, we provide the full calibrated photo-$z$ posterior for all sources for which the Hierarchical Bayesian procedure outlined above could be performed (as well as working examples of how to extract and use this information).
Secondly, we provide summary values for the posteriors in a format suitable for catalogues and single-value based quality statistics.
We follow the approach outlined in \citet{Duncan:2019a}, which aims to provide an accurate representation of the redshift posteriors, regardless of whether the posterior is uni- or multi-modal (as is often the case for photo-$z$s).
In summary, for each calibrated redshift posterior the primary (and secondary if present) peak above the 80\% highest probability density (HPD) credible interval (CI) is identified based on the redshifts at which the redshift posterior, $P(z)$, crosses this threshold.
For each peak, the median redshift within the boundaries of the 80\% HPD CI is calculated to produce our point-estimate of the photo-$z$ (hereafter $z_{1,\textup{median}}$ or $z_{2,\textup{median}}$). 
To present a measure of redshift uncertainty within the HELP catalogues we also then present the lower and upper boundaries of the 80\% HPD CI peaks (i.e. where the $P(z)$ crosses the threshold).
In the HELP catalogues, database and the subsequent analysis (e.g. Section~\ref{sec:physical_modelling}), photo-$z$s values are taken to be $z_{1,\textup{median}}$.

\subsubsection{Literature Photometric redshifts}
For the fields AKARI-NEP, AKARI-SEP, ELAIS-N2, HDF-N, SA13, SPIRE-NEP, xFLS, and XMM-13hr we use the photometric redshifts presented in \cite{Zou:2019} based on Legacy Survey $grz$ fluxes and Wise W1, and W2. These are fields without additional data sets to those presented there and therefore where recalculating them was of little additional benefit. These are matched into the \masterlist\ using a positional cross match with a radius of 0.4~arcsec. These redshifts are subject to a cut of $r < 23$~mag. After processing they also impose a redshift cut of $z<1$ and stellar mass cut $8.4 <$~log$(M_*)< 11.9$. These cuts impose limits on studies that can be conducted with these areas but lead to a well defined selection function. This redshift selection function can be automatically handled using the photometric redhsift depth maps. 

\subsubsection{Photo-z Validation}
\begin{figure} 
	\centering
	\includegraphics[width=\columnwidth]{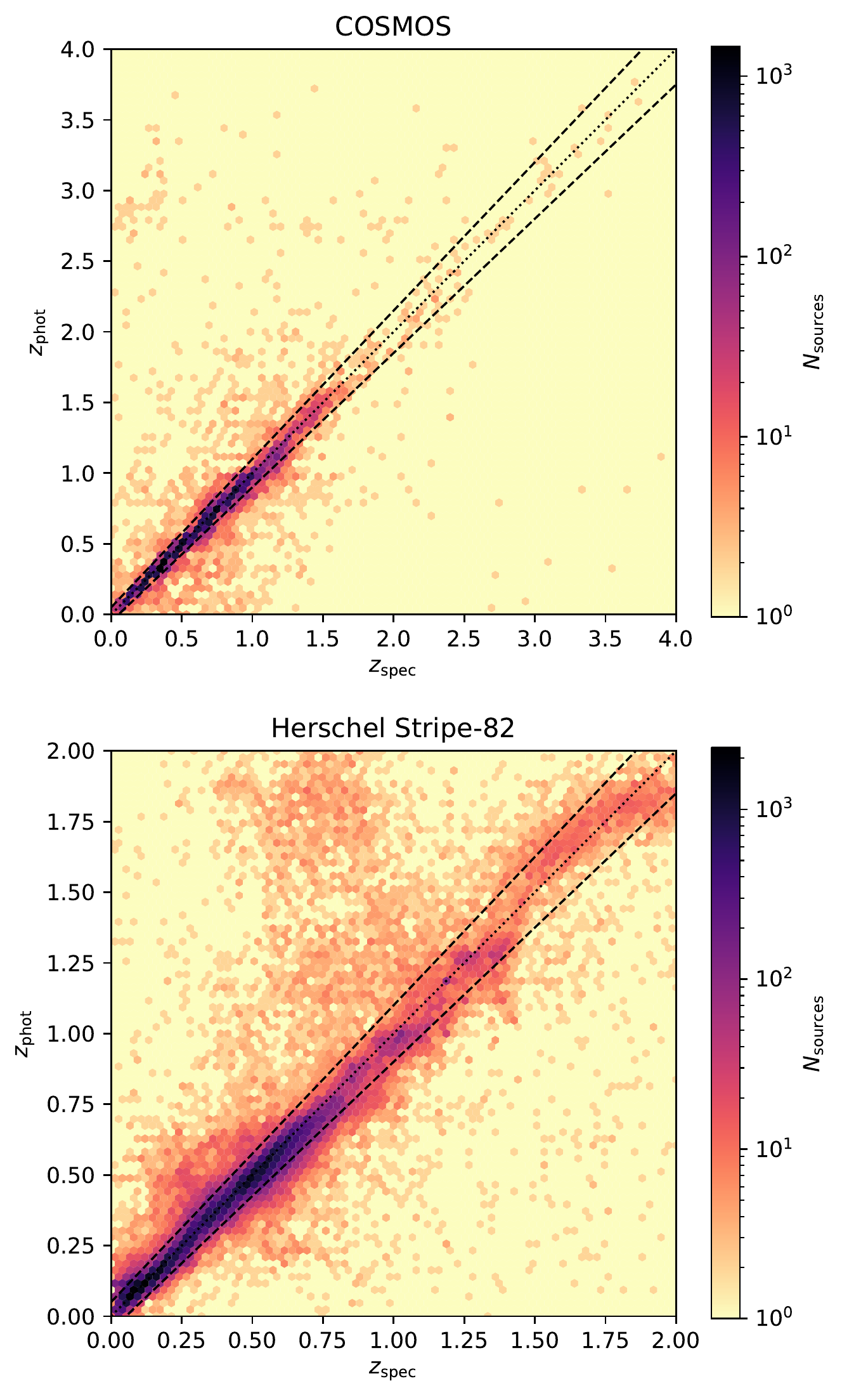}
	\caption{Consensus photometric redshift estimates as a function of spectroscopic redshift for two HELP fields at scales of $\sim10^{0}$ and $\sim10^{2}$ deg$^{2}$ (COSMOS and \Herschel\ Stripe 82 respectively). The dotted black line corresponds to the desired 1:1 relation while the dashed lines correspond to $\pm 0.05 \times (1+z_{spec})$, illustrating the typical scatter within the samples.}
    \label{fig:pz_specz_example}
\end{figure}

Due to the range in photometric data quality and spectroscopic training samples available in each field, there is significant variation in the photo-$z$ quality across the HELP footprint.
Fig.~\ref{fig:pz_specz_example} presents a qualitative illustration of the accuracy of photo-$z$ in fields that demonstrate the dynamic range in parameter space probed by HELP; the deep but small COSMOS field and the \Herschel\ Stripe 82 field that spans over 360 deg$^2$.
In both fields, we limit the sample to sources with reliable spectroscopic redshifts and detections in at least the optical and NIR regimes.
With additional selection criteria (such as magnitude selection and redshift limits), samples with reliable and precise photo-$z$ can be easily produced for each field.  
To facilitate this, as part of the photo-$z$ validation steps we generate a number of diagnostic and illustrative plots to enable assessment of the photo-$z$ quality within each field. 
However, we note that given the limited availability of spectroscopic data in some fields (and the overall variation in spectroscopic coverage), it is not possible to provide homogenous and complete assessment across the full HELP photo-$z$ sample.

\subsubsection{Photo-z Selection Functions}
Following the HELP goals and philosophy, we have also endeavoured to provide informative data products and tools for understanding and accounting for both the explicit and implicit photo-$z$ selection functions.

Given the in-homegeneity across both the full HELP sky and across each individual survey field, quantifying the spatially varying selection function is critical. 
However, due to the complicated nature of the optical selection function within deep fields that typically have very heterogeneous depth and filter coverage the selection will be highly multi-dimensional.
Additionally, as the exact selection function corresponding to a given sample are user and science-case dependent (i.e. depending on required redshift or photometric quality criteria), a novel and flexible approach is required.

Building upon the HEALPix based optical depth maps produced in the production of the optical \masterlist\ \citep{Shirley:2019}, we provide a set of tools to calculate photo-$z$ completeness across a given HELP field as a function of any desired \masterlist\ magnitude within the field.
Specifically, these tools allow simple calculation of the area within a field where desired photo-$z$ completeness is met (given the measured photometric quality as a function of magnitude). 
Or alternatively, the same tools can be used to calculate the magnitude selection at every position in the field that is required to meet a desired photo-$z$ selection criteria - accounting for variable photo-$z$ quality across a field due to heterogeneous photometric coverage.
Full details of the motivation and method will be presented in Duncan et al. (in prep.).
As part of HELP Data Release 1, we provide working example notebooks for both the generation and exploitation of these HEALPix based photo-$z$ selection functions.

\subsubsection{Future HELP Photo-z Plans}

The photo-$z$ estimation methodology applied to majority of HELP fields and sources \citep[i.e.][]{Duncan:2018a, Duncan:2018b} was developed with the aim of providing the optimum photo-$z$ estimate given the available data, regardless of source type.
However, the requirement for template fitting (and machine learning), hierarchical Bayesian combination and detailed posterior.
This difficulty is particularly acute in the largest and most complicated datasets (e.g. \Herschel\ Stripe 82) where significant high-performance computing resources were required for the runs. 

Going forward, we will therefore aim to move to a more scaleable and automated approach that can exploit the infrastructure provided by HELP.
The ingestion of the homogenised optical datasets into the HELP virtual observatory now opens up possibilities for optimised machine-learning based photo-$z$s that combine all spectroscopic redshifts available for a given combination of optical to near-IR photometry, regardless of field.
By combining this unified resource with, for example, updated GPz algorithm that naturally allow for redshift predictions in the case of missing data (Almosallam et al. in prep), it will be possible to provide comparable high quality photo-$z$ estimates in a more scalable manner.
Additionally, on-demand computation within the database could be provided following the approach presented in \citet{Beck:2017}.

\subsection[Physical Modelling]{Physical Modelling}\label{sec:physical_modelling}

Physical modelling of \masterlist\ sources with both a redshift estimate and FIR data was carried out using Code Investigating GALaxy Emission\footnote{\url{https: //cigale.lam.fr}} \cite[CIGALE,][]{Boquien:2019}. 
We refer to \cite{Burgarella:2005,Noll:2009,Boquien:2019} for a detailed description of the code, and \cite{Malek:2018} for a detailed description of the range of parameters used within the HELP fits but describe the key features here. 

CIGALE conserves the energy balance between dust absorption in the UV to near infrared domain and emission in the mid and far~IR when generating SEDs. The stellar emission is constructed
from composite stellar populations from simple stellar populations (SSP) combined with flexible star formation histories (SFH). Attenuation curves are then applied to estimate the fraction of energy from stars and gas absorbed and re-emitted by the dust using a dust emission template. 
CIGALE is a highly flexible code, containing multiple different modules for SSP, SFH, dust attenuation, dust emission and AGN component for users to apply or they can add their own modules. 

For the HELP project we selected low resolution  \cite{Bruzal:2003} single stellar population models, assuming a \cite{Chabrier:2003} initial mass function.
SFHs are chosen to have the so-called delayed $\tau$ parametrisation ($\propto t~e^{-t/\tau}$) form, with additional bursts to allow the SFH to model starburst populations on top of older (potentially quiescent) populations. 
The SFH is defined as:
\begin{equation}
\rm SFR(t) \propto \begin{cases}
SFR_{delayed}(t) & \text{if $t<t_0$}.\\
SFR_{delayed}(t)+SFR_{burst}(t) , & \text{if $t \geq t_0$}.
\end{cases}
\end{equation}
with $t_0$ being the age of onset of the second episode of star formation.

We perform SED-fitting runs with \cite{Charlot:2000} model as the dust attenuation recipe. 
The \cite{Charlot:2000} law assumed that stars are formed in interstellar birth clouds (BC), and after $\rm 10^7$~yr, young stars disrupt their ``nursery" and migrate into the ambient inter stellar medium (ISM).  
Both regions, BC and ISM, are characterised by a different power law -- one representing dust attenuation in the BC and the second in the ISM for older stars.  
As a result, the emission from young stellar population is attenuated first in the BC and then it goes through the dust in the ISM. 
Stars older than $\rm 10^7$~yr are attenuated only in the ISM. 
Different power slopes for BC and ISM can be used, but here we chose to keep both power law slopes of the attenuation fixed at
-0.7, as in the original \cite{Charlot:2000} work. 

To model the IR SEDs of the HELP galaxies, we use a dust emission model where the majority of the dust is heated by a radiation field from the diffuse interstellar medium, while a much smaller fraction of dust is illuminated by the starlight.
AGNs can substantially contribute to the mid-IR emission. To improve the accuracy of derived galaxy properties and because we have data coverage from optical to FIR, we add an Active Galactic Nuclei (AGN) component to the SED modelling, using the dusty torus models of \cite{fritz:2006}.

Based on the five components a large grid of models are fitted to the data. The number of parameter values depends on the properties of the field containing between 50 and 100 million individual models.
The physical properties published as a part of HELP DR1; dust luminosity ($\rm L_{dust}$), stellar mass ($\rm M_{star}$), and SFR, are built from the probability distribution function and for each parameter, the likelihood-weighted mean and standard deviations are calculated. These measurements have already been used for science purposes providing validation of the method against the literature \citep{Ocran2021}.

\begin{figure} 
	\centering
	\includegraphics[width=0.5\textwidth]{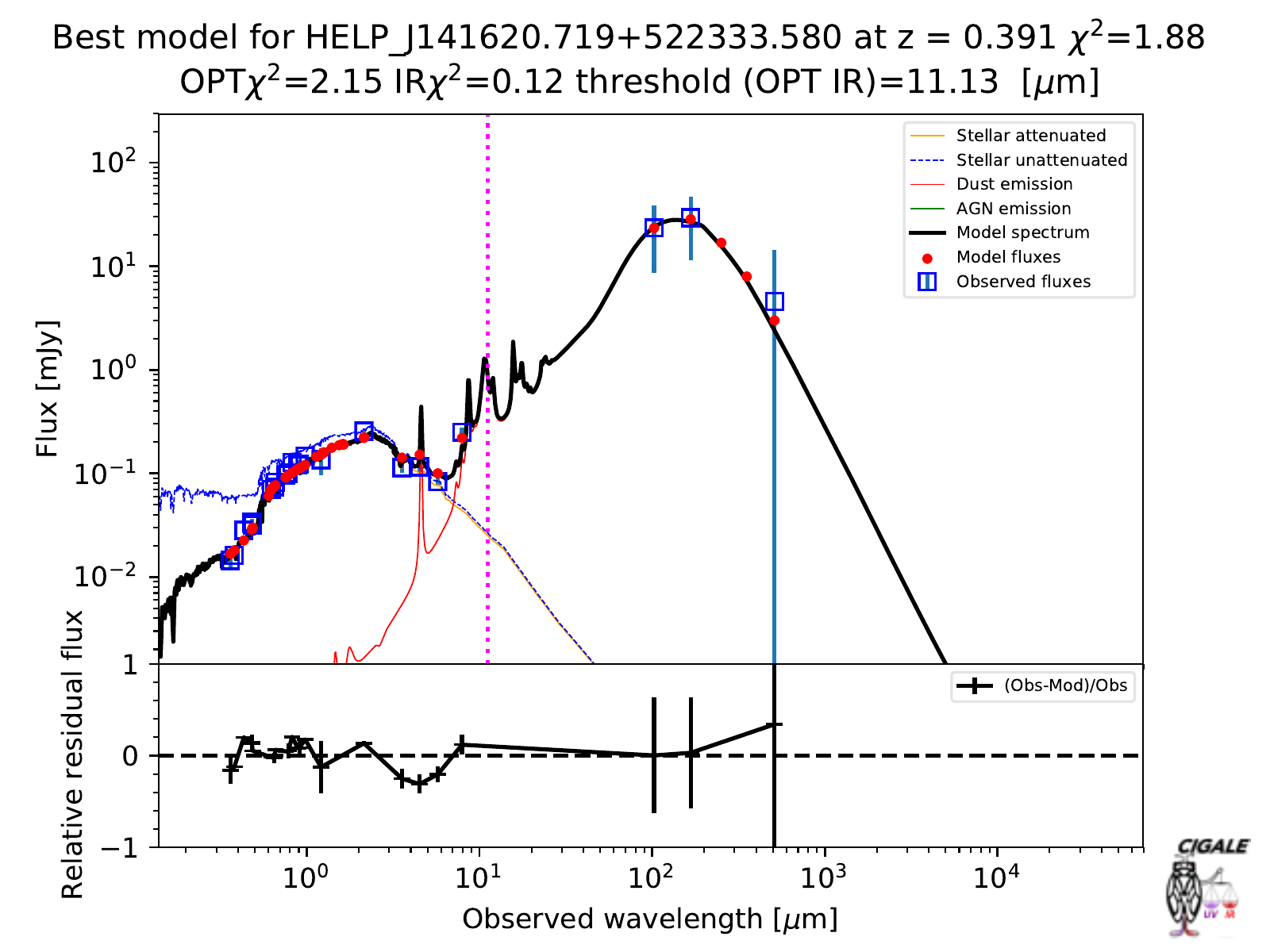}
	\caption{An exemplary fit of a galaxy from the HELP-EGS field. Observed fluxes are plotted with open blue squares. Filled red circles correspond to the model fluxes. The final model is plotted as a solid black line. Attenuated stellar component is plotted as a solid yellow line, while the unattenuated - as the dashed blue line. Red line mimics the dust emission.
	The relative residual fluxes, calculated as (observed flux - best model flux)/observed flux, are plotted at the bottom of each spectrum. 
	Magenta dotted line represents 8~$\mu$m rest frame wavelength (here 11.13~$\mu$m for redshift 0.391).
    Obtained physical properties of the presented galaxy are following: $\rm log(M_{star})=10.69\pm0.14[M_{\odot}]$, $\rm log(L_{dust})=11.15\pm0.14[L_{\odot}]$, and log(SFR)=$1.27\pm0.17[M_{\odot}yr^{-1}]$.}
    	\label{fig:CIGALE_KM}
\end{figure}

For each fitted galaxy we calculate two values of the reduced $\chi^2$ for the best-fit template and photometric measurements: $\rm OPT_{\chi^2}$ (for wavelengths lower than 8~$\mu$m restframe) and $\rm IR_{\chi^2}$ (for wavelengths larger than 8~$\mu$m restframe). Using these two
$\chi^2$ values enables identification of SED-fitting failures or peculiar objects \citep[see][]{Malek:2018}. An example CIGALE fit is presented in Fig.~\ref{fig:CIGALE_KM}.

CIGALE is run for all HELP galaxies with at least two optical and at least two near-IR detections. 
Additionally, to select the sample for which physical properties are estimated, we keep only sources with at least 2 far-IR measurements (signal to noise ratio $\geq$2). 
As the redshift is essential for the physical modelling process we used the photo-$z$s (as presented in Section~\ref{sec:photoz}). 
As SED fitting codes are sensitive to having one wavelength region overly weighted due to the presence of multiple measurements on a single passband.  Constraining power from multiple measurements at similar passbands will dominate over other bands. When multiple measurements are available in similar passbands we take the deepest only as determined by the signal to noise ratio. 

On average, based on the selection described above, we estimated physical properties of 1.7 million galaxies or 1\% off all objects in the \masterlist . 
Table~\ref{table:dr1_overview} shows the summary of the catalogues in each field, including the number of sources for which $\rm M_{star}$, $\rm L_{dust}$ and SFR were estimated. 
To make HELP's SED fitting easily reproducible, we published a dedicated version of CIGALE,  \texttt{cigalon}\footnote{\url{https://gitlab.lam.fr/cigale/cigale.git}} which contains the modules and parameters used to fit HELP's galaxies. This version can be found on the main CIGALE page.

As it was shown in \cite{Malek:2018}, CIGALE's implementation of the energy balance enables predictions of  $\rm L_{dust}$ for standard IR galaxies, which preserve energy budget, based on the UV to near infrared data only. 
Here we would like to stress that CIGALE cannot estimate monochromatic fluxes with reliable uncertainties, but only the total value of $\rm L_{dust}$. 
We used the predictions as priors for the IR extraction pipeline XID+. 
A similar method was used by \cite{Pearson:2018} for the star forming galaxies beyond the \Herschel\ confusion limit. 
To predict the $\rm L_{dust}$ for galaxies without IR measurements, we run \texttt{cigalon} using the same parameters and methodology as described above but without the AGN module as without mid-IR data we are not able to constrain a reliable AGN component.

\subsection[Database]{Database structure and access}

HELP data is distributed through the \emph{Herschel Database in Marseille}\footnote{\url{https://hedam.lam.fr}} (HeDaM) in addition to the \emph{Virtual Observatory at susseX}\footnote{\url{https://herschel-vos.phys.sussex.ac.uk}} (VOX). The former allows access to all raw data for reprocessing or direct handling. The latter permits fast queries over the full HELP area to generate samples for scientific analysis. Data accessed by code on GitHub can be found via its relative link on HeDaM such that the user can download the entire database and perform a full rerun. We also include meta data files in the Git repository with links to the corresponding data files. 

\subsubsection{HeDaM catalogues and images}

On HeDaM each data product is organised by field. For each field there is a final catalogue containing all the information from the optical fluxes to the infrared, the redshifts and the physical parameters derived with SED fitting. The HELP \Herschel\ SPIRE and PACS images are also present in addition to the blind sources associated with each. The database is designed to run across the entire HELP sky. Many scientific users will be interested in an individual field of interest. We have therefore provided overviews of each field to help a new user become familiar with the data presented there.

HeDaM also provides everything to reprocess the data exactly as described here. This facilitates rerunning the HELP work while changing some parameters or, for instance, adding a new optical catalogue to the process. As we have emphasised throughout, all our code is available on GitHub as Python modules and Jupyter notebooks.  For storage reasons, the data files are not included in the Github repository.  HeDaM contains a file storage with the exact same structure but with the data files within. For many science cases the final merged catalogues with summary information on every galaxy will be sufficient. 

\subsubsection{Virtual Observatory}

The Virtual Observatory at susseX (VOX) is a virtual observatory server built using the German Astrophysical Virtual
Observatory (GAVO) DaCHS software: \emph{Data Center Helper Suite}\footnote{\url{https://dachs-doc.readthedocs.io}} \citep{2014A&C.....7...27D, 2018ascl.soft04005D}. VOX contains both the images and the catalogue data. 

Images are available through the \emph{Simple Image Access Protocol} (SIAP). In particular, VOX makes is possible to get image cutouts at a given position.

The catalogue data is gathered into a single table across all the coverage that
users can query using the \emph{Table Access Protocol} (TAP) with compliant
software like TOPCAT \citep{2005ASPC..347...29T}, STILTS \citep{2006ASPC..351..666T} or
PyVO\footnote{https://pyvo.readthedocs.io}. This allows users to make
sophisticated queries or to remotely crossmatch their catalogues with HELP data.

The total catalogue containing all photometry measurements across all fields has around 500 columns
where each source may only have flux information in a small subset of these
bands. We therefore also provide a `best' photometry catalogue which contains the lowest error measurement in each $ugrizyJHKK_s$ band in addition to the far infrared fluxes. This reduces the number of columns to around 50. Due to the reduction in size it is also possible to index every column allowing fast queries. If the user requires the full photometry measurement they can then join their selection to the main table. For people working on a specific field they might prefer to download the full catalogue table on the field from HeDaM. VOX is particularly helpful when looking at samples scattered across a large area where it is unfeasible to download the full catalogue to perform cross matching.

\section[Results]{Results}\label{sec:results}

In this section we summarise the quality and sensitivities of the DR1 catalogue products. Table~\ref{table:dr1_overview} gives an overview of the numbers of processed objects and areas associated with each field. All fields have been fully processed through the HELP pipeline. Each field has different features which determine the depths and quality of forced far infrared fluxes and calculated physical parameters. Our aim is to allow these features to be modelled automatically across the whole HELP area as much as possible when users are constructing samples. Depending on the scientific question at hand, the desired sample properties can range from complete but heterogeneous samples over large areas/multiple fields to homogeneous samples within individual fields (and any permutation in between). Constructing the selection functions associated with these varying samples can be facilitated using the depth maps described and presented in \cite{Shirley:2019} and here with the additional far infrared bands.


%

\begin{table*}
\caption{Summary of the HELP catalogue numbers on each field. The blind sources are measured completely independently of the \emph{master list}. HDF-N, a very small field with extremely deep priors, has no objects selected for CIGALE fitting due to no objects being sufficiently bright in SPIRE bands to pass the signal to noise thresholds. This is reflected in the absence of a single blind detection on the field. Specialist fields such as this are subject to further development of more sophisticated priors. Starred photometric redshifts are from \protect\cite{Zou:2019}.}

\label{table:dr1_overview}
\begin{tabular}{l rrrrrrr}
\hline
Field & Objects & area deg.$^2$ & {\sc XID+} & photo-z & CIGALE & Blind & spec-z\\
\hline
AKARI-NEP & 531 746 & 9.2 & 31 441 & *107 228 & 1 239 & 9 848 & 1 243\\
AKARI-SEP & 844 172 & 8.7 & 108 119 & *139 059 & 566 & 20 169 & 362\\
Bo\"otes & 3 398 098 & 11.4 & 495 159 & 1 570 512 & 38 980 & 30 566 & 23 424\\
CDFS-SWIRE & 2 171 051 & 13.0 & 283 406 & 136 944 & 9 308 & 40 880 & 29 063\\
COSMOS & 2 599 374 & 5.1 & 25 898 & 691 502 & 15 747 & 12 603 & 36 686\\
EGS & 1 412 613 & 3.6 & 223 598 & 1 182 503 & 4 159 & 9 551 & 19 799\\
ELAIS-N1 & 4 026 292 & 13.5 & 269 611 & 2 714 686 & 49 985 & 34 501 & 4 619\\
ELAIS-N2 & 1 783 240 & 9.2 & 86 591 & *120 723 & 6 798 & 19 483 & 2 471\\
ELAIS-S1 & 1 655 564 & 9.0 & 194 276 & 1 013 582 & 25 393 & 22 743 & 10 396\\
GAMA-09 & 12 937 982 & 62.0 & 1 386 659 & 8 833 874 & 130 293 & 112 461 & 38 407\\
GAMA-12 & 12 369 415 & 62.7 & 1 099 477 & 8 569 951 & 108 139 & 112 471 & 41 149\\
GAMA-15 & 14 232 880 & 61.7 & 1 236 395 & 10 083 210 & 117 234 & 116 436 & 81 413\\
HATLAS-NGP & 6 759 591 & 177.7 & 1 233 547 & 3 166 952 & 185 290 & 344 635 & 58 476\\
HATLAS-SGP & 29 790 690 & 294.6 & 3 511 594 & 17 054 138 & 352 804 & 497 501 & 47 213\\
HDF-N & 130 679 & 0.67 & 834 & *7 435 & 0 & 0 & 3 360\\
Herschel-Stripe-82 & 50 196 455 & 363.2 & 2 976 447 & 21 509 448 & 250 644 & 232 589 & 132 358\\
Lockman-SWIRE & 4 366 298 & 22.4 & 242 065 & 1 377 139 & 46 719 & 54 106 & 7 243\\
SA13 & 9 799 & 0.27 & 812 & *2 884 & 70 & 315 & 188\\
SPIRE-NEP & 2 674 & 0.13 & 562 & *935 & 71 & 374 & 1\\
SSDF & 12 661 903 & 111.1 & 4 395 253 & 9 250 727 & 305 576 & 196 895 & 1 417\\
XMM-13hr & 38 629 & 0.76 & 3 563 & *10 773 & 670 & 1 218 & 365\\
XMM-LSS & 8 705 837 & 21.8 & 360 500 & 6 124 027 & 61 888 & 50 362 & 78 192\\
xFLS & 977 148 & 7.4 & 52 187 & *100 993 & 5 944 & 19 757 & 3 562\\
\hline
Totals: & 171 602 130 & 1269.1 & 18 217 994 & 93 769 225 & 1 717 517 & 1 939 464 & 621 407\\
Percentages: &  &  & 10.6\% & 54.6\% & 1.0\% & & 0.4\%\\
\hline

\end{tabular}
\end{table*}

\begin{figure*} 
\centering
\includegraphics[width=1.0\textwidth]{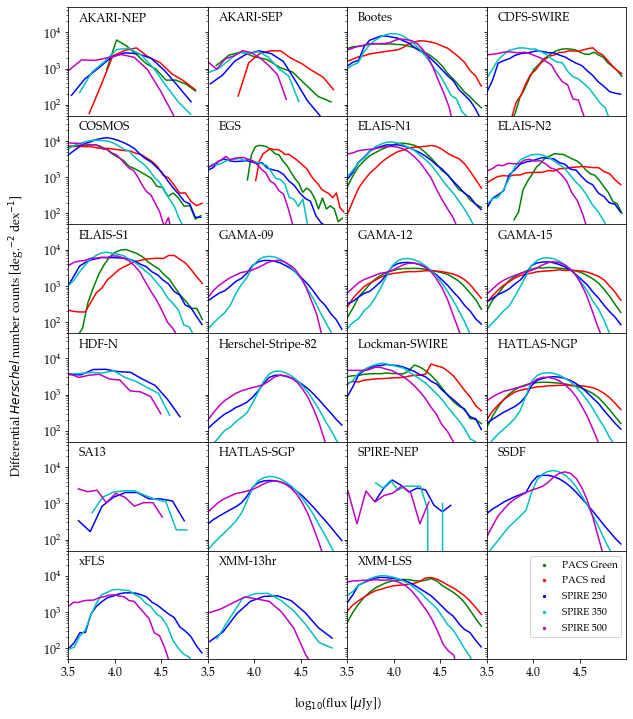}
\caption{\label{fig:numbers_allfields} The differential number counts in \Herschel\ bands on each field. Some of the small fields suffer from small number statistics. The normalisation factors for each field are computed based on the area over which there are any XID+ priors. The size of this area is dependent on what XID+ prior used for each field, and can cover less than half of the field depending (e.g. IRAC based priors).}
\end{figure*}

\begin{figure*}
\centering \includegraphics[width=18cm]{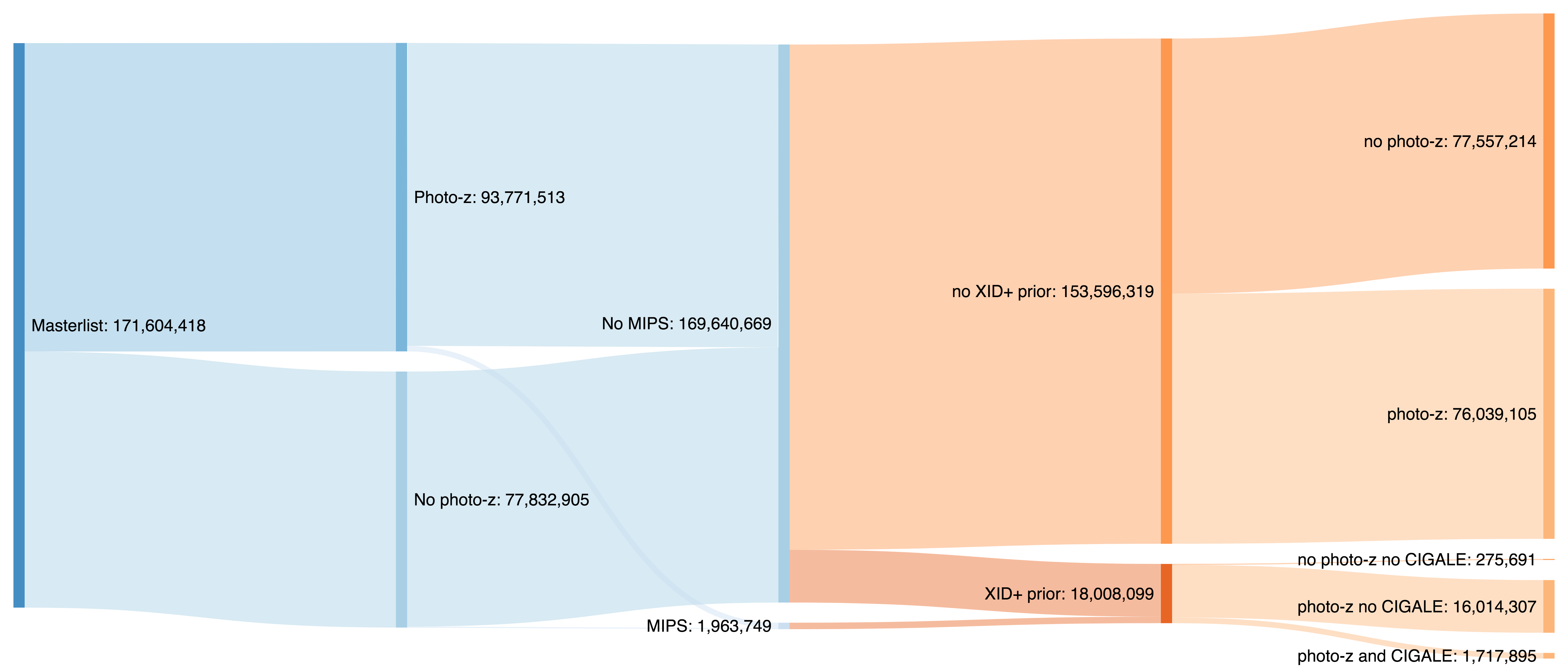}
\caption[Masterlist fractions]{Overview of the final numbers of objects showing the fraction that have a given measurement. There are broadly two types of object that are in the XID+ SPIRE prior; those with a MIPS detection in the MIPS area and those with an Ldust prediction prior in the regions without MIPS. The final stage shows the relative numbers which will constitute typical samples. }.\label{fig:masterlist_fractions}
\end{figure*}

Figure~\ref{fig:numbers_allfields} shows the differential number counts in the \emph{Herschel} bands PACS 100, PACS 160, SPIRE 250, SPIRE 350, and SPIRE 500. Figure~\ref{fig:masterlist_fractions} shows the relative fractions and numbers of each type of object in the final catalogue. Together these figures give an overview of the galaxy numbers as they pass through the pipeline. 

\subsection{Summary of \masterlist\ and depths}
The \masterlist\ number counts as a function of the optical and near infrared band magnitude on each field are summarised in \cite{Shirley:2019}. We also show the basic catalogue statistics here in Table~\ref{table:dr1_overview}. Defining depth for the forced photometry is dependent on the various factors contributing to the selection function. Here we take a similar approach to the depth maps described in \citet{Shirley:2019} and compute mean errors for objects with signal to noise above 2. This gives a metric analogous to the traditional notion of depth for optical and NIR detection images and allows us to compare limits on the faintness of objects across the full wavelength range. Nevertheless the effective far infrared flux depth is in turn dependent on the depths of all the bands contributing to the determination of the \emph{prior list}.

\begin{figure} 
	\centering
	\includegraphics[width=0.4\textwidth]{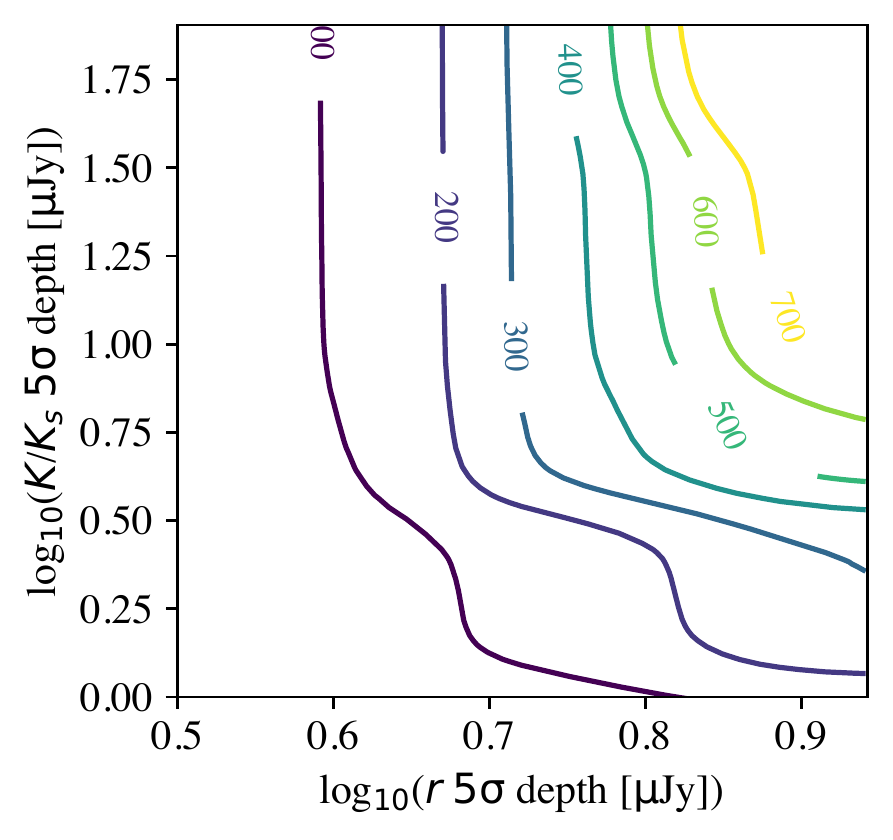}
	\caption{\label{fig:2d_r_k_depth} Area on the sky available to a given depth or deeper as a function of $r$ band and $K/K_s$ band depth.}
\end{figure}

\begin{figure} 
	\centering
	\includegraphics[width=0.4\textwidth]{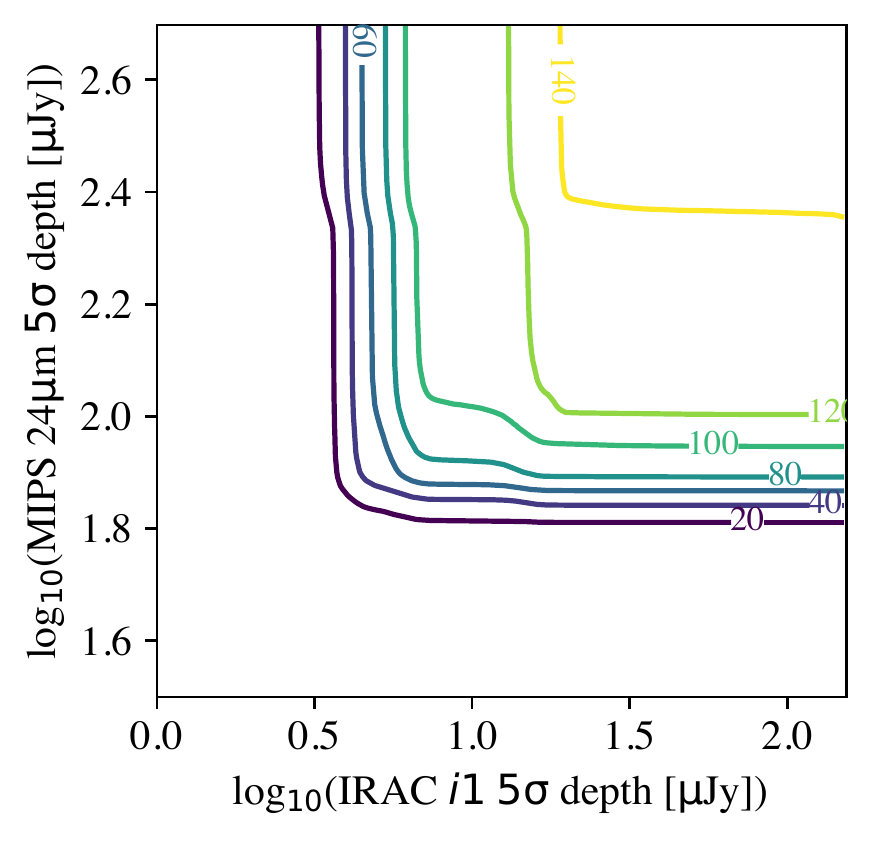}
	\caption{\label{fig:2d_i1_mips_depth} Area on the sky available to a given depth or deeper as a function of IRAC $i1$ band and MIPS 24$\mu$m band depth. }
\end{figure}

\begin{figure} 
	\centering
	\includegraphics[width=0.4\textwidth]{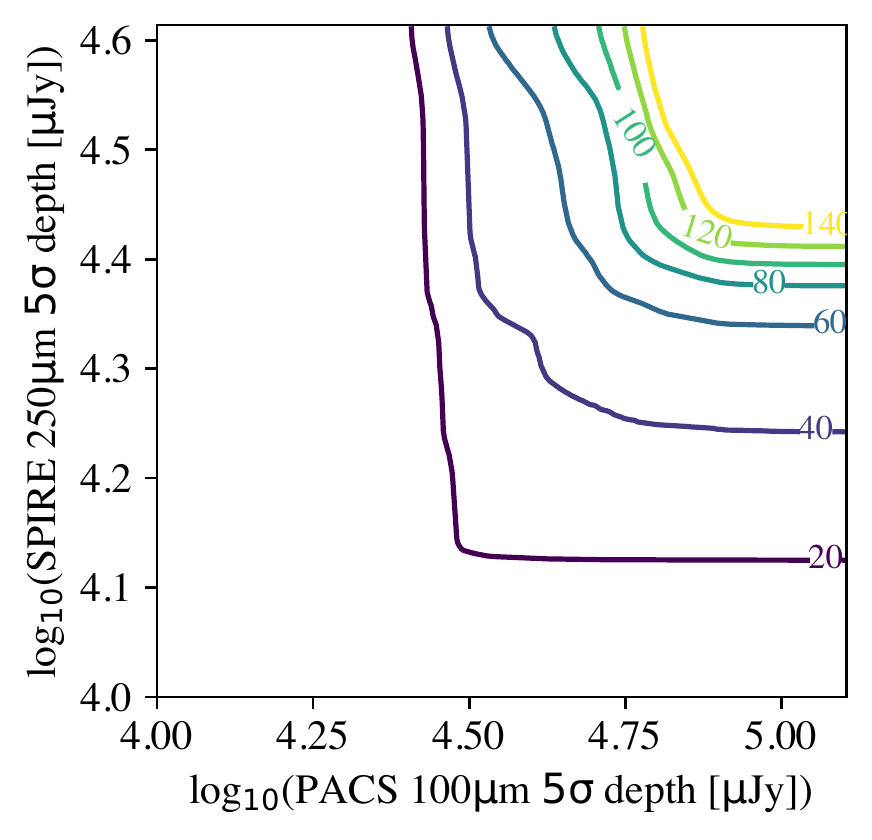}
	\caption{\label{fig:2d_pacs_spire_depth} Area on the sky available to a given depth or deeper as a function of PACS 100$\mu$m band and SPIRE 250$\mu$m band depth. }
\end{figure}

\begin{figure} 
	\centering
	\includegraphics[width=0.4\textwidth]{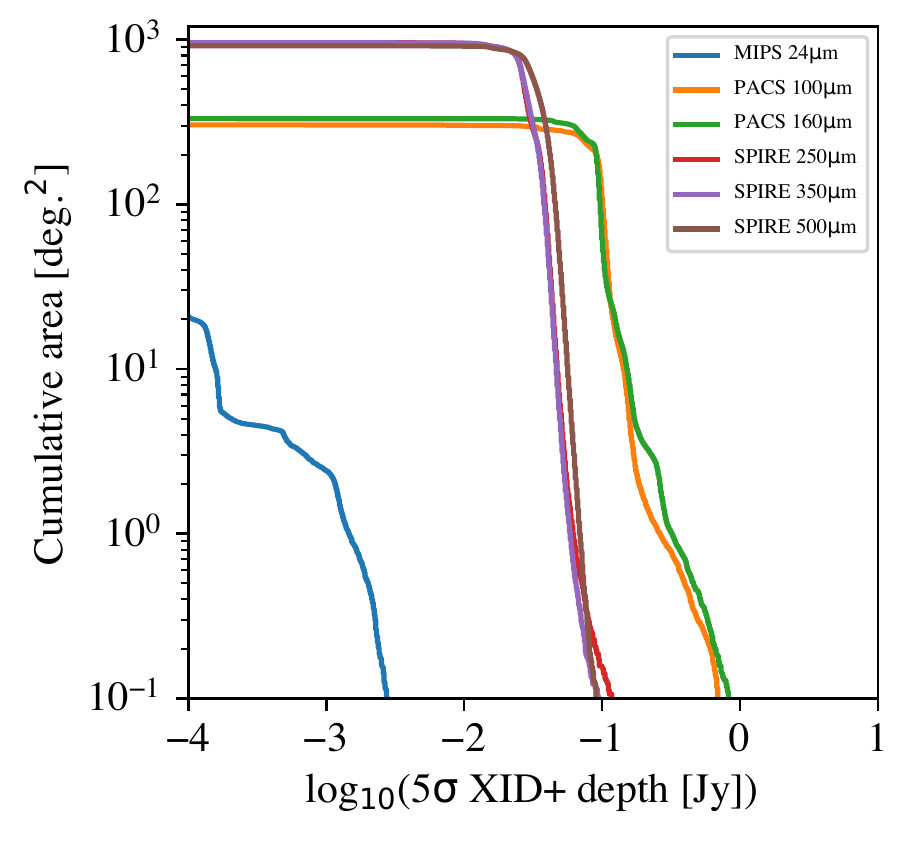}
	\caption{\label{fig:fir_cumulative_area_depth} Cumulative area to a given depth or deeper for the XID+ forced photometry for far infrared fluxes in the MIPS 24, PACS Green, Red, SPIRE 250, 350, and 500 bands. }
\end{figure}

\begin{figure*}
\centering
\includegraphics[width=0.9\textwidth]{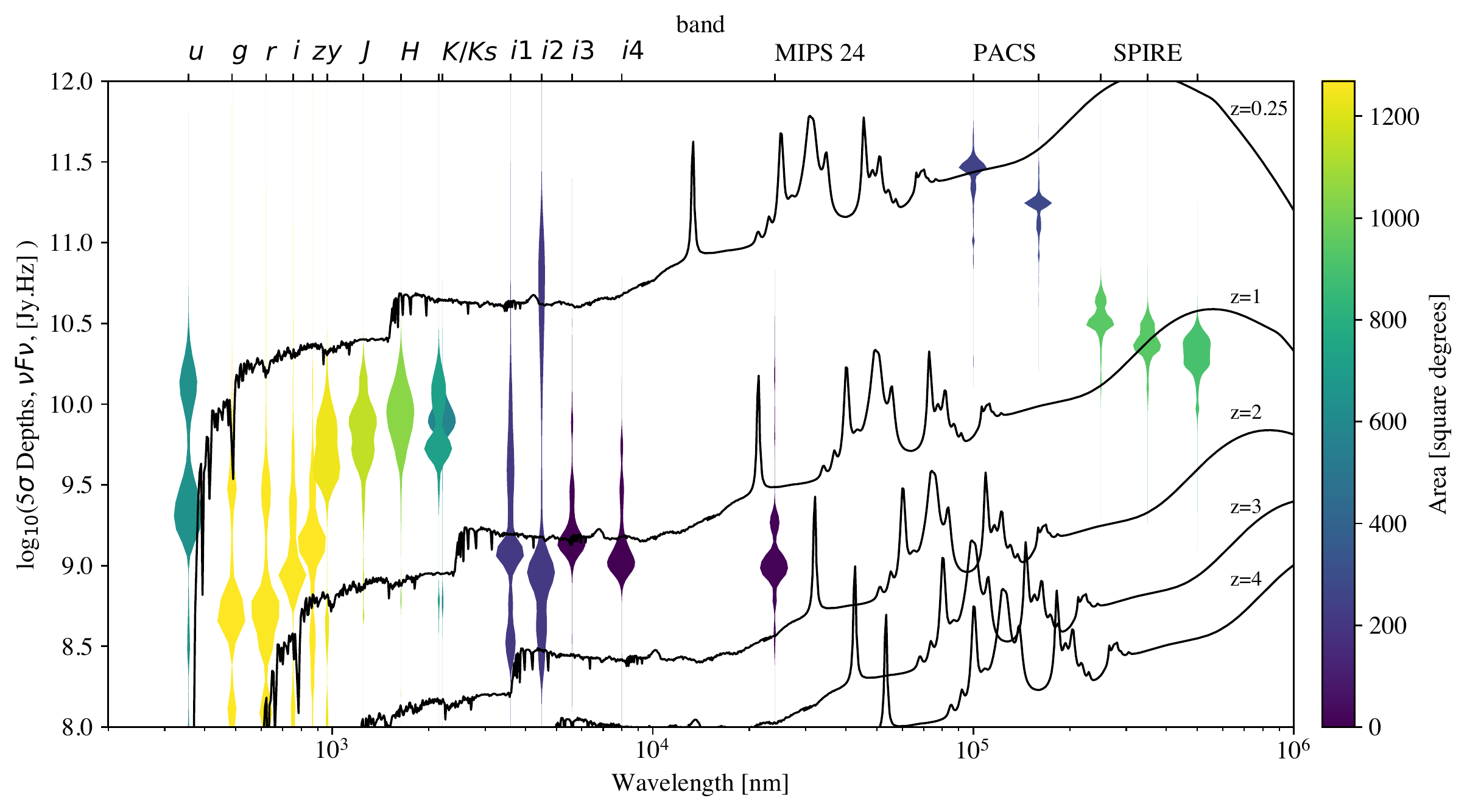}
\caption{\label{fig:bands_depths}The distribution of area on the sky to a given depth, $\nu F \nu$, shown via a violin plot, for each broad band type (taking the deepest specific band available in a given HEALPix cell). Optical and NIR depths are 5$\sigma$ depths in a 2 arcsec aperture. MIPS, PACS, and SPIRE depths are 5$\sigma$ depths for the {\sc XID+} forced photometry values. The colour of each area is determined by the total area that data for that band is available. We also plot a HELP spectral energy distribution for a ULIRG galaxy with star formation rate of 200~M$_\odot$/yr and a stellar mass of $10^{10}$~M$_\odot$ at various redshifts.}
\end{figure*}

Figure~\ref{fig:fir_cumulative_area_depth} shows the cumulative depths of MIPS, PACS and SPIRE coverage. Figure~\ref{fig:bands_depths} shows an overview of depths for each band compared to a typical  HELP ULIRG galaxy SED.
Figures~\ref{fig:2d_r_k_depth} to \ref{fig:2d_pacs_spire_depth} show areas available to a given depth or deeper as a function of various optical to FIR band depths to illustrate how band depth are correlated with each other. These figures show the complexity of selection effects and its high dimensionality, often being dependent on over five band depths in addition to the shape of the SED. The additional data products presented here facilitate selection function modelling. The method used to model selection effects must be targeted to a given science goal since a given physical property of interest will be impacted by different selection effects depending on how it is correlated with the relevant detection bands and requirements made by each processing stage.

\subsection{Summary of XID+ catalogues}

\begin{table*}
	\setlength{\tabcolsep}{.3em}
\caption{Flux cuts where Gaussian approximation to uncertainties is valid. (*)Ldust was used as prior.}
    \label{tab:xid+flux_cuts}
\begin{center}
	\begin{tabular}{c c c c c}
	\hline
		field          &  MIPS  &  SPIRE (250,350,500$\mathrm{\mu m}$) &  PACS (100,160$\mathrm{\mu m}$)\\ 
		\hline
        AKARI-NEP      &  30$\mathrm{\mu Jy}$ &   5, 5, 6 $\mathrm{m Jy}$  &   12.5, 17.5$\mathrm{m Jy}$   \\    
        AKARI-SEP      &  40$\mathrm{\mu Jy}$ &   --, --, --                 &   --, -- \\    
        Bootes         &  20$\mathrm{\mu Jy}$ &   5, 5, 10 $\mathrm{m Jy}$ &   10, 12.5$\mathrm{m Jy}$    \\    
        CDFS-SWIRE     &  20$\mathrm{\mu Jy}$ &   4, 4, 6 $\mathrm{m Jy}$  &   30, 30 $\mathrm{m Jy}$    \\    
        COSMOS         &  --$\mathrm{\mu Jy}$ &   ~$\mathrm{m Jy}$         &   ~$\mathrm{m Jy}$    \\      
        EGS(*)         &  --             &   --, --, --                 &   10, 10 $\mathrm{m Jy}$    \\      
        ELAIS-N1       &  20$\mathrm{\mu Jy}$ &   4, 4, 4$\mathrm{m Jy}$ & 12.5, 17.5 $\mathrm{m Jy}$    \\
        ELAIS-N2       &  20$\mathrm{\mu Jy}$ &   4, 4, 4 $\mathrm{m Jy}$ &    12.5, 17.5 $\mathrm{m Jy}$    \\
        ELAIS-S1       &  30$\mathrm{\mu Jy}$ &   4, 4, 6 $\mathrm{m Jy}$ &    20, 30 $\mathrm{m Jy}$ \\     
        GAMA-09(*)     &   --            &   4, 4, 6 $\mathrm{m Jy}$ &    20, 30 $\mathrm{m Jy}$ \\        
        GAMA-12(*)     &   --            &   4, 4, 6 $\mathrm{m Jy}$ &    20, 30 $\mathrm{m Jy}$ \\        
        GAMA-15(*)     &   --            &   4, 6, 10 $\mathrm{m Jy}$ &   20, 30 $\mathrm{m Jy}$ \\    
        HDF-N(*)       &   --            &   4, 4, 4 $\mathrm{m Jy}$ &   --                  \\    
        Herschel-Stripe-82(*) &   --    &   10, 10, 12 $\mathrm{m Jy}$ &   --                  \\    
        Lockman-SWIRE  &  20$\mathrm{\mu Jy}$ &   4, 4, 6 $\mathrm{m Jy}$ &    16, 25 $\mathrm{m Jy}$ \\    
        NGP(*)         &   --          &   6, 6, 10$\mathrm{m Jy}$ &   25, 25 $\mathrm{m Jy}$ \\    
        SA13(*)        &   --          &   4, 4, 4 $\mathrm{m Jy}$ &   --                  \\    
        SGP(*)         &   --          &   6, 6, 9 $\mathrm{m Jy}$ &   --                  \\    
        SPIRE-NEP     &  20$\mathrm{\mu Jy}$ &   6, 6, 6 $\mathrm{m Jy}$ &   --                  \\    
        SSDF(*)       &   --          &   10, 10, 10 $\mathrm{m Jy}$ &   --                  \\    
        xFLS          &  20$\mathrm{\mu Jy}$ &   4, 4, 4 $\mathrm{m Jy}$ &   --                  \\    
        XMM-13hr(*)   &  --           &   4, 4, 4 $\mathrm{m Jy}$ &   --                  \\    
        XMM-LSS       &  SWIRE: 20$\mathrm{\mu Jy}$; SPUDS: 10$\mathrm{\mu Jy}$ &    4, 4, 4 $\mathrm{m Jy}$ &   12.5, 17.5 $\mathrm{m Jy}$ \\    
                                                                                                                       
		\hline
	\end{tabular}
\end{center}

\end{table*}

The number of objects that have been deblended with {\sc XID+} (for MIPS, PACS and or SPIRE) in each field can be seen in Table \ref{table:dr1_overview}.
We provide the number counts in the PACS and SPIRE bands produced by the forced photometry presented here on each field in Figure~\ref{fig:numbers_allfields}. These number counts are jointly determined by the depth of the images and the \emph{prior list}. These numbers go beyond the confusion limit that determine the number counts for blind source extracted catalogues. The flux cuts for each field, described in section \ref{sec:xid+pipeline} can be found in Table \ref{tab:xid+flux_cuts}.

\subsection{Summary of blind catalogues}
The number of blind SPIRE objects in each field can be seen in Table~\ref{table:dr1_overview}. In total there are 1.9 million blind sources across the HELP fields. We have compared these objects to previous blind catalogues \citep{Smith:2012lr,Wang:2013lr,Herschel-spire-point2017} and find over 90\% overlap between them. The previous blind catalogues from HerMES and HATLAS have been used to identify numerous high star-forming high redshift galaxies \citep[e.g.][]{Riechers:2013lr, Asboth:2015lr} and candidates for follow-up programs \citep{Duivenvoorden:2018}. The HELP blind catalogues, across all 1270 deg.$^2$ are a data product for which many more of the rare, highly star forming, high redshift candidates can be identified. They are also a useful comparison with the forced {\sc XID+} catalogues in understanding the multiplicity of SPIRE sources \citep{Scudder:2016} and the impact of the prior catalogues used for forced photometry. 

\subsection{Summary of CIGALE Physical Properties}
The main physical properties were estimated for all galaxies  with known redshift (spectroscopic or photometric), with at least two detections in the optical range, two detections in NIR range and at least two FIR measurements (PACS and/or SPIRE) with SNR$>$2. We call this the \emph{A list}.
The CIGALE code was used to estimate dust luminosity, star formation rate, stellar mass and the AGN contribution to dust luminosity. 
Alongside the physical properties obtained from the full UV to far infrared SED fitting, three different  values of $\chi^2$s are provided: reduced $\chi_r^2$, quantifying the global quality of the SED fit for each galaxy, and $OPT_{\chi^2}$ and $IR_{\chi^2}$, described in section \ref{sec:physical_modelling}  \citep[detailed description can be found in ][]{Malek:2018}.
Those quantities can help to identify possible interesting sources or modelling failures. 

\begin{figure*}
\subfloat[\label{fig:redshift_cigale_dustlumin} Dust luminosity and star formation rate]{\includegraphics[height=0.35\linewidth]{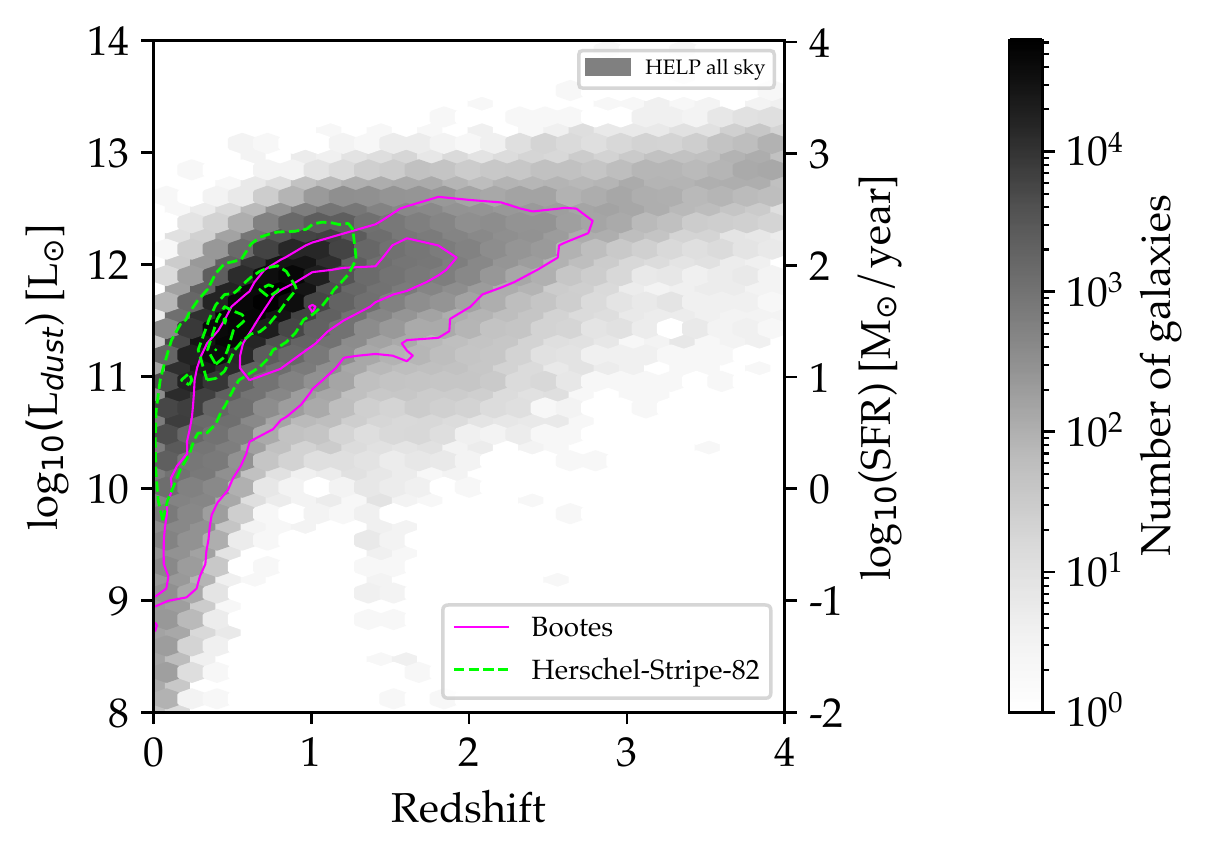}}
\subfloat[\label{fig:redshift_cigale_mstar} Stellar mass]{\includegraphics[height=0.35\linewidth]{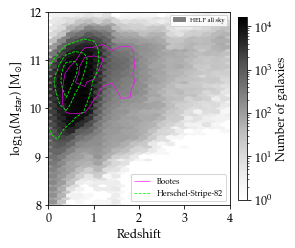}}
\caption{\label{fig:cigale_overviews} Overview of dust luminosity, star formation rate and stellar mass estimated using CIGALE SED fitting tool as a function of redshift across the full HELP sky. As an example we overlay contours between the wide \Herschel\ Stripe 82 field and the narrow deep field Bo\"{o}tes. Star formation rates are scaled to dust luminosity using the median ratio.}
\end{figure*}

Figure~\ref{fig:cigale_overviews} shows the distribution of estimated physical properties: dust luminosity, star formation rate, and stellar mass  as a function of redshift. These all-sky samples are drawn from vastly different areas of sky in terms of depths and areas.  
To illustrate this variation in dynamic range. we overlay contours between the wide and deep \Herschel\ Stripe 82 field and the narrow deep field Bo\"otes in Figure~\ref{fig:cigale_overviews}. 
Those figures show the uniqueness of HELP in providing data for extremely bright IR sources at high redshift, and normal star forming galaxies at low redshift. 
Around 90\% of HELP galaxies with SED fits have dust luminosity larger than 10$^{11}$~L$_{\odot}$. 
The majority of IR galaxies belong to Luminous Infra Red Galaxies (75\%), 15\% are classified as Ultra Luminous Infra Red Galaxies. 
Using our SED fitting procedure, we found more than 3\,500 (0.2\%) IR galaxies with dust luminosity larger than 10$^{13}$~L$_{\odot}$ -- those extraordinarily bright and active objects are still rare and not well understood \citep[e.g. ][]{MRR:2000,Farrah:2002,Wang:2020}.
The most extreme IR sources are also very active in star formation processes, especially at high redshift. This selection effect can also be seen in fig.~\ref{fig:redshift_cigale_dustlumin}. 


\section{Discussion and Conclusion}\label{sec:discussion}

Understanding galaxy formation and evolution requires measurements from many facilities to trace the different physical processes such as star formation and AGN activity in galaxies over cosmic time. These data need to cover large areas of the sky to obtain the large samples to characterise the population properties. These data sets need to be homogenous and well understood to enable the statistical studies required.

Construction of such a dataset has been the main goal of the HELP project. The corresponding data release documented in this paper describes HELP's first attempt at solving the challenges in the collation and merging of large, disparate, complex multi-wavelength datasets.  Focusing on the \Herschel\ fields for our first data release, DR1, HELP collates and homogenous optical and near-infrared catalogues over 1270 deg.$^2$ and applies novel methods to provide reliable photometry from the inherently source-confused far infrared images. Making as much use as possible of the \Herschel\ data remains an important consideration, given this data captures most of the CIRB over a wavelength range for which no new instrument is planned for the foreseeable future. 

\subsection{Enabling Science}
The ultimate validation of the HELP data will be through its utilisation for new scientific analysis.

The collation of data over many fields enables scientific investigations with larger statistical samples. \cite{Duivenvoorden:2020} used the collated \masterlist\ and \Herschel\ SPIRE images to investigate how the contribution to the CIRB varies with galaxies selected at different wavelengths.

The novel methods of extracting information from low resolution far infrared data also enables new investigations. \cite{Scudder:2016, Scudder:2018} used the full posterior probability distribution provided by {\sc XID+} to show how the multiplicity of \Herschel\ SPIRE sources changes with flux.

The large-area means that rare objects are included, and the wide variety of data enables these to be discovered. An example can be found in the hyper luminous obscured quasar recently discovered at z $\sim$ 4 \citep{Efstathiou:2021}. 

The far infrared data, uniquely available in the DR1 fields, permits studies of massive dusty star-forming galaxies \citep[DSFGs, e.g. ][]{Weiss:2013,Casey:2014}.  
These galaxies play an important role in understanding galaxy evolution because the dust-obscured star formation activity becomes more important at higher redshift \citep{Donevski:2020}. They are also crucial to understand how massive galaxies assembled \citep{Hamed:2021}. 
Furthermore, a large sample of infrared galaxies spanning a wide redshift range allow us to study the effect of dust attenuation on the physical properties of galaxies \citep{LoFaro:2017,Buat:2018, Buat:2019}

The HELP data products cover many significant extra-galactic survey fields, they are a valuable resource for existing surveys such as the Spitzer Extragalactic Representative Volume Survey (SERVS) and the DeepDrill survey \citep{servs:2012}. Ongoing and future surveys will also be able to exploit and build on the HELP data products. The LOFAR deep field team have used the HELP products directly \citep[e.g.][]{Gloudemans:2020,Smith:2020,Wang:2020} and together with HELP team, exploring the far-infrared radio correlation \cite{McCheyne:2021}. The MeerKAT International GHz Tiered Extragalactic Exploration (MIGHTEE, \citep{mightee}), is another ongoing radio survey with fields that overlaps with HELP and so products like the deblended FIR {\sc XID+} fluxes will be a valuable resource e.g. in deeper exploration of the far infrared-radio correlation.

The wide spectrum of physical properties can be also used to design or simulate upcoming large surveys such as the Legacy Survey of Space and Time (LSST) \citep{Riccio:2021}. The selected LSST Deep Drilling fields (ELAIS~S1, XMM-LSS, Extended Chandra Deep Field-South and COSMOS) are also covered by HELP and so data products these fields could provide immediate inputs to planning and direct contributions to early science.

The large area and depth of HELP can enable more robust analysis of the fundamental statistical relations. However such analyses must take into account the associated selection functions. Modelling the full selection functions for physical properties is now the focus of HELP research. Modelling how fluxes propagate through to redshifts and other physical properties. We have aimed to provide the necessary information throughout the HELP pipeline to make this more tractable. \cite{CamposVarillas:2021} will develop empirical methods to estimate the HELP selection function in order to probe the bright end of the stellar mass function of galaxies in \Herschel\ Stripe 82.

\subsection{Future data releases and how to contribute}
HELP's open approach means decisions made during the collation and production of DR1 are transparent and can be reproduced by other teams. It also provides a pathway for any astronomers to contribute and add to the HELP dataset in the future. As  optical, infrared and radio surveys produce new datasets, these will need to be combined with legacy surveys. The methods, pipelines and tools described in this paper provide a platform to enable their ingestion.

As further data sets are incorporated it will also be possible to improve the \emph{prior list} and the subsequent photometry and the SED fitting. The open source nature of the HELP pipelines and tools means these improvements can be done by anyone and are not dependent on the current HELP team. For DR1 we decided to generate the most versatile data products, with a non-informative prior, but specific science may benefit from using the full optical to mid infrared photometry and SED fitting to use more informative flux priors. 

For HELP to continue being a valuable resource there will need to be a transition from a centrally managed project to becoming a distributed community endeavour. Using version control systems such as GitHub which have enabled many collaborators to contribute to open source projects will be one way in which such a community can be fostered. We are working with other survey teams and training them in the use of the HELP pipelines and tools. We welcome new teams interested in combining their observational data, or value-added related datasets (e.g. alternative physical parameter catalogues, galaxy cluster catalogues etc) to contribute and integrate their data.

\subsection{Summary}
We present the HELP project which collates extragalactic surveys from the optical to \Herschel\ far infrared bands. This includes an open source software pipeline as well as the resultant data products. This first data release, DR1, can be used to study an unprecedented wide area of \emph{Herschel} sky. Some key highlights of this new data set are:

\begin{itemize}
\item We have collated data for 170 million objects from optical to far infrared over 1270 deg$^2$ of the prime extragalactic fields, with boundaries defined by the mapping of the \emph{Herschel} Space observatory.

\item We present far-infrared photometry for 18 million objects in an optical to mid infrared selected \emph{prior list} chosen to be tightly correlated with far infrared bright objects. We calculate a posterior distribution on the flux for all objects using Bayesian inference. 
\item We publish the main physical properties; stellar mass, dust luminosity and star formation rate based on spectral energy distribution modelling with the CIGALE code. This is done for all galaxies with at least two detections in each wavelength region for a total of 1.7 million objects, being 1\% of the total HELP catalogue.

\item The new catalogue is presented alongside an array of other data products, including newly homogenised images, supplementary catalogues and extensive tools for accessing and analysing these new data sets.

\end{itemize}


\section*{Author Contributions}
R. Shirley made the largest contribution to the data products, analysis and led the writing of text in this paper and is thus identified as first author.  Oliver defined the project, secured the funding and led the team  in the development and execution of the project and is identified in the senior author (last) position. K.~Duncan, P.D.~Hurley, K.~Ma\l{}ek, M.C.~Campos-Varillas,  Y.~Roehlly and M.W.L.~Smith contributed very significantly and equally to the production of the data products, the analysis and the writing. The other authors all contributed significantly to either the design of the project, some data products, validation, analysis or writing of the paper.

\section*{Acknowledgements}

The authors would like to acknowledge Manda~Banerji, Michelle~Cluver, Duncan~Farrah, Peter~Hatfield, Tom~Jarrett, Emeric~Le~Floc'h, Richard~McMahon,  Mark~Sargent,   Roberto~Scipioni, Dan~Smith, Elisabetta~Valiante,  Ivan~Valtchanov, Stephen~Wilkins and Louise~Winters for useful conversations or important contributions to the project that do not form a specific part of this paper. We are grateful to Stephen Serjeant for useful comments on the manuscript while under review.


The research leading to these results has received funding from the European Union Seventh Framework Programme FP7/2007-2013/ under grant agreement No. 607254. This publication reflects only the authors' view and the European Union is not responsible for any use that may be made of the information contained therein.

Seb Oliver acknowledges support from the Science and Technology Facilities Council (grant number ST/L000652/1)

KM has been supported by the National Science Centre (UMO-2018/30/E/ST9/00082).

MJ, LM, MP and MV acknowledge support from the South African Department of Science and Technology (DST/CON 0134/2014).

MP and MV acknowledge financial support from the Inter-University Institute for Data Intensive Astronomy (IDIA). IDIA is a partnership of the University of Cape Town, the University of Pretoria and the University of the Western Cape.

VB has received funding from the Excellence Initiative of Aix-Marseille University-AMIDEX, a French ‘Investissement d’Avenir’ programme.

HCSS / HSpot / HIPE are joint developments by the \Herschel\ Science Ground Segment Consortium, consisting of ESA, the NASA \Herschel\ Science Center, and the HIFI, PACS and SPIRE consortia.

This research made extensive use of TOPCAT \citep{2005ASPC..347...29T}, STILTS \citep{2006ASPC..351..666T} and Astropy \citep{2018AJ....156..123A}.

This research has made use of the NASA/IPAC Extragalactic Database (NED) which is operated by the Jet Propulsion Laboratory, California Institute of Technology, under contract with the National Aeronautics and Space Administration.

This work has made use of the COSMOS2015 catalogue \citep{cosmos} based on data products from observations made with ESO Telescopes at the La Silla Paranal Observatory under ESO programme ID 179.A-2005 and on data products produced by TERAPIX and the Cambridge Astronomy Survey Unit on behalf of the UltraVISTA consortium.

SPIRE has been developed by a consortium of institutes led by Cardiff Univ. (UK) and including Univ. Lethbridge (Canada); NAOC (China); CEA, LAM (France); IFSI, Univ. Padua (Italy); IAC (Spain); Stockholm Observatory (Sweden); Imperial
College London, RAL, UCL-MSSL, UKATC, Univ. Sussex (UK); Caltech, JPL, NHSC, Univ. Colorado (USA). This development has been supported by national funding agencies: CSA (Canada); NAOC (China); CEA, CNES, CNRS (France); ASI (Italy); MCINN (Spain); SNSB (Sweden); STFC, UKSA (UK); and NASA (USA).

HELP would like to thank the HELP Scientific Advisory Board members past and present for invaluable advice in defining the project: Simon~Driver (chair), Loretta~Dunne, Carol~Lonsdale, Mark~Lacy, Peter~Capak, Takashi~Onaka, Mara~Salvato, Brent~Groves, G\"{o}ran~Pilbratt, and David~Elbaz.

The data presented in this paper is released through the \Herschel\ Database in Marseille HeDaM ({\url{http://hedam.lam.fr/HELP}})

Huge thanks also to our Project Manager, Louise Winters, for keeping us on-track and on-time with good humour.

Finally, we would like to take this opportunity to remember Natalie Christopher who passed away in August 2019. Natalie worked on HELP from the start and her contribution to astronomy is sorely missed. Many have described how Natalie's unique vision of making astronomy accessible to all, coupled to her recognition of the power of astronomy to bring people together through their shared common interest, have had a profound impact. 

\section*{Data availability}
The data underlying this article are available in HeDaM at \url{http://hedam.lam.fr/HELP/}, and can be accessed publicly.

\bibliographystyle{mnras}
\bibliography{./HELP_bib}

\begin{thebibliography}{}
\makeatletter
\relax
\def\mn@urlcharsother{\let\do\@makeother \do\$\do\&\do\#\do\^\do\_\do\%\do\~}
\def\mn@doi{\begingroup\mn@urlcharsother \@ifnextchar [ {\mn@doi@}
  {\mn@doi@[]}}
\def\mn@doi@[#1]#2{\def\@tempa{#1}\ifx\@tempa\@empty \href
  {http://dx.doi.org/#2} {doi:#2}\else \href {http://dx.doi.org/#2} {#1}\fi
  \endgroup}
\def\mn@eprint#1#2{\mn@eprint@#1:#2::\@nil}
\def\mn@eprint@arXiv#1{\href {http://arxiv.org/abs/#1} {{\tt arXiv:#1}}}
\def\mn@eprint@dblp#1{\href {http://dblp.uni-trier.de/rec/bibtex/#1.xml}
  {dblp:#1}}
\def\mn@eprint@#1:#2:#3:#4\@nil{\def\@tempa {#1}\def\@tempb {#2}\def\@tempc
  {#3}\ifx \@tempc \@empty \let \@tempc \@tempb \let \@tempb \@tempa \fi \ifx
  \@tempb \@empty \def\@tempb {arXiv}\fi \@ifundefined
  {mn@eprint@\@tempb}{\@tempb:\@tempc}{\expandafter \expandafter \csname
  mn@eprint@\@tempb\endcsname \expandafter{\@tempc}}}

\bibitem[\protect\citeauthoryear{Almosallam, Lindsay, Jarvis  \&
  Roberts}{Almosallam et~al.}{2016a}]{Almosallam:2016a}
Almosallam I.~A.,  Lindsay S.~N.,  Jarvis M.~J.,   Roberts S.~J.,  2016a,
  MNRAS, 455, 2387

\bibitem[\protect\citeauthoryear{Almosallam, Jarvis  \& Roberts}{Almosallam
  et~al.}{2016b}]{Almosallam:2016b}
Almosallam I.~A.,  Jarvis M.~J.,   Roberts S.~J.,  2016b, MNRAS, 462, 726

\bibitem[\protect\citeauthoryear{{Asboth}, {etc.}, {etc}  \& C.}{{Asboth}
  et~al.}{2015}]{Asboth:2015lr}
{Asboth} V.,  {etc.} A.,  {etc} B.,   C. e.,  2015, \mnras, sub.

\bibitem[\protect\citeauthoryear{{Astropy Collaboration} et~al.,}{{Astropy
  Collaboration} et~al.}{2018}]{2018AJ....156..123A}
{Astropy Collaboration} et~al., 2018, \mn@doi [\aj] {10.3847/1538-3881/aabc4f},
  \href {https://ui.adsabs.harvard.edu/abs/2018AJ....156..123A} {156, 123}

\bibitem[\protect\citeauthoryear{{Baldry} et~al.,}{{Baldry}
  et~al.}{2018}]{Baldry2018}
{Baldry} I.~K.,  et~al., 2018, \mn@doi [\mnras] {10.1093/mnras/stx3042}, \href
  {https://ui.adsabs.harvard.edu/abs/2018MNRAS.474.3875B} {474, 3875}

\bibitem[\protect\citeauthoryear{{Balestra} et~al.,}{{Balestra}
  et~al.}{2010}]{Balestra2010}
{Balestra} I.,  et~al., 2010, \mn@doi [\aap] {10.1051/0004-6361/200913626},
  \href {https://ui.adsabs.harvard.edu/abs/2010A&A...512A..12B} {512, A12}

\bibitem[\protect\citeauthoryear{{Barger}, {Cowie}  \& {Wang}}{{Barger}
  et~al.}{2008}]{Barger2008}
{Barger} A.~J.,  {Cowie} L.~L.,   {Wang} W.~H.,  2008, \mn@doi [\apj]
  {10.1086/592735}, \href
  {https://ui.adsabs.harvard.edu/abs/2008ApJ...689..687B} {689, 687}

\bibitem[\protect\citeauthoryear{{Beck}, {Dobos}, {Budav{\'a}ri}, {Szalay}  \&
  {Csabai}}{{Beck} et~al.}{2017}]{Beck:2017}
{Beck} R.,  {Dobos} L.,  {Budav{\'a}ri} T.,  {Szalay} A.~S.,   {Csabai} I.,
  2017, \mn@doi [Astronomy and Computing] {10.1016/j.ascom.2017.03.002}, \href
  {https://ui.adsabs.harvard.edu/abs/2017A&C....19...34B} {19, 34}

\bibitem[\protect\citeauthoryear{{Berta} et~al.,}{{Berta}
  et~al.}{2007}]{Berta2007}
{Berta} S.,  et~al., 2007, \mn@doi [\aap] {10.1051/0004-6361:20066795}, \href
  {https://ui.adsabs.harvard.edu/abs/2007A&A...467..565B} {467, 565}

\bibitem[\protect\citeauthoryear{{Blanton} et~al.,}{{Blanton}
  et~al.}{2017}]{sdss}
{Blanton} M.~R.,  et~al., 2017, \mn@doi [\aj] {10.3847/1538-3881/aa7567}, \href
  {https://ui.adsabs.harvard.edu/abs/2017AJ....154...28B} {154, 28}

\bibitem[\protect\citeauthoryear{{Boller}, {Freyberg}, {Tr{\"u}mper}, {Haberl},
  {Voges}  \& {Nandra}}{{Boller} et~al.}{2016}]{Boller:2016}
{Boller} T.,  {Freyberg} M.~J.,  {Tr{\"u}mper} J.,  {Haberl} F.,  {Voges} W.,
  {Nandra} K.,  2016, \mn@doi [\aap] {10.1051/0004-6361/201525648}, \href
  {https://ui.adsabs.harvard.edu/abs/2016A&A...588A.103B} {588, A103}

\bibitem[\protect\citeauthoryear{{Boquien}, {Burgarella}, {Roehlly}, {Buat},
  {Ciesla}, {Corre}, {Inoue}  \& {Salas}}{{Boquien}
  et~al.}{2019}]{Boquien:2019}
{Boquien} M.,  {Burgarella} D.,  {Roehlly} Y.,  {Buat} V.,  {Ciesla} L.,
  {Corre} D.,  {Inoue} A.~K.,   {Salas} H.,  2019, \mn@doi [\aap]
  {10.1051/0004-6361/201834156}, \href
  {https://ui.adsabs.harvard.edu/abs/2019A&A...622A.103B} {622, A103}

\bibitem[\protect\citeauthoryear{{Bradshaw} et~al.,}{{Bradshaw}
  et~al.}{2013}]{Bradshaw2013}
{Bradshaw} E.~J.,  et~al., 2013, \mn@doi [\mnras] {10.1093/mnras/stt715}, \href
  {https://ui.adsabs.harvard.edu/abs/2013MNRAS.433..194B} {433, 194}

\bibitem[\protect\citeauthoryear{Brammer, van Dokkum  \& Coppi}{Brammer
  et~al.}{2008}]{Brammer:2008}
Brammer G.~B.,  van Dokkum P.~G.,   Coppi P.,  2008, ApJ, 686, 1503

\bibitem[\protect\citeauthoryear{Brown et~al.,}{Brown
  et~al.}{2014}]{Brown:2014}
Brown M. J.~I.,  et~al., 2014, ApJS, 212, 18

\bibitem[\protect\citeauthoryear{Brown et~al.,}{Brown
  et~al.}{2016}]{brown2016gaia}
Brown A.~G.,  et~al., 2016, Astronomy \& Astrophysics, 595, A2

\bibitem[\protect\citeauthoryear{{Bruzual} \& {Charlot}}{{Bruzual} \&
  {Charlot}}{2003}]{Bruzal:2003}
{Bruzual} G.,  {Charlot} S.,  2003, \mn@doi [\mnras]
  {10.1046/j.1365-8711.2003.06897.x}, \href
  {http://adsabs.harvard.edu/abs/2003MNRAS.344.1000B} {344, 1000}

\bibitem[\protect\citeauthoryear{{Buat}, {Boquien}, {Ma{\l}ek}, {Corre},
  {Salas}, {Roehlly}, {Shirley}  \& {Efstathiou}}{{Buat}
  et~al.}{2018}]{Buat:2018}
{Buat} V.,  {Boquien} M.,  {Ma{\l}ek} K.,  {Corre} D.,  {Salas} H.,  {Roehlly}
  Y.,  {Shirley} R.,   {Efstathiou} A.,  2018, \mn@doi [\aap]
  {10.1051/0004-6361/201833841}, \href
  {https://ui.adsabs.harvard.edu/abs/2018A&A...619A.135B} {619, A135}

\bibitem[\protect\citeauthoryear{{Buat}, {Ciesla}, {Boquien}, {Ma{\l}ek}  \&
  {Burgarella}}{{Buat} et~al.}{2019}]{Buat:2019}
{Buat} V.,  {Ciesla} L.,  {Boquien} M.,  {Ma{\l}ek} K.,   {Burgarella} D.,
  2019, \mn@doi [\aap] {10.1051/0004-6361/201936643}, \href
  {https://ui.adsabs.harvard.edu/abs/2019A&A...632A..79B} {632, A79}

\bibitem[\protect\citeauthoryear{{Burgarella}, {Buat}  \&
  {Iglesias-P{\'a}ramo}}{{Burgarella} et~al.}{2005}]{Burgarella:2005}
{Burgarella} D.,  {Buat} V.,   {Iglesias-P{\'a}ramo} J.,  2005, \mn@doi
  [\mnras] {10.1111/j.1365-2966.2005.09131.x}, \href
  {https://ui.adsabs.harvard.edu/abs/2005MNRAS.360.1413B} {360, 1413}

\bibitem[\protect\citeauthoryear{{Campos~Varillas}}{{Campos~Varillas}}{2021}]{CamposVarillas:2021}
{Campos~Varillas} M.,  2021, Stellar Mass function of HS82, Unpublished
  Manuscript

\bibitem[\protect\citeauthoryear{{Cardamone} et~al.,}{{Cardamone}
  et~al.}{2010}]{Cardamone2010}
{Cardamone} C.~N.,  et~al., 2010, \mn@doi [\apjs]
  {10.1088/0067-0049/189/2/270}, \href
  {https://ui.adsabs.harvard.edu/abs/2010ApJS..189..270C} {189, 270}

\bibitem[\protect\citeauthoryear{{Casey} et~al.,}{{Casey}
  et~al.}{2014}]{Casey:2014}
{Casey} C.~M.,  et~al., 2014, \mn@doi [\apj] {10.1088/0004-637X/796/2/95},
  \href {https://ui.adsabs.harvard.edu/abs/2014ApJ...796...95C} {796, 95}

\bibitem[\protect\citeauthoryear{{Castellano} et~al.,}{{Castellano}
  et~al.}{2016}]{astrodeep_2}
{Castellano} M.,  et~al., 2016, \mn@doi [\aap] {10.1051/0004-6361/201527514},
  \href {https://ui.adsabs.harvard.edu/abs/2016A&A...590A..31C} {590, A31}

\bibitem[\protect\citeauthoryear{{Chabrier}}{{Chabrier}}{2003}]{Chabrier:2003}
{Chabrier} G.,  2003, \mn@doi [\pasp] {10.1086/376392}, \href
  {http://adsabs.harvard.edu/abs/2003PASP..115..763C} {115, 763}

\bibitem[\protect\citeauthoryear{{Chapin} et~al.,}{{Chapin}
  et~al.}{2011}]{Chapin:2011lr}
{Chapin} E.~L.,  et~al., 2011, \mn@doi [\mnras]
  {10.1111/j.1365-2966.2010.17697.x}, \href
  {http://adsabs.harvard.edu/abs/2011MNRAS.411..505C} {411, 505}

\bibitem[\protect\citeauthoryear{{Chapman}, {Smail}, {Blain}  \&
  {Ivison}}{{Chapman} et~al.}{2004}]{Chapman2004}
{Chapman} S.~C.,  {Smail} I.,  {Blain} A.~W.,   {Ivison} R.~J.,  2004, \mn@doi
  [\apj] {10.1086/423833}, \href
  {https://ui.adsabs.harvard.edu/abs/2004ApJ...614..671C} {614, 671}

\bibitem[\protect\citeauthoryear{{Chapman}, {Blain}, {Smail}  \&
  {Ivison}}{{Chapman} et~al.}{2005}]{Chapman2005}
{Chapman} S.~C.,  {Blain} A.~W.,  {Smail} I.,   {Ivison} R.~J.,  2005, \mn@doi
  [\apj] {10.1086/428082}, \href
  {https://ui.adsabs.harvard.edu/abs/2005ApJ...622..772C} {622, 772}

\bibitem[\protect\citeauthoryear{{Charlot} \& {Fall}}{{Charlot} \&
  {Fall}}{2000}]{Charlot:2000}
{Charlot} S.,  {Fall} S.~M.,  2000, \mn@doi [\apj] {10.1086/309250}, \href
  {http://adsabs.harvard.edu/abs/2000ApJ...539..718C} {539, 718}

\bibitem[\protect\citeauthoryear{{Childress} et~al.,}{{Childress}
  et~al.}{2017}]{Childress2017}
{Childress} M.~J.,  et~al., 2017, \mn@doi [\mnras] {10.1093/mnras/stx1872},
  \href {https://ui.adsabs.harvard.edu/abs/2017MNRAS.472..273C} {472, 273}

\bibitem[\protect\citeauthoryear{{Chincarini}, {Tarenghi}, {Sol}, {Crane},
  {Manousoyannaki}  \& {Materne}}{{Chincarini} et~al.}{1984}]{Chincarini1984}
{Chincarini} G.,  {Tarenghi} M.,  {Sol} H.,  {Crane} P.,  {Manousoyannaki} I.,
   {Materne} J.,  1984, \aaps, \href
  {https://ui.adsabs.harvard.edu/abs/1984A&AS...57....1C} {57, 1}

\bibitem[\protect\citeauthoryear{{Cohen}, {Hogg}, {Blandford}, {Cowie}, {Hu},
  {Songaila}, {Shopbell}  \& {Richberg}}{{Cohen} et~al.}{2000}]{Cohen2000}
{Cohen} J.~G.,  {Hogg} D.~W.,  {Blandford} R.,  {Cowie} L.~L.,  {Hu} E.,
  {Songaila} A.,  {Shopbell} P.,   {Richberg} K.,  2000, \mn@doi [\apj]
  {10.1086/309096}, \href
  {https://ui.adsabs.harvard.edu/abs/2000ApJ...538...29C} {538, 29}

\bibitem[\protect\citeauthoryear{{Coil} et~al.,}{{Coil}
  et~al.}{2011}]{Coil2011}
{Coil} A.~L.,  et~al., 2011, \mn@doi [\apj] {10.1088/0004-637X/741/1/8}, \href
  {https://ui.adsabs.harvard.edu/abs/2011ApJ...741....8C} {741, 8}

\bibitem[\protect\citeauthoryear{{Colless} et~al.,}{{Colless}
  et~al.}{2001}]{Colless2001}
{Colless} M.,  et~al., 2001, \mn@doi [\mnras]
  {10.1046/j.1365-8711.2001.04902.x}, \href
  {https://ui.adsabs.harvard.edu/abs/2001MNRAS.328.1039C} {328, 1039}

\bibitem[\protect\citeauthoryear{{Comparat} et~al.,}{{Comparat}
  et~al.}{2015}]{Comparat2015}
{Comparat} J.,  et~al., 2015, \mn@doi [\aap] {10.1051/0004-6361/201424767},
  \href {https://ui.adsabs.harvard.edu/abs/2015A&A...575A..40C} {575, A40}

\bibitem[\protect\citeauthoryear{{Cooper} et~al.,}{{Cooper}
  et~al.}{2011}]{Cooper2011}
{Cooper} M.~C.,  et~al., 2011, \mn@doi [\apjs] {10.1088/0067-0049/193/1/14},
  \href {https://ui.adsabs.harvard.edu/abs/2011ApJS..193...14C} {193, 14}

\bibitem[\protect\citeauthoryear{{Cooper} et~al.,}{{Cooper}
  et~al.}{2012}]{Cooper2012}
{Cooper} M.~C.,  et~al., 2012, \mn@doi [\mnras]
  {10.1111/j.1365-2966.2012.21524.x}, \href
  {https://ui.adsabs.harvard.edu/abs/2012MNRAS.425.2116C} {425, 2116}

\bibitem[\protect\citeauthoryear{{Cowie}, {Songaila}, {Hu}  \& {Cohen}}{{Cowie}
  et~al.}{1996}]{Cowie1996}
{Cowie} L.~L.,  {Songaila} A.,  {Hu} E.~M.,   {Cohen} J.~G.,  1996, \mn@doi
  [\aj] {10.1086/118058}, \href
  {https://ui.adsabs.harvard.edu/abs/1996AJ....112..839C} {112, 839}

\bibitem[\protect\citeauthoryear{{Cowie}, {Barger}, {Hu}, {Capak}  \&
  {Songaila}}{{Cowie} et~al.}{2004}]{Cowie2004}
{Cowie} L.~L.,  {Barger} A.~J.,  {Hu} E.~M.,  {Capak} P.,   {Songaila} A.,
  2004, \mn@doi [\aj] {10.1086/420997}, \href
  {https://ui.adsabs.harvard.edu/abs/2004AJ....127.3137C} {127, 3137}

\bibitem[\protect\citeauthoryear{{Cristiani} \& {D'Odorico}}{{Cristiani} \&
  {D'Odorico}}{2000}]{Cristiani2000}
{Cristiani} S.,  {D'Odorico} V.,  2000, \mn@doi [\aj] {10.1086/301575}, \href
  {https://ui.adsabs.harvard.edu/abs/2000AJ....120.1648C} {120, 1648}

\bibitem[\protect\citeauthoryear{{Croom}, {Warren}  \& {Glazebrook}}{{Croom}
  et~al.}{2001}]{Croom2001}
{Croom} S.~M.,  {Warren} S.~J.,   {Glazebrook} K.,  2001, \mn@doi [\mnras]
  {10.1046/j.1365-8711.2001.04846.x}, \href
  {https://ui.adsabs.harvard.edu/abs/2001MNRAS.328..150C} {328, 150}

\bibitem[\protect\citeauthoryear{{Croom}, {Smith}, {Boyle}, {Shanks}, {Miller},
  {Outram}  \& {Loaring}}{{Croom} et~al.}{2004}]{Croom2004}
{Croom} S.~M.,  {Smith} R.~J.,  {Boyle} B.~J.,  {Shanks} T.,  {Miller} L.,
  {Outram} P.~J.,   {Loaring} N.~S.,  2004, \mn@doi [\mnras]
  {10.1111/j.1365-2966.2004.07619.x}, \href
  {https://ui.adsabs.harvard.edu/abs/2004MNRAS.349.1397C} {349, 1397}

\bibitem[\protect\citeauthoryear{{Croom} et~al.,}{{Croom}
  et~al.}{2009}]{Croom2009}
{Croom} S.~M.,  et~al., 2009, \mn@doi [\mnras]
  {10.1111/j.1365-2966.2008.14052.x}, \href
  {https://ui.adsabs.harvard.edu/abs/2009MNRAS.392...19C} {392, 19}

\bibitem[\protect\citeauthoryear{Dahlen et~al.,}{Dahlen
  et~al.}{2013}]{Dahlen:2013}
Dahlen T.,  et~al., 2013, ApJ, 775, 93

\bibitem[\protect\citeauthoryear{{Dawson}, {Stern}, {Bunker}, {Spinrad}  \&
  {Dey}}{{Dawson} et~al.}{2001}]{Dawson2001}
{Dawson} S.,  {Stern} D.,  {Bunker} A.~J.,  {Spinrad} H.,   {Dey} A.,  2001,
  \mn@doi [\aj] {10.1086/321160}, \href
  {https://ui.adsabs.harvard.edu/abs/2001AJ....122..598D} {122, 598}

\bibitem[\protect\citeauthoryear{{Demleitner}}{{Demleitner}}{2018}]{2018ascl.soft04005D}
{Demleitner} M.,  2018, {DaCHS: Data Center Helper Suite} (\mn@eprint {ascl}
  {1804.005})

\bibitem[\protect\citeauthoryear{{Demleitner}, {Neves}, {Rothmaier}  \&
  {Wambsganss}}{{Demleitner} et~al.}{2014}]{2014A&C.....7...27D}
{Demleitner} M.,  {Neves} M.~C.,  {Rothmaier} F.,   {Wambsganss} J.,  2014,
  \mn@doi [Astronomy and Computing] {10.1016/j.ascom.2014.08.003}, \href
  {https://ui.adsabs.harvard.edu/abs/2014A&C.....7...27D} {7, 27}

\bibitem[\protect\citeauthoryear{{Dickinson} et~al.,}{{Dickinson}
  et~al.}{2004}]{Dickinson2004}
{Dickinson} M.,  et~al., 2004, \mn@doi [\apjl] {10.1086/381119}, \href
  {https://ui.adsabs.harvard.edu/abs/2004ApJ...600L..99D} {600, L99}

\bibitem[\protect\citeauthoryear{{Donevski} et~al.,}{{Donevski}
  et~al.}{2020}]{Donevski:2020}
{Donevski} D.,  et~al., 2020, \mn@doi [\aap] {10.1051/0004-6361/202038405},
  \href {https://ui.adsabs.harvard.edu/abs/2020A&A...644A.144D} {644, A144}

\bibitem[\protect\citeauthoryear{{Donley} et~al.,}{{Donley}
  et~al.}{2012}]{Donley:2012}
{Donley} J.~L.,  et~al., 2012, \mn@doi [\apj] {10.1088/0004-637X/748/2/142},
  \href {https://ui.adsabs.harvard.edu/abs/2012ApJ...748..142D} {748, 142}

\bibitem[\protect\citeauthoryear{Dressler, Smail, Poggianti, Butcher, Couch,
  Ellis  \& Oemler~Jr}{Dressler et~al.}{1999}]{dressler1999spectroscopic}
Dressler A.,  Smail I.,  Poggianti B.~M.,  Butcher H.,  Couch W.~J.,  Ellis
  R.~S.,   Oemler~Jr A.,  1999, The Astrophysical Journal Supplement Series,
  122, 51

\bibitem[\protect\citeauthoryear{{Drinkwater} et~al.,}{{Drinkwater}
  et~al.}{2010}]{Drinkwater2010}
{Drinkwater} M.~J.,  et~al., 2010, \mn@doi [\mnras]
  {10.1111/j.1365-2966.2009.15754.x}, \href
  {https://ui.adsabs.harvard.edu/abs/2010MNRAS.401.1429D} {401, 1429}

\bibitem[\protect\citeauthoryear{{Driver} et~al.,}{{Driver}
  et~al.}{2009}]{gama_1}
{Driver} S.~P.,  et~al., 2009, \mn@doi [Astronomy and Geophysics]
  {10.1111/j.1468-4004.2009.50512.x}, \href
  {https://ui.adsabs.harvard.edu/abs/2009A&G....50e..12D} {50, 5.12}

\bibitem[\protect\citeauthoryear{{Driver} et~al.,}{{Driver}
  et~al.}{2011}]{gama_2}
{Driver} S.~P.,  et~al., 2011, \mn@doi [\mnras]
  {10.1111/j.1365-2966.2010.18188.x}, \href
  {https://ui.adsabs.harvard.edu/abs/2011MNRAS.413..971D} {413, 971}

\bibitem[\protect\citeauthoryear{{Duivenvoorden} et~al.,}{{Duivenvoorden}
  et~al.}{2016}]{Duivenoorden:2016}
{Duivenvoorden} S.,  et~al., 2016, \mn@doi [\mnras] {10.1093/mnras/stw1466},
  \href {https://ui.adsabs.harvard.edu/abs/2016MNRAS.462..277D} {462, 277}

\bibitem[\protect\citeauthoryear{{Duivenvoorden} et~al.,}{{Duivenvoorden}
  et~al.}{2018}]{Duivenvoorden:2018}
{Duivenvoorden} S.,  et~al., 2018, \mn@doi [\mnras] {10.1093/mnras/sty691},
  \href {https://ui.adsabs.harvard.edu/abs/2018MNRAS.477.1099D} {477, 1099}

\bibitem[\protect\citeauthoryear{{Duivenvoorden} et~al.,}{{Duivenvoorden}
  et~al.}{2020}]{Duivenvoorden:2020}
{Duivenvoorden} S.,  et~al., 2020, \mn@doi [\mnras] {10.1093/mnras/stz3110},
  \href {https://ui.adsabs.harvard.edu/abs/2020MNRAS.491.1355D} {491, 1355}

\bibitem[\protect\citeauthoryear{{Duncan} et~al.,}{{Duncan}
  et~al.}{2018a}]{Duncan:2018a}
{Duncan} K.~J.,  et~al., 2018a, \mn@doi [\mnras] {10.1093/mnras/stx2536}, \href
  {http://adsabs.harvard.edu/abs/2018MNRAS.473.2655D} {473, 2655}

\bibitem[\protect\citeauthoryear{Duncan, Jarvis, Brown  \&
  R{\"o}ttgering}{Duncan et~al.}{2018b}]{Duncan:2018b}
Duncan K.~J.,  Jarvis M.~J.,  Brown M.~J.,   R{\"o}ttgering H.~J.,  2018b,
  Monthly Notices of the Royal Astronomical Society, 477, 5177

\bibitem[\protect\citeauthoryear{{Duncan} et~al.,}{{Duncan}
  et~al.}{2019}]{Duncan:2019a}
{Duncan} K.~J.,  et~al., 2019, \mn@doi [\aap] {10.1051/0004-6361/201833562},
  \href {https://ui.adsabs.harvard.edu/abs/2019A&A...622A...3D} {622, A3}

\bibitem[\protect\citeauthoryear{{ESA}}{{ESA}}{2017}]{Herschel-spire-point2017}
{ESA} 2017, {Herschel-SPIRE Point Source catalogue},
  \url{http://archives.esac.esa.int/hsa/legacy/HPDP/SPIRE/SPIRE-P/SPSC/SPIREPointSourceCatalogExplanatorySupplementFull20170203.pdf}

\bibitem[\protect\citeauthoryear{{Eales} et~al.,}{{Eales}
  et~al.}{2010}]{Eales:2010lr}
{Eales} S.,  et~al., 2010, \mn@doi [\pasp] {10.1086/653086}, \href
  {http://adsabs.harvard.edu/abs/2010PASP..122..499E} {122, 499}

\bibitem[\protect\citeauthoryear{Efstathiou et~al.,}{Efstathiou
  et~al.}{2021}]{Efstathiou:2021}
Efstathiou A.,  et~al., 2021, Monthly Notices of the Royal Astronomical
  Society: Letters, 503, L11

\bibitem[\protect\citeauthoryear{{Falco} et~al.,}{{Falco}
  et~al.}{1999}]{Falco1999}
{Falco} E.~E.,  et~al., 1999, \mn@doi [\pasp] {10.1086/316343}, \href
  {https://ui.adsabs.harvard.edu/abs/1999PASP..111..438F} {111, 438}

\bibitem[\protect\citeauthoryear{{Farrah}, {Verma}, {Oliver}, {Rowan-Robinson}
  \& {McMahon}}{{Farrah} et~al.}{2002}]{Farrah:2002}
{Farrah} D.,  {Verma} A.,  {Oliver} S.,  {Rowan-Robinson} M.,   {McMahon} R.,
  2002, \mn@doi [\mnras] {10.1046/j.1365-8711.2002.04991.x}, \href
  {https://ui.adsabs.harvard.edu/abs/2002MNRAS.329..605F} {329, 605}

\bibitem[\protect\citeauthoryear{{Fernique}, {Boch}, {Donaldson}, {Durand},
  {O'Mullane}, {Reinecke}  \& {Taylor}}{{Fernique} et~al.}{2019}]{MOC}
{Fernique} P.,  {Boch} T.,  {Donaldson} T.,  {Durand} D.,  {O'Mullane} W.,
  {Reinecke} M.,   {Taylor} M.,  2019, Technical report, {MOC – HEALPix
  Multi-Order Coverage map, Version 1.1}, \url
  {http://www.ivoa.net/Documents/}.
\url {http://www.ivoa.net/Documents/}

\bibitem[\protect\citeauthoryear{{Feruglio} et~al.,}{{Feruglio}
  et~al.}{2008}]{Feruglio2008}
{Feruglio} C.,  et~al., 2008, \mn@doi [\aap] {10.1051/0004-6361:200809571},
  \href {https://ui.adsabs.harvard.edu/abs/2008A&A...488..417F} {488, 417}

\bibitem[\protect\citeauthoryear{{Flesch}}{{Flesch}}{2015}]{Flesch:2015}
{Flesch} E.~W.,  2015, \mn@doi [\pasa] {10.1017/pasa.2015.10}, \href
  {https://ui.adsabs.harvard.edu/abs/2015PASA...32...10F} {32, e010}

\bibitem[\protect\citeauthoryear{{Fritz}, {Franceschini}  \&
  {Hatziminaoglou}}{{Fritz} et~al.}{2006}]{fritz:2006}
{Fritz} J.,  {Franceschini} A.,   {Hatziminaoglou} E.,  2006, \mn@doi [\mnras]
  {10.1111/j.1365-2966.2006.09866.x}, \href
  {http://adsabs.harvard.edu/abs/2006MNRAS.366..767F} {366, 767}

\bibitem[\protect\citeauthoryear{{Garcet} et~al.,}{{Garcet}
  et~al.}{2007}]{Garcet2007}
{Garcet} O.,  et~al., 2007, \mn@doi [\aap] {10.1051/0004-6361:20077778}, \href
  {https://ui.adsabs.harvard.edu/abs/2007A&A...474..473G} {474, 473}

\bibitem[\protect\citeauthoryear{Gelman, Meng  \& Stern}{Gelman
  et~al.}{1996}]{gelman1996posterior}
Gelman A.,  Meng X.-L.,   Stern H.,  1996, Statistica sinica, pp 733--760

\bibitem[\protect\citeauthoryear{Gelman, Carlin, Stern, Dunson, Vehtari  \&
  Rubin}{Gelman et~al.}{2013}]{gelman2013bayesian}
Gelman A.,  Carlin J.~B.,  Stern H.~S.,  Dunson D.~B.,  Vehtari A.,   Rubin
  D.~B.,  2013, Bayesian data analysis.
CRC press

\bibitem[\protect\citeauthoryear{{Gloudemans, A.J.} et~al.,}{{Gloudemans, A.J.}
  et~al.}{2020}]{Gloudemans:2020}
{Gloudemans, A.J.} et~al., 2020, \mn@doi [A\&A] {10.1051/0004-6361/202038819}

\bibitem[\protect\citeauthoryear{{Griffin} et~al.,}{{Griffin}
  et~al.}{2010}]{Griffin:2010lr}
{Griffin} M.~J.,  et~al., 2010, \mn@doi [\aap] {10.1051/0004-6361/201014519},
  \href {http://adsabs.harvard.edu/abs/2010A%26A...518L...3G} {518, L3}

\bibitem[\protect\citeauthoryear{{Hamed}, {Ciesla}, {B{\'e}thermin},
  {Ma{\l}ek}, {Daddi}, {Sargent}  \& {Gobat}}{{Hamed}
  et~al.}{2021}]{Hamed:2021}
{Hamed} M.,  {Ciesla} L.,  {B{\'e}thermin} M.,  {Ma{\l}ek} K.,  {Daddi} E.,
  {Sargent} M.~T.,   {Gobat} R.,  2021, arXiv e-prints, \href
  {https://ui.adsabs.harvard.edu/abs/2021arXiv210107724H} {p. arXiv:2101.07724}

\bibitem[\protect\citeauthoryear{{Herenz} et~al.,}{{Herenz}
  et~al.}{2017}]{Herenz2017}
{Herenz} E.~C.,  et~al., 2017, \mn@doi [\aap] {10.1051/0004-6361/201731055},
  \href {https://ui.adsabs.harvard.edu/abs/2017A&A...606A..12H} {606, A12}

\bibitem[\protect\citeauthoryear{{Holder} et~al.,}{{Holder}
  et~al.}{2013}]{2013ApJ...771L..16H}
{Holder} G.~P.,  et~al., 2013, \mn@doi [\apjl] {10.1088/2041-8205/771/1/L16},
  \href {https://ui.adsabs.harvard.edu/abs/2013ApJ...771L..16H} {771, L16}

\bibitem[\protect\citeauthoryear{{Hopkins} et~al.,}{{Hopkins}
  et~al.}{2013}]{Hopkins2013}
{Hopkins} A.~M.,  et~al., 2013, \mn@doi [\mnras] {10.1093/mnras/stt030}, \href
  {https://ui.adsabs.harvard.edu/abs/2013MNRAS.430.2047H} {430, 2047}

\bibitem[\protect\citeauthoryear{{Houck} et~al.,}{{Houck}
  et~al.}{2005}]{Houck2005}
{Houck} J.~R.,  et~al., 2005, \mn@doi [\apjl] {10.1086/429405}, \href
  {https://ui.adsabs.harvard.edu/abs/2005ApJ...622L.105H} {622, L105}

\bibitem[\protect\citeauthoryear{{Huang} et~al.,}{{Huang}
  et~al.}{2009}]{Huang2009}
{Huang} J.~S.,  et~al., 2009, \mn@doi [\apj] {10.1088/0004-637X/700/1/183},
  \href {https://ui.adsabs.harvard.edu/abs/2009ApJ...700..183H} {700, 183}

\bibitem[\protect\citeauthoryear{{Huchra} et~al.,}{{Huchra}
  et~al.}{2012}]{Huchra2012}
{Huchra} J.~P.,  et~al., 2012, \mn@doi [\apjs] {10.1088/0067-0049/199/2/26},
  \href {https://ui.adsabs.harvard.edu/abs/2012ApJS..199...26H} {199, 26}

\bibitem[\protect\citeauthoryear{{Hurley} et~al.,}{{Hurley}
  et~al.}{2017}]{Hurley:2017lr}
{Hurley} P.~D.,  et~al., 2017, \mn@doi [\mnras] {10.1093/mnras/stw2375}, \href
  {https://ui.adsabs.harvard.edu/abs/2017MNRAS.464..885H} {464, 885}

\bibitem[\protect\citeauthoryear{{Ilbert} et~al.,}{{Ilbert}
  et~al.}{2013}]{cosmos_2}
{Ilbert} O.,  et~al., 2013, \mn@doi [\aap] {10.1051/0004-6361/201321100}, \href
  {https://ui.adsabs.harvard.edu/abs/2013A&A...556A..55I} {556, A55}

\bibitem[\protect\citeauthoryear{{Jarvis} et~al.,}{{Jarvis}
  et~al.}{2016}]{mightee}
{Jarvis} M.,  et~al., 2016, in MeerKAT Science: On the Pathway to the SKA. p.~6
  (\mn@eprint {arXiv} {1709.01901})

\bibitem[\protect\citeauthoryear{{Jeltema}, {Mulchaey}, {Lubin}  \&
  {Fassnacht}}{{Jeltema} et~al.}{2007}]{Jeltema2007}
{Jeltema} T.~E.,  {Mulchaey} J.~S.,  {Lubin} L.~M.,   {Fassnacht} C.~D.,  2007,
  \mn@doi [\apj] {10.1086/511852}, \href
  {https://ui.adsabs.harvard.edu/abs/2007ApJ...658..865J} {658, 865}

\bibitem[\protect\citeauthoryear{{Jones} et~al.,}{{Jones}
  et~al.}{2004}]{Jones2004}
{Jones} D.~H.,  et~al., 2004, \mn@doi [\mnras]
  {10.1111/j.1365-2966.2004.08353.x}, \href
  {https://ui.adsabs.harvard.edu/abs/2004MNRAS.355..747J} {355, 747}

\bibitem[\protect\citeauthoryear{{Kochanek} et~al.,}{{Kochanek}
  et~al.}{2012}]{Kochanek2012}
{Kochanek} C.~S.,  et~al., 2012, \mn@doi [\apjs] {10.1088/0067-0049/200/1/8},
  \href {https://ui.adsabs.harvard.edu/abs/2012ApJS..200....8K} {200, 8}

\bibitem[\protect\citeauthoryear{{Kriek} et~al.,}{{Kriek}
  et~al.}{2008}]{Kriek2008}
{Kriek} M.,  et~al., 2008, \mn@doi [\apj] {10.1086/528945}, \href
  {https://ui.adsabs.harvard.edu/abs/2008ApJ...677..219K} {677, 219}

\bibitem[\protect\citeauthoryear{{Kurk} et~al.,}{{Kurk}
  et~al.}{2013}]{Kurk2013}
{Kurk} J.,  et~al., 2013, \mn@doi [\aap] {10.1051/0004-6361/201117847}, \href
  {https://ui.adsabs.harvard.edu/abs/2013A&A...549A..63K} {549, A63}

\bibitem[\protect\citeauthoryear{{Lacy}, {Petric}, {Sajina}, {Canalizo},
  {Storrie-Lombardi}, {Armus}, {Fadda}  \& {Marleau}}{{Lacy}
  et~al.}{2007}]{Lacy2007}
{Lacy} M.,  {Petric} A.~O.,  {Sajina} A.,  {Canalizo} G.,  {Storrie-Lombardi}
  L.~J.,  {Armus} L.,  {Fadda} D.,   {Marleau} F.~R.,  2007, \mn@doi [\aj]
  {10.1086/509617}, \href
  {https://ui.adsabs.harvard.edu/abs/2007AJ....133..186L} {133, 186}

\bibitem[\protect\citeauthoryear{{Lacy} et~al.,}{{Lacy}
  et~al.}{2013}]{Lacy2013}
{Lacy} M.,  et~al., 2013, \mn@doi [\apjs] {10.1088/0067-0049/208/2/24}, \href
  {https://ui.adsabs.harvard.edu/abs/2013ApJS..208...24L} {208, 24}

\bibitem[\protect\citeauthoryear{{Laigle} et~al.,}{{Laigle}
  et~al.}{2016}]{cosmos_3}
{Laigle} C.,  et~al., 2016, \mn@doi [\apjs] {10.3847/0067-0049/224/2/24}, \href
  {https://ui.adsabs.harvard.edu/abs/2016ApJS..224...24L} {224, 24}

\bibitem[\protect\citeauthoryear{{Le F{\`e}vre} et~al.,}{{Le F{\`e}vre}
  et~al.}{2013}]{LeFevre2013}
{Le F{\`e}vre} O.,  et~al., 2013, \mn@doi [\aap] {10.1051/0004-6361/201322179},
  \href {https://ui.adsabs.harvard.edu/abs/2013A&A...559A..14L} {559, A14}

\bibitem[\protect\citeauthoryear{{Levenson} et~al.,}{{Levenson}
  et~al.}{2010}]{Levenson:2010lr}
{Levenson} L.,  et~al., 2010, \mn@doi [\mnras]
  {10.1111/j.1365-2966.2010.17771.x}, \href
  {http://adsabs.harvard.edu/abs/2010MNRAS.409...83L} {409, 83}

\bibitem[\protect\citeauthoryear{{Lidman} et~al.,}{{Lidman}
  et~al.}{2013}]{Lidman2013}
{Lidman} C.,  et~al., 2013, \mn@doi [\pasa] {10.1017/pasa.2012.001}, \href
  {https://ui.adsabs.harvard.edu/abs/2013PASA...30....1L} {30, e001}

\bibitem[\protect\citeauthoryear{Lindegren et~al.,}{Lindegren
  et~al.}{2016}]{lindegren2016gaia}
Lindegren L.,  et~al., 2016, Astronomy \& Astrophysics, 595, A4

\bibitem[\protect\citeauthoryear{{Liske}, {Lemon}, {Driver}, {Cross}  \&
  {Couch}}{{Liske} et~al.}{2003}]{Liske2003}
{Liske} J.,  {Lemon} D.~J.,  {Driver} S.~P.,  {Cross} N.~J.~G.,   {Couch}
  W.~J.,  2003, \mn@doi [\mnras] {10.1046/j.1365-8711.2003.06826.x}, \href
  {https://ui.adsabs.harvard.edu/abs/2003MNRAS.344..307L} {344, 307}

\bibitem[\protect\citeauthoryear{{Liu}, {Petry}, {Impey}  \& {Foltz}}{{Liu}
  et~al.}{1999}]{Liu1999}
{Liu} C.~T.,  {Petry} C.~E.,  {Impey} C.~D.,   {Foltz} C.~B.,  1999, \mn@doi
  [\aj] {10.1086/301100}, \href
  {https://ui.adsabs.harvard.edu/abs/1999AJ....118.1912L} {118, 1912}

\bibitem[\protect\citeauthoryear{{Lo Faro}, {Buat}, {Roehlly},
  {Alvarez-Marquez}, {Burgarella}, {Silva}  \& {Efstathiou}}{{Lo Faro}
  et~al.}{2017}]{LoFaro:2017}
{Lo Faro} B.,  {Buat} V.,  {Roehlly} Y.,  {Alvarez-Marquez} J.,  {Burgarella}
  D.,  {Silva} L.,   {Efstathiou} A.,  2017, \mn@doi [\mnras]
  {10.1093/mnras/stx1901}, \href
  {https://ui.adsabs.harvard.edu/abs/2017MNRAS.472.1372L} {472, 1372}

\bibitem[\protect\citeauthoryear{{Loveday}, {Peterson}, {Maddox}  \&
  {Efstathiou}}{{Loveday} et~al.}{1996}]{Loveday1996}
{Loveday} J.,  {Peterson} B.~A.,  {Maddox} S.~J.,   {Efstathiou} G.,  1996,
  \mn@doi [\apjs] {10.1086/192360}, \href
  {https://ui.adsabs.harvard.edu/abs/1996ApJS..107..201L} {107, 201}

\bibitem[\protect\citeauthoryear{{Lutz} et~al.,}{{Lutz} et~al.}{2011}]{pep}
{Lutz} D.,  et~al., 2011, \mn@doi [\aap] {10.1051/0004-6361/201117107}, \href
  {http://adsabs.harvard.edu/abs/2011A%26A...532A..90L} {532, A90}

\bibitem[\protect\citeauthoryear{{Maddox} et~al.,}{{Maddox}
  et~al.}{2018}]{Maddox2016}
{Maddox} S.~J.,  et~al., 2018, \mn@doi [\apjs] {10.3847/1538-4365/aab8fc},
  \href {https://ui.adsabs.harvard.edu/abs/2018ApJS..236...30M} {236, 30}

\bibitem[\protect\citeauthoryear{Ma{\l}ek et~al.,}{Ma{\l}ek
  et~al.}{2018}]{Malek:2018}
Ma{\l}ek K.,  et~al., 2018, Astronomy \& Astrophysics, 620, A50

\bibitem[\protect\citeauthoryear{{Mao} et~al.,}{{Mao} et~al.}{2012}]{Mao2012}
{Mao} M.~Y.,  et~al., 2012, \mn@doi [\mnras]
  {10.1111/j.1365-2966.2012.21913.x}, \href
  {https://ui.adsabs.harvard.edu/abs/2012MNRAS.426.3334M} {426, 3334}

\bibitem[\protect\citeauthoryear{{Marleau}, {Fadda}, {Appleton},
  {Noriega-Crespo}, {Im}  \& {Clancy}}{{Marleau} et~al.}{2007}]{Marleau2007}
{Marleau} F.~R.,  {Fadda} D.,  {Appleton} P.~N.,  {Noriega-Crespo} A.,  {Im}
  M.,   {Clancy} D.,  2007, \mn@doi [\apj] {10.1086/518114}, \href
  {https://ui.adsabs.harvard.edu/abs/2007ApJ...663..218M} {663, 218}

\bibitem[\protect\citeauthoryear{{Masters} et~al.,}{{Masters}
  et~al.}{2019}]{Masters2019}
{Masters} D.~C.,  et~al., 2019, \mn@doi [\apj] {10.3847/1538-4357/ab184d},
  \href {https://ui.adsabs.harvard.edu/abs/2019ApJ...877...81M} {877, 81}

\bibitem[\protect\citeauthoryear{{Mauduit} et~al.,}{{Mauduit}
  et~al.}{2012}]{servs:2012}
{Mauduit} J.~C.,  et~al., 2012, \mn@doi [\pasp] {10.1086/666945}, \href
  {https://ui.adsabs.harvard.edu/abs/2012PASP..124..714M} {124, 714}

\bibitem[\protect\citeauthoryear{{Maza}, {Ortiz}, {Wischnjewsky}, {Antezana}
  \& {Gonz{\'a}lez}}{{Maza} et~al.}{1995}]{Maza1995}
{Maza} J.,  {Ortiz} P.~F.,  {Wischnjewsky} M.,  {Antezana} R.,   {Gonz{\'a}lez}
  L.~E.,  1995, \rmxaa, \href
  {https://ui.adsabs.harvard.edu/abs/1995RMxAA..31..159M} {31, 159}

\bibitem[\protect\citeauthoryear{{McCheyne}}{{McCheyne}}{2021}]{McCheyne:2021}
{McCheyne} I.,  2021, FIRC of Herschel and LOFAR sources, Unpublished
  Manuscript

\bibitem[\protect\citeauthoryear{{McLure} et~al.,}{{McLure}
  et~al.}{2013}]{McLure2013}
{McLure} R.~J.,  et~al., 2013, \mn@doi [\mnras] {10.1093/mnras/sts092}, \href
  {https://ui.adsabs.harvard.edu/abs/2013MNRAS.428.1088M} {428, 1088}

\bibitem[\protect\citeauthoryear{{McLure} et~al.,}{{McLure}
  et~al.}{2018}]{McLure2018}
{McLure} R.~J.,  et~al., 2018, \mn@doi [\mnras] {10.1093/mnras/sty1213}, \href
  {https://ui.adsabs.harvard.edu/abs/2018MNRAS.479...25M} {479, 25}

\bibitem[\protect\citeauthoryear{{Merlin} et~al.,}{{Merlin}
  et~al.}{2016}]{astrodeep_1}
{Merlin} E.,  et~al., 2016, \mn@doi [\aap] {10.1051/0004-6361/201527513}, \href
  {https://ui.adsabs.harvard.edu/abs/2016A&A...590A..30M} {590, A30}

\bibitem[\protect\citeauthoryear{{Mignoli} et~al.,}{{Mignoli}
  et~al.}{2005}]{Mignoli2005}
{Mignoli} M.,  et~al., 2005, \mn@doi [\aap] {10.1051/0004-6361:20042434}, \href
  {https://ui.adsabs.harvard.edu/abs/2005A&A...437..883M} {437, 883}

\bibitem[\protect\citeauthoryear{{Momcheva} et~al.,}{{Momcheva}
  et~al.}{2016}]{Momcheva2016}
{Momcheva} I.~G.,  et~al., 2016, \mn@doi [\apjs] {10.3847/0067-0049/225/2/27},
  \href {https://ui.adsabs.harvard.edu/abs/2016ApJS..225...27M} {225, 27}

\bibitem[\protect\citeauthoryear{{Mountrichas}, {Buat}, {Yang}, {Boquien},
  {Burgarella}  \& {Ciesla}}{{Mountrichas} et~al.}{2021}]{Mountrichas2021}
{Mountrichas} G.,  {Buat} V.,  {Yang} G.,  {Boquien} M.,  {Burgarella} D.,
  {Ciesla} L.,  2021, \mn@doi [\aap] {10.1051/0004-6361/202039401}, \href
  {https://ui.adsabs.harvard.edu/abs/2021A&A...646A..29M} {646, A29}

\bibitem[\protect\citeauthoryear{{Nandra} et~al.,}{{Nandra}
  et~al.}{2015}]{Nandra2015}
{Nandra} K.,  et~al., 2015, \mn@doi [\apjs] {10.1088/0067-0049/220/1/10}, \href
  {https://ui.adsabs.harvard.edu/abs/2015ApJS..220...10N} {220, 10}

\bibitem[\protect\citeauthoryear{{Newman} et~al.,}{{Newman}
  et~al.}{2013}]{Newman2013}
{Newman} J.~A.,  et~al., 2013, \mn@doi [\apjs] {10.1088/0067-0049/208/1/5},
  \href {https://ui.adsabs.harvard.edu/abs/2013ApJS..208....5N} {208, 5}

\bibitem[\protect\citeauthoryear{{Nguyen} et~al.,}{{Nguyen}
  et~al.}{2010}]{Nguyen:2010lr}
{Nguyen} H.~T.,  et~al., 2010, \mn@doi [\aap] {10.1051/0004-6361/201014680},
  \href {http://adsabs.harvard.edu/abs/2010A%26A...518L...5N} {518, L5}

\bibitem[\protect\citeauthoryear{{Noll}, {Burgarella}, {Giovannoli}, {Buat},
  {Marcillac}  \& {Mu{\~n}oz-Mateos}}{{Noll} et~al.}{2009}]{Noll:2009}
{Noll} S.,  {Burgarella} D.,  {Giovannoli} E.,  {Buat} V.,  {Marcillac} D.,
  {Mu{\~n}oz-Mateos} J.~C.,  2009, \mn@doi [\aap]
  {10.1051/0004-6361/200912497}, \href
  {https://ui.adsabs.harvard.edu/abs/2009A&A...507.1793N} {507, 1793}

\bibitem[\protect\citeauthoryear{{Ocran}, {Taylor}, {Vaccari},
  {Ishwara-Chandra}, {Prandoni}, {Prescott}  \& {Mancuso}}{{Ocran}
  et~al.}{2021}]{Ocran2021}
{Ocran} E.~F.,  {Taylor} A.~R.,  {Vaccari} M.,  {Ishwara-Chandra} C.~H.,
  {Prandoni} I.,  {Prescott} M.,   {Mancuso} C.,  2021, \mn@doi [\mnras]
  {10.1093/mnras/staa3538}, \href
  {https://ui.adsabs.harvard.edu/abs/2021MNRAS.500.4685O} {500, 4685}

\bibitem[\protect\citeauthoryear{{Oliver} et~al.,}{{Oliver}
  et~al.}{2012}]{Oliver:2012}
{Oliver} S.~J.,  et~al., 2012, \mn@doi [\mnras]
  {10.1111/j.1365-2966.2012.20912.x}, \href
  {http://adsabs.harvard.edu/abs/2012MNRAS.424.1614O} {424, 1614}

\bibitem[\protect\citeauthoryear{{Ono}, {Ouchi}, {Shimasaku}, {Dunlop},
  {Farrah}, {McLure}  \& {Okamura}}{{Ono} et~al.}{2010}]{Ono2010}
{Ono} Y.,  {Ouchi} M.,  {Shimasaku} K.,  {Dunlop} J.,  {Farrah} D.,  {McLure}
  R.,   {Okamura} S.,  2010, \mn@doi [\apj] {10.1088/0004-637X/724/2/1524},
  \href {https://ui.adsabs.harvard.edu/abs/2010ApJ...724.1524O} {724, 1524}

\bibitem[\protect\citeauthoryear{{Onodera} et~al.,}{{Onodera}
  et~al.}{2015}]{Onodera2015}
{Onodera} M.,  et~al., 2015, \mn@doi [\apj] {10.1088/0004-637X/808/2/161},
  \href {https://ui.adsabs.harvard.edu/abs/2015ApJ...808..161O} {808, 161}

\bibitem[\protect\citeauthoryear{{Ott}}{{Ott}}{2010}]{Ott:2010}
{Ott} S.,  2010, in {Mizumoto} Y.,  {Morita} K.~I.,   {Ohishi} M.,  eds,
  Astronomical Society of the Pacific Conference Series Vol. 434, Astronomical
  Data Analysis Software and Systems XIX. p.~139 (\mn@eprint {arXiv}
  {1011.1209})

\bibitem[\protect\citeauthoryear{{Ouchi} et~al.,}{{Ouchi}
  et~al.}{2008}]{Ouchi2008}
{Ouchi} M.,  et~al., 2008, \mn@doi [\apjs] {10.1086/527673}, \href
  {https://ui.adsabs.harvard.edu/abs/2008ApJS..176..301O} {176, 301}

\bibitem[\protect\citeauthoryear{{Page} et~al.,}{{Page}
  et~al.}{2006}]{Page2006}
{Page} M.~J.,  et~al., 2006, \mn@doi [\mnras]
  {10.1111/j.1365-2966.2006.10278.x}, \href
  {https://ui.adsabs.harvard.edu/abs/2006MNRAS.369..156P} {369, 156}

\bibitem[\protect\citeauthoryear{{Papovich} et~al.,}{{Papovich}
  et~al.}{2006}]{Papovich2006}
{Papovich} C.,  et~al., 2006, \mn@doi [\aj] {10.1086/504598}, \href
  {https://ui.adsabs.harvard.edu/abs/2006AJ....132..231P} {132, 231}

\bibitem[\protect\citeauthoryear{{Pearson} et~al.,}{{Pearson}
  et~al.}{2017a}]{akari-nep}
{Pearson} C.,  et~al., 2017a, \mn@doi [Publication of Korean Astronomical
  Society] {10.5303/PKAS.2017.32.1.219}, \href
  {https://ui.adsabs.harvard.edu/abs/2017PKAS...32..219P} {32, 219}

\bibitem[\protect\citeauthoryear{{Pearson}, {Wang}, {van der Tak}, {Hurley},
  {Burgarella}  \& {Oliver}}{{Pearson} et~al.}{2017b}]{Pearson:2017}
{Pearson} W.~J.,  {Wang} L.,  {van der Tak} F.~F.~S.,  {Hurley} P.~D.,
  {Burgarella} D.,   {Oliver} S.~J.,  2017b, \mn@doi [\aap]
  {10.1051/0004-6361/201630105}, \href
  {https://ui.adsabs.harvard.edu/abs/2017A&A...603A.102P} {603, A102}

\bibitem[\protect\citeauthoryear{{Pearson} et~al.,}{{Pearson}
  et~al.}{2018}]{Pearson:2018}
{Pearson} W.~J.,  et~al., 2018, \mn@doi [\aap] {10.1051/0004-6361/201832821},
  \href {https://ui.adsabs.harvard.edu/abs/2018A&A...615A.146P} {615, A146}

\bibitem[\protect\citeauthoryear{{Piazzo}}{{Piazzo}}{2017}]{unimap_7}
{Piazzo} L.,  2017, \mn@doi [IEEE Transactions on Image Processing]
  {10.1109/TIP.2017.2736421}, 26, 5232

\bibitem[\protect\citeauthoryear{Piazzo, Ikhenaode, Natoli, Pestalozzi,
  Piacentini  \& Traficante}{Piazzo et~al.}{2012}]{unimap_2}
Piazzo L.,  Ikhenaode D.,  Natoli P.,  Pestalozzi M.,  Piacentini F.,
  Traficante A.,  2012, \mn@doi [IEEE transactions on image processing : a
  publication of the IEEE Signal Processing Society]
  {10.1109/TIP.2012.2197009}, 21

\bibitem[\protect\citeauthoryear{Piazzo, Panuzzo  \& Pestalozzi}{Piazzo
  et~al.}{2015a}]{unimap_3}
Piazzo L.,  Panuzzo P.,   Pestalozzi M.,  2015a, \mn@doi [Signal Processing]
  {10.1016/j.sigpro.2014.09.039}, 108, 430

\bibitem[\protect\citeauthoryear{{Piazzo}, {Calzoletti}, {Faustini},
  {Pestalozzi}, {Pezzuto}, {Elia}, {di Giorgio}  \& {Molinari}}{{Piazzo}
  et~al.}{2015b}]{unimap_4}
{Piazzo} L.,  {Calzoletti} L.,  {Faustini} F.,  {Pestalozzi} M.,  {Pezzuto} S.,
   {Elia} D.,  {di Giorgio} A.,   {Molinari} S.,  2015b, \mn@doi [\mnras]
  {10.1093/mnras/stu2453}, \href
  {https://ui.adsabs.harvard.edu/abs/2015MNRAS.447.1471P} {447, 1471}

\bibitem[\protect\citeauthoryear{{Piazzo}, {Raguso}, {Carpio}  \&
  {Altieri}}{{Piazzo} et~al.}{2016a}]{unimap_5}
{Piazzo} L.,  {Raguso} M.~C.,  {Carpio} J.~G.,   {Altieri} B.,  2016a, in 2016
  24th European Signal Processing Conference (EUSIPCO). pp 1553--1557,
  \mn@doi{10.1109/EUSIPCO.2016.7760509}

\bibitem[\protect\citeauthoryear{{Piazzo}, {Raguso}, {Calzoletti}, {Seu}  \&
  {Altieri}}{{Piazzo} et~al.}{2016b}]{unimap_6}
{Piazzo} L.,  {Raguso} M.~C.,  {Calzoletti} L.,  {Seu} R.,   {Altieri} B.,
  2016b, \mn@doi [IEEE Transactions on Image Processing]
  {10.1109/TIP.2016.2592700}, \href
  {https://ui.adsabs.harvard.edu/abs/2016ITIP...25.4458P} {25, 4458}

\bibitem[\protect\citeauthoryear{{Piffaretti}, {Arnaud}, {Pratt},
  {Pointecouteau}  \& {Melin}}{{Piffaretti} et~al.}{2011}]{Piffaretti2011}
{Piffaretti} R.,  {Arnaud} M.,  {Pratt} G.~W.,  {Pointecouteau} E.,   {Melin}
  J.~B.,  2011, \mn@doi [\aap] {10.1051/0004-6361/201015377}, \href
  {https://ui.adsabs.harvard.edu/abs/2011A&A...534A.109P} {534, A109}

\bibitem[\protect\citeauthoryear{{Pilbratt} et~al.,}{{Pilbratt}
  et~al.}{2010}]{Pilbratt:2010lr}
{Pilbratt} G.~L.,  et~al., 2010, \mn@doi [\aap] {10.1051/0004-6361/201014759},
  \href {http://adsabs.harvard.edu/abs/2010A%26A...518L...1P} {518, L1}

\bibitem[\protect\citeauthoryear{{Planck Collaboration} et~al.,}{{Planck
  Collaboration} et~al.}{2014}]{Planck:2014lr}
{Planck Collaboration} et~al., 2014, \mn@doi [\aap]
  {10.1051/0004-6361/201321580}, \href
  {http://adsabs.harvard.edu/abs/2014A%26A...571A..12P} {571, A12}

\bibitem[\protect\citeauthoryear{{Poglitsch} et~al.,}{{Poglitsch}
  et~al.}{2010}]{Poglitsch:2010lr}
{Poglitsch} A.,  et~al., 2010, \mn@doi [\aap] {10.1051/0004-6361/201014535},
  \href {http://adsabs.harvard.edu/abs/2010A%26A...518L...2P} {518, L2}

\bibitem[\protect\citeauthoryear{{Pope} et~al.,}{{Pope}
  et~al.}{2008}]{Pope2008}
{Pope} A.,  et~al., 2008, \mn@doi [\apj] {10.1086/527030}, \href
  {https://ui.adsabs.harvard.edu/abs/2008ApJ...675.1171P} {675, 1171}

\bibitem[\protect\citeauthoryear{{Ravikumar} et~al.,}{{Ravikumar}
  et~al.}{2007}]{Ravikumar2007}
{Ravikumar} C.~D.,  et~al., 2007, \mn@doi [\aap] {10.1051/0004-6361:20065358},
  \href {https://ui.adsabs.harvard.edu/abs/2007A&A...465.1099R} {465, 1099}

\bibitem[\protect\citeauthoryear{{Reddy}, {Steidel}, {Erb}, {Shapley}  \&
  {Pettini}}{{Reddy} et~al.}{2006}]{Reddy2006}
{Reddy} N.~A.,  {Steidel} C.~C.,  {Erb} D.~K.,  {Shapley} A.~E.,   {Pettini}
  M.,  2006, \mn@doi [\apj] {10.1086/508851}, \href
  {https://ui.adsabs.harvard.edu/abs/2006ApJ...653.1004R} {653, 1004}

\bibitem[\protect\citeauthoryear{{Riccio}}{{Riccio}}{2021}]{Riccio:2021}
{Riccio} G.,  2021, Getting ready for the LSST data - estimating the physical
  parameters of main sequence galaxies, Unpublished Manuscript

\bibitem[\protect\citeauthoryear{{Riechers}}{{Riechers}}{2013}]{Riechers:2013lr}
{Riechers} D.~A.,  2013, \mn@doi [\nat] {10.1038/502459a}, \href
  {http://adsabs.harvard.edu/abs/2013Natur.502..459R} {502, 459}

\bibitem[\protect\citeauthoryear{{Rieke} et~al.,}{{Rieke}
  et~al.}{2004}]{Rieke:2004}
{Rieke} G.~H.,  et~al., 2004, \mn@doi [\apjs] {10.1086/422717}, \href
  {https://ui.adsabs.harvard.edu/abs/2004ApJS..154...25R} {154, 25}

\bibitem[\protect\citeauthoryear{Rodighiero et~al.,}{Rodighiero
  et~al.}{2006}]{Rodighiero2006}
Rodighiero G.,  et~al., 2006, \mn@doi [Monthly Notices of the Royal
  Astronomical Society] {10.1111/j.1365-2966.2006.10844.x}, 371, 1891

\bibitem[\protect\citeauthoryear{{Roseboom} et~al.,}{{Roseboom}
  et~al.}{2012}]{Roseboom2012}
{Roseboom} I.~G.,  et~al., 2012, \mn@doi [\mnras]
  {10.1111/j.1365-2966.2012.21777.x}, \href
  {https://ui.adsabs.harvard.edu/abs/2012MNRAS.426.1782R} {426, 1782}

\bibitem[\protect\citeauthoryear{{Rowan-Robinson}}{{Rowan-Robinson}}{2000}]{MRR:2000}
{Rowan-Robinson} M.,  2000, \mn@doi [\mnras]
  {10.1046/j.1365-8711.2000.03588.x}, \href
  {https://ui.adsabs.harvard.edu/abs/2000MNRAS.316..885R} {316, 885}

\bibitem[\protect\citeauthoryear{{Rowan-Robinson}, {Gonzalez-Solares},
  {Vaccari}  \& {Marchetti}}{{Rowan-Robinson}
  et~al.}{2013}]{Rowan-Robinson2013}
{Rowan-Robinson} M.,  {Gonzalez-Solares} E.,  {Vaccari} M.,   {Marchetti} L.,
  2013, \mn@doi [\mnras] {10.1093/mnras/sts163}, \href
  {https://ui.adsabs.harvard.edu/abs/2013MNRAS.428.1958R} {428, 1958}

\bibitem[\protect\citeauthoryear{{Sacchi} et~al.,}{{Sacchi}
  et~al.}{2009}]{Sacchi2009}
{Sacchi} N.,  et~al., 2009, \mn@doi [\apj] {10.1088/0004-637X/703/2/1778},
  \href {https://ui.adsabs.harvard.edu/abs/2009ApJ...703.1778S} {703, 1778}

\bibitem[\protect\citeauthoryear{Salvato et~al.,}{Salvato
  et~al.}{2008}]{Salvato:2008}
Salvato M.,  et~al., 2008, ApJ, 690, 1250

\bibitem[\protect\citeauthoryear{Salvato et~al.,}{Salvato
  et~al.}{2011}]{Salvato:2011}
Salvato M.,  et~al., 2011, ApJ, 742, 61

\bibitem[\protect\citeauthoryear{{Salvato} et~al.,}{{Salvato}
  et~al.}{2018}]{Salvato:2018}
{Salvato} M.,  et~al., 2018, \mn@doi [\mnras] {10.1093/mnras/stx2651}, \href
  {https://ui.adsabs.harvard.edu/abs/2018MNRAS.473.4937S} {473, 4937}

\bibitem[\protect\citeauthoryear{{Sargsyan} \& {Weedman}}{{Sargsyan} \&
  {Weedman}}{2009}]{Sargsyan2009}
{Sargsyan} L.~A.,  {Weedman} D.~W.,  2009, \mn@doi [\apj]
  {10.1088/0004-637X/701/2/1398}, \href
  {https://ui.adsabs.harvard.edu/abs/2009ApJ...701.1398S} {701, 1398}

\bibitem[\protect\citeauthoryear{{Saunders} et~al.,}{{Saunders}
  et~al.}{2000}]{Saunders2000}
{Saunders} W.,  et~al., 2000, \mn@doi [\mnras]
  {10.1046/j.1365-8711.2000.03528.x}, \href
  {https://ui.adsabs.harvard.edu/abs/2000MNRAS.317...55S} {317, 55}

\bibitem[\protect\citeauthoryear{{Scodeggio} et~al.,}{{Scodeggio}
  et~al.}{2018}]{Scodeggio2018}
{Scodeggio} M.,  et~al., 2018, \mn@doi [\aap] {10.1051/0004-6361/201630114},
  \href {https://ui.adsabs.harvard.edu/abs/2018A&A...609A..84S} {609, A84}

\bibitem[\protect\citeauthoryear{{Scoville} et~al.,}{{Scoville}
  et~al.}{2007}]{cosmos}
{Scoville} N.,  et~al., 2007, \mn@doi [\apjs] {10.1086/516585}, \href
  {https://ui.adsabs.harvard.edu/abs/2007ApJS..172....1S} {172, 1}

\bibitem[\protect\citeauthoryear{{Scudder}, {Oliver}, {Hurley}, {Griffin},
  {Sargent}, {Scott}, {Wang}  \& {Wardlow}}{{Scudder}
  et~al.}{2016}]{Scudder:2016}
{Scudder} J.~M.,  {Oliver} S.,  {Hurley} P.~D.,  {Griffin} M.,  {Sargent}
  M.~T.,  {Scott} D.,  {Wang} L.,   {Wardlow} J.~L.,  2016, \mn@doi [\mnras]
  {10.1093/mnras/stw1044}, \href
  {https://ui.adsabs.harvard.edu/abs/2016MNRAS.460.1119S} {460, 1119}

\bibitem[\protect\citeauthoryear{{Scudder}, {Oliver}, {Hurley}, {Wardlow},
  {Wang}  \& {Farrah}}{{Scudder} et~al.}{2018}]{Scudder:2018}
{Scudder} J.~M.,  {Oliver} S.,  {Hurley} P.~D.,  {Wardlow} J.~L.,  {Wang} L.,
  {Farrah} D.,  2018, \mn@doi [\mnras] {10.1093/mnras/sty2009}, \href
  {https://ui.adsabs.harvard.edu/abs/2018MNRAS.480.4124S} {480, 4124}

\bibitem[\protect\citeauthoryear{{Sedgwick} et~al.,}{{Sedgwick}
  et~al.}{2017}]{Sedgwick2017}
{Sedgwick} C.,  et~al., 2017, \mn@doi [Publication of Korean Astronomical
  Society] {10.5303/PKAS.2017.32.1.281}, \href
  {https://ui.adsabs.harvard.edu/abs/2017PKAS...32..281S} {32, 281}

\bibitem[\protect\citeauthoryear{{Shim} et~al.,}{{Shim}
  et~al.}{2013}]{Shim2013}
{Shim} H.,  et~al., 2013, \mn@doi [\apjs] {10.1088/0067-0049/207/2/37}, \href
  {https://ui.adsabs.harvard.edu/abs/2013ApJS..207...37S} {207, 37}

\bibitem[\protect\citeauthoryear{Shirley et~al.,}{Shirley
  et~al.}{2019}]{Shirley:2019}
Shirley R.,  et~al., 2019, Monthly Notices of the Royal Astronomical Society,
  490, 634

\bibitem[\protect\citeauthoryear{{Silverman} et~al.,}{{Silverman}
  et~al.}{2010}]{Silverman2010}
{Silverman} J.~D.,  et~al., 2010, \mn@doi [\apjs]
  {10.1088/0067-0049/191/1/124}, \href
  {https://ui.adsabs.harvard.edu/abs/2010ApJS..191..124S} {191, 124}

\bibitem[\protect\citeauthoryear{{Smail}, {Chapman}, {Blain}  \&
  {Ivison}}{{Smail} et~al.}{2004}]{Smail2004}
{Smail} I.,  {Chapman} S.~C.,  {Blain} A.~W.,   {Ivison} R.~J.,  2004, \mn@doi
  [\apj] {10.1086/424896}, \href
  {https://ui.adsabs.harvard.edu/abs/2004ApJ...616...71S} {616, 71}

\bibitem[\protect\citeauthoryear{{Smith, D. J. B.} et~al.,}{{Smith, D. J. B.}
  et~al.}{2020}]{Smith:2020}
{Smith, D. J. B.} et~al., 2020, \mn@doi [A\&A] {10.1051/0004-6361/202039343}

\bibitem[\protect\citeauthoryear{{Smith} et~al.,}{{Smith}
  et~al.}{2012}]{Smith:2012lr}
{Smith} A.~J.,  et~al., 2012, \mn@doi [\mnras]
  {10.1111/j.1365-2966.2011.19709.x}, \href
  {http://adsabs.harvard.edu/abs/2012MNRAS.419..377S} {419, 377}

\bibitem[\protect\citeauthoryear{{Smith} et~al.,}{{Smith}
  et~al.}{2017}]{MSmith:2017}
{Smith} M. W.~L.,  et~al., 2017, \mn@doi [\apjs] {10.3847/1538-4365/aa9b35},
  \href {https://ui.adsabs.harvard.edu/abs/2017ApJS..233...26S} {233, 26}

\bibitem[\protect\citeauthoryear{{Stalin}, {Petitjean}, {Srianand}, {Fox},
  {Coppolani}  \& {Schwope}}{{Stalin} et~al.}{2010}]{Stalin2010}
{Stalin} C.~S.,  {Petitjean} P.,  {Srianand} R.,  {Fox} A.~J.,  {Coppolani} F.,
    {Schwope} A.,  2010, \mn@doi [\mnras] {10.1111/j.1365-2966.2009.15636.x},
  \href {https://ui.adsabs.harvard.edu/abs/2010MNRAS.401..294S} {401, 294}

\bibitem[\protect\citeauthoryear{{Steffen}, {Barger}, {Capak}, {Cowie},
  {Mushotzky}  \& {Yang}}{{Steffen} et~al.}{2004}]{Steffen2004}
{Steffen} A.~T.,  {Barger} A.~J.,  {Capak} P.,  {Cowie} L.~L.,  {Mushotzky}
  R.~F.,   {Yang} Y.,  2004, \mn@doi [\aj] {10.1086/423998}, \href
  {https://ui.adsabs.harvard.edu/abs/2004AJ....128.1483S} {128, 1483}

\bibitem[\protect\citeauthoryear{{Steidel}, {Adelberger}, {Shapley}, {Pettini},
  {Dickinson}  \& {Giavalisco}}{{Steidel} et~al.}{2003}]{Steidel2003}
{Steidel} C.~C.,  {Adelberger} K.~L.,  {Shapley} A.~E.,  {Pettini} M.,
  {Dickinson} M.,   {Giavalisco} M.,  2003, \mn@doi [\apj] {10.1086/375772},
  \href {https://ui.adsabs.harvard.edu/abs/2003ApJ...592..728S} {592, 728}

\bibitem[\protect\citeauthoryear{{Strolger} et~al.,}{{Strolger}
  et~al.}{2004}]{Strolger2004}
{Strolger} L.-G.,  et~al., 2004, \mn@doi [\apj] {10.1086/422901}, \href
  {https://ui.adsabs.harvard.edu/abs/2004ApJ...613..200S} {613, 200}

\bibitem[\protect\citeauthoryear{{Swinbank}, {Smail}, {Chapman}, {Blain},
  {Ivison}  \& {Keel}}{{Swinbank} et~al.}{2004}]{Swinbank2004}
{Swinbank} A.~M.,  {Smail} I.,  {Chapman} S.~C.,  {Blain} A.~W.,  {Ivison}
  R.~J.,   {Keel} W.~C.,  2004, \mn@doi [\apj] {10.1086/425171}, \href
  {https://ui.adsabs.harvard.edu/abs/2004ApJ...617...64S} {617, 64}

\bibitem[\protect\citeauthoryear{{Swinbank} et~al.,}{{Swinbank}
  et~al.}{2005}]{Swinbank2005}
{Swinbank} A.~M.,  et~al., 2005, \mn@doi [\mnras]
  {10.1111/j.1365-2966.2005.08901.x}, \href
  {https://ui.adsabs.harvard.edu/abs/2005MNRAS.359..401S} {359, 401}

\bibitem[\protect\citeauthoryear{{Swinbank} et~al.,}{{Swinbank}
  et~al.}{2007}]{Swinbank2007}
{Swinbank} A.~M.,  et~al., 2007, \mn@doi [\mnras]
  {10.1111/j.1365-2966.2007.12037.x}, \href
  {https://ui.adsabs.harvard.edu/abs/2007MNRAS.379.1343S} {379, 1343}

\bibitem[\protect\citeauthoryear{{Szokoly} et~al.,}{{Szokoly}
  et~al.}{2004}]{Szokoly2004}
{Szokoly} G.~P.,  et~al., 2004, \mn@doi [\apjs] {10.1086/424707}, \href
  {https://ui.adsabs.harvard.edu/abs/2004ApJS..155..271S} {155, 271}

\bibitem[\protect\citeauthoryear{{Taylor}}{{Taylor}}{2005}]{2005ASPC..347...29T}
{Taylor} M.~B.,  2005, in {Shopbell} P.,  {Britton} M.,   {Ebert} R.,  eds,
  Astronomical Society of the Pacific Conference Series Vol. 347, Astronomical
  Data Analysis Software and Systems XIV. p.~29

\bibitem[\protect\citeauthoryear{{Taylor}}{{Taylor}}{2006a}]{Taylor2006}
{Taylor} M.~B.,  2006a, {STILTS - A Package for Command-Line Processing of
  Tabular Data}.
p.~666

\bibitem[\protect\citeauthoryear{{Taylor}}{{Taylor}}{2006b}]{2006ASPC..351..666T}
{Taylor} M.~B.,  2006b, in {Gabriel} C.,  {Arviset} C.,  {Ponz} D.,   {Enrique}
  S.,  eds,  Astronomical Society of the Pacific Conference Series Vol. 351,
  Astronomical Data Analysis Software and Systems XV. p.~666

\bibitem[\protect\citeauthoryear{{Treister} et~al.,}{{Treister}
  et~al.}{2009}]{Treister2009}
{Treister} E.,  et~al., 2009, \mn@doi [\apj] {10.1088/0004-637X/693/2/1713},
  \href {https://ui.adsabs.harvard.edu/abs/2009ApJ...693.1713T} {693, 1713}

\bibitem[\protect\citeauthoryear{{Trichas} et~al.,}{{Trichas}
  et~al.}{2010}]{Trichas2010}
{Trichas} M.,  et~al., 2010, \mn@doi [\mnras]
  {10.1111/j.1365-2966.2010.16632.x}, \href
  {https://ui.adsabs.harvard.edu/abs/2010MNRAS.405.2243T} {405, 2243}

\bibitem[\protect\citeauthoryear{{Tully}, {Shaya}, {Karachentsev}, {Courtois},
  {Kocevski}, {Rizzi}  \& {Peel}}{{Tully} et~al.}{2008}]{Tully2008}
{Tully} R.~B.,  {Shaya} E.~J.,  {Karachentsev} I.~D.,  {Courtois} H.~M.,
  {Kocevski} D.~D.,  {Rizzi} L.,   {Peel} A.,  2008, \mn@doi [\apj]
  {10.1086/527428}, \href
  {https://ui.adsabs.harvard.edu/abs/2008ApJ...676..184T} {676, 184}

\bibitem[\protect\citeauthoryear{{Vaccari}}{{Vaccari}}{2015}]{Vaccari2015}
{Vaccari} M.,  2015, in Proceedings of ``The many facets of extragalactic radio
  surveys: towards new scientific challenges'' (EXTRA-RADSUR2015). 20-23
  October 2015. Bologna. p.~27 (\mn@eprint {arXiv} {1604.02353})

\bibitem[\protect\citeauthoryear{{Vanden Berk}, {Stoughton}, {Crotts}, {Tytler}
   \& {Kirkman}}{{Vanden Berk} et~al.}{2000}]{VandenBerk2000}
{Vanden Berk} D.~E.,  {Stoughton} C.,  {Crotts} A. P.~S.,  {Tytler} D.,
  {Kirkman} D.,  2000, \mn@doi [\aj] {10.1086/301404}, \href
  {https://ui.adsabs.harvard.edu/abs/2000AJ....119.2571V} {119, 2571}

\bibitem[\protect\citeauthoryear{{Vanzella} et~al.,}{{Vanzella}
  et~al.}{2008}]{Vanzella2008}
{Vanzella} E.,  et~al., 2008, \mn@doi [\aap] {10.1051/0004-6361:20078332},
  \href {https://ui.adsabs.harvard.edu/abs/2008A&A...478...83V} {478, 83}

\bibitem[\protect\citeauthoryear{{Viero} et~al.,}{{Viero}
  et~al.}{2013}]{Viero:2013rt}
{Viero} M.~P.,  et~al., 2013, \mn@doi [\apj] {10.1088/0004-637X/772/1/77},
  \href {http://adsabs.harvard.edu/abs/2013ApJ...772...77V} {772, 77}

\bibitem[\protect\citeauthoryear{{Viero} et~al.,}{{Viero}
  et~al.}{2014a}]{Viero:2014lr}
{Viero} M.~P.,  et~al., 2014a, \mn@doi [\apjs] {10.1088/0067-0049/210/2/22},
  \href {http://adsabs.harvard.edu/abs/2014ApJS..210...22V} {210, 22}

\bibitem[\protect\citeauthoryear{{Viero} et~al.,}{{Viero}
  et~al.}{2014b}]{2014ApJS..210...22V}
{Viero} M.~P.,  et~al., 2014b, \mn@doi [\apjs] {10.1088/0067-0049/210/2/22},
  \href {https://ui.adsabs.harvard.edu/abs/2014ApJS..210...22V} {210, 22}

\bibitem[\protect\citeauthoryear{{Wang, L.} et~al.,}{{Wang, L.}
  et~al.}{2020}]{Wang:2020}
{Wang, L.} et~al., 2020, \mn@doi [A\&A] {10.1051/0004-6361/202038811}

\bibitem[\protect\citeauthoryear{{Wang} et~al.,}{{Wang}
  et~al.}{2014}]{Wang:2013lr}
{Wang} L.,  et~al., 2014, \mn@doi [\mnras] {10.1093/mnras/stu1569}, \href
  {http://adsabs.harvard.edu/abs/2014MNRAS.444.2870W} {444, 2870}

\bibitem[\protect\citeauthoryear{{Weedman} \& {Houck}}{{Weedman} \&
  {Houck}}{2009}]{Weedman2009}
{Weedman} D.~W.,  {Houck} J.~R.,  2009, \mn@doi [\apj]
  {10.1088/0004-637X/693/1/370}, \href
  {https://ui.adsabs.harvard.edu/abs/2009ApJ...693..370W} {693, 370}

\bibitem[\protect\citeauthoryear{{Wei{\ss}} et~al.,}{{Wei{\ss}}
  et~al.}{2013}]{Weiss:2013}
{Wei{\ss}} A.,  et~al., 2013, \mn@doi [\apj] {10.1088/0004-637X/767/1/88},
  \href {https://ui.adsabs.harvard.edu/abs/2013ApJ...767...88W} {767, 88}

\bibitem[\protect\citeauthoryear{{Wirth} et~al.,}{{Wirth}
  et~al.}{2004}]{Wirth2004}
{Wirth} G.~D.,  et~al., 2004, \mn@doi [\aj] {10.1086/420999}, \href
  {https://ui.adsabs.harvard.edu/abs/2004AJ....127.3121W} {127, 3121}

\bibitem[\protect\citeauthoryear{{Yamada} et~al.,}{{Yamada}
  et~al.}{2005}]{Yamada2005}
{Yamada} T.,  et~al., 2005, \mn@doi [\apj] {10.1086/496954}, \href
  {https://ui.adsabs.harvard.edu/abs/2005ApJ...634..861Y} {634, 861}

\bibitem[\protect\citeauthoryear{{Yan} et~al.,}{{Yan} et~al.}{2007}]{Yan2007}
{Yan} L.,  et~al., 2007, \mn@doi [\apj] {10.1086/511516}, \href
  {https://ui.adsabs.harvard.edu/abs/2007ApJ...658..778Y} {658, 778}

\bibitem[\protect\citeauthoryear{{York} et~al.,}{{York}
  et~al.}{2000}]{York2000}
{York} D.~G.,  et~al., 2000, \mn@doi [\aj] {10.1086/301513}, \href
  {https://ui.adsabs.harvard.edu/abs/2000AJ....120.1579Y} {120, 1579}

\bibitem[\protect\citeauthoryear{Zou, Gao, Zhou  \& Kong}{Zou
  et~al.}{2019}]{Zou:2019}
Zou H.,  Gao J.,  Zhou X.,   Kong X.,  2019, The Astrophysical Journal
  Supplement Series, 242, 8

\bibitem[\protect\citeauthoryear{{da Costa} et~al.,}{{da Costa}
  et~al.}{1998}]{daCosta1998}
{da Costa} L.~N.,  et~al., 1998, \mn@doi [\aj] {10.1086/300410}, \href
  {https://ui.adsabs.harvard.edu/abs/1998AJ....116....1D} {116, 1}

\bibitem[\protect\citeauthoryear{{van der Wel}, {Franx}, {van Dokkum}  \&
  {Rix}}{{van der Wel} et~al.}{2004}]{VanderWel2004}
{van der Wel} A.,  {Franx} M.,  {van Dokkum} P.~G.,   {Rix} H.~W.,  2004,
  \mn@doi [\apjl] {10.1086/381887}, \href
  {https://ui.adsabs.harvard.edu/abs/2004ApJ...601L...5V} {601, L5}

\makeatother
\end{thebibliography}


\appendix

\section{OBSIDs}\label{appendix:obsids}

All the \Herschel\ observation identification numbers or OBSIDs. are available on the GitHub pages for PACS here:\\

\url{https://github.com/H-E-L-P/dmu_products/blob/master/dmu18/dmu18_HELP-PACS-maps/pacs_obsid.csv}\\
and for SPIRE here:\\

\url{https://github.com/H-E-L-P/dmu_products/blob/master/dmu19/dmu19_HELP-SPIRE-maps/spire_obsids.csv}\\

All products available through the HELP www pages \url{herschel.sussex.ac.uk}.


\section{Database structure and access}\label{appendix:database}

The data presented here is all publicly available including all input data sets that were used to produce it. All the code used for the processing is also publicly available on GitHub here:\\

\url{https://github.com/H-E-L-P/dmu_products/}\\

The raw data as fits tables and images is available at the \Herschel\ database at Marseille here:\\

\url{http://hedam.lam.fr/HELP/dataproducts/}\\

Those large files can be inconvenient so we also supply a Virtual Observatory server for automated querying of the full data set here:\\

\url{https://herschel-vos.phys.sussex.ac.uk/}\\

All the code used to produce the figures presented in this paper is available here:\\

\url{https://github.com/H-E-L-P/dmu_products/tree/master/dmu31/dmu31_Examples}\\

In order to access the flat files you will need to navigate the structure of Data Management Units products defined in Table~\ref{table:dmu_overview}. At each stage of the pipeline the working files are saved with the final merged catalogues stored in DMU32. We have produced and continue to work on extensive documentation to aid navigating the database. Starting at the front page of the GitHub repository should allow the reader to locate detailed descriptions of each section of the database. Alternatively we also provide per field summaries in addition to imaging of each field in order to inspect individual areas or objects.

\section{Spectroscopic Redshift Sources}
\label{app:specz}

Spectroscopic redshifts are compiled from numerous sources. Table~\ref{table:specZ} gives all the relevant references and can be used to find the source for a given object in the \emph{masterlist}.

\onecolumn

\begin{table*}
\caption{Overview of the Data Management Unit (DMU) database structure which is used in processing and web databases.}

\label{table:dmu_overview}
\begin{tabular}{l l}
\hline
Data Management Unit (DMU) name & DMU content \\
\hline
DMU0 &	Pristine catalogues \\
DMU1 &	Masterlist data \\
DMU2 &	Field definitions \\
DMU3 &	Morphologies (under development) \\
DMU4 &	Bright Star Mask \\
DMU5 &	Known Star Flag \\
DMU6 &	Optical photometry validation \\
DMU7 &	Optical photometry (under development) \\
DMU8 &	Radio data - LOFAR \& FIRST/NVSS/TGSS \\
DMU9 &	Radio data - JVLA-DEEP \& GMRT-DEEP \\
DMU10 &	Data Fusion \\
DMU11 &	Cross matching MIPS/PACS/SPIRE \\
DMU12 &	Cross Matching LOFAR \& FIRST/NVSS/TGSS \\
DMU13 &	Cross Matching JVLA-DEEP \& GMRT-DEEP \\
DMU14 &	GALEX data \\
DMU15 &	X-Ray data (under development) \\
DMU16 &	WISE Photometry \\
DMU17 &	MIPS images \\
DMU18 &	PACS images \\
DMU19 &	SPIRE images \\
DMU20 &	MIPS blind photometry (under development) \\
DMU21 &	PACS blind photometry (under development) \\
DMU22 &	SPIRE blind photometry \\
DMU23 &	Spec-z data \\
DMU24 &	Photo-z \\
DMU25 &	Prior model \\
DMU26 &	XID+ \\
DMU27 &	Empirical models / templates \\
DMU28 &	SED fitting / CIGALE \\
DMU29 &	Radiative transfer models (under development) \\
DMU30 &	Missing (supplementary) Sources \\
DMU31 &	Tools \\
DMU32 &	Final merged catalogues \\
\hline
\end{tabular}
\end{table*}

\pagebreak

\begin{longtable}{lrl}
\caption{The individual spectroscopic catalogues used for each field}\\

\label{table:specZ}

\\\hline
Field                & Identifier    & Source  \\
\hline
\endfirsthead
\hline
Field                & Identifier    & Source  \\
\hline
\endhead

\hline
\multicolumn{3}{r}{Continued on Next Page\ldots}
\endfoot
\hline\\
\multicolumn{3}{l}{\footnotesize $^\dagger$ Available from \url{https://www.nottingham.ac.uk/astronomy/UDS/data/data.html}}\\
\multicolumn{3}{l}{\footnotesize $^\star$ Available from \url{http://localgroup.ps.uci.edu/cooper/IMACS/zcatalog.html}}
\endlastfoot

~                    & 1 & \citet{Shim2013} \\
AKARI-NEP            & 2 & 2MASS Redshift Survey \citep[2MRS,][]{Huchra2012}\\
~                    & 4 & IRAS Point Source Catalog Redshift Catalog \citep[IRASPSCZ,][]{Saunders2000}\\
~                    & 8 & Updated Zwicky Catalog \citep[UZC][]{Falco1999}\\
~                    & 16 & NED sources compiled by M. Vaccari \citep{Vaccari2015}\\
\hline
~                    & 1 &  \citet{Sedgwick2017}\\
AKARI-SEP            & 2 & 6dF Galaxy Survey \citep{Jones2004}\\
~                    & 4 & 2MASS Redshift Survey \citep[2MRS,][]{Huchra2012}\\
~                    & 8 & IRAS Point Source Catalog Redshift Catalog \citep[IRASPSCZ,][]{Saunders2000}\\
~                    & 16 & \citet{Chincarini1984}\\
~                    & 32 & \citet{dressler1999spectroscopic}\\
~                    & 64 & \citet{Loveday1996}\\
~                    & 128 & \citet{Maza1995}\\
~                    & 256 & \citet{Tully2008}\\
\hline
~                    & 1 & AGES: The AGN and Galaxy Evolution Survey \citep{Kochanek2012}\\
Boötes               & 2 & SDSS DR12 \citep{York2000}\\
~                    & 4 & \citet{Houck2005}\\
~                    & 8 & \citet{Weedman2009}\\
\hline
~                    & 1 & 2dF Galaxy Redshift Survey \citep{Colless2001}\\
CDFS-SWIRE           & 2 & 6dF Galaxy Survey \citep{Jones2004}\\
~                    & 4 & VIMOS VLT Deep Survey \citep{LeFevre2013}\\
~                    & 8 & VLT/FORS2 spectroscopy in the GOODS-South Field \citep{Vanzella2008}\\
CDFS-SWIRE (cont.)   & 16 & GOODS - VLT/VIMOS Spectroscopy DR 2.0.1 \citep{Balestra2010}\\
~                    & 32 & IMAGES spectroscopy \citep{Ravikumar2007}\\
~                    & 64 & GMASS Ultradeep Spectroscopy \citep{Kurk2013}\\
~                    & 128 & \citet{Szokoly2004}, via MUSYC catalogue \citep{Cardamone2010}\\
~                    & 256 & \citet{Croom2001}, via MUSYC catalogue \citep{Cardamone2010}\\
~                    & 512 & \citet{VanderWel2004}, via2004 MUSYC catalogue \citep{Cardamone2010}\\
~                    & 1024 & \citet{Cristiani2000}, via MUSYC catalogue \citep{Cardamone2010}\\
~                    & 2048 & \citet{Strolger2004}, via MUSYC catalogue \citep{Cardamone2010}\\
~                    & 4096 & Lira et al. \textit{in prep},  via MUSYC catalogue \citep{Cardamone2010}\\
~                    & 8192 & \citet{Treister2009}, via MUSYC catalogue \citep{Cardamone2010}\\
~                    & 16384 & \citet{Kriek2008}, via MUSYC catalogue \citep{Cardamone2010}\\
~                    & 32768 & K20 Survey \citep{Mignoli2005}\\
~                    & 65536 & \citet{Silverman2010}\\
~                    & 131072 & \citet{Dickinson2004}\\
~                    & 262144 & PRIMUS \citep{Coil2011}\\
~                    & 524288 & 2MASS Redshift Survey \citep[2MRS,][]{Huchra2012}\\
~                    & 1048576 & IRAS Point Source Catalog Redshift Catalog \citep[IRASPSCZ,][]{Saunders2000}\\
~                    & 2097152 & The Arizona CDFS Environment Survey \citep[ACES,][]{Cooper2012}\\
~                    & 4194304 & \citet{Lacy2013}\\
~                    & 8388608 & OzDES DR1 \citep{Childress2017}\\
~                    & 16777216 & MUSE-Wide Survey \citep{Herenz2017}\\
~                    & 33554432 & VANDELS DR2 \citep{McLure2018}\\
\hline
~                    & 1 & SDSS DR15 \citep{York2000}\\
COSMOS               & 2 & 2MASS Redshift Survey \citep[2MRS,][]{Huchra2012}\\
~                    & 4 & PRIMUS \citep{Coil2011}\\
~                    & 8 & COSMOS spec-z catalogue$^\dagger$ (public redshifts), source 3DHST \citep{Momcheva2016}\\
~                    & 16 & COSMOS spec-z catalogue$^\dagger$ (public redshifts), source \citet{Onodera2015}\\
~                    & 32 & COSMOS spec-z catalogue$^\dagger$ (public redshifts), source FAST (N.Wright, F Civano)\\
~                    & 64 & COSMOS spec-z catalogue$^\dagger$ (public redshifts), source AZTEC (N.Wright, F Civano)\\
~                    & 128 & COSMOS spec-z catalogue$^\dagger$ (public redshifts), source \citep{Roseboom2012}\\
~                    & 256 & COSMOS spec-z catalogue$^\dagger$ (public redshifts), source \citep{Comparat2015}\\
~                    & 512 & COSMOS spec-z catalogue$^\dagger$ (public redshifts), source 2MRS \citep{Huchra2012}\\
~                    & 1024 & COSMOS spec-z catalogue$^\dagger$ (public redshifts), source GEMINI-S (M. Balogh)\\
~                    & 2048 & COSMOS spec-z catalogue$^\dagger$ (public redshifts), source HST GRISM (K. Kartaltepe, M. Brusa)\\
\hline
~                    & 1 & SDSS DR15 \citep{York2000}\\
EGS                  & 2 & DEEP 2 \& 3 \citep{Newman2013, Cooper2011}\\
~                    & 4 & 3DHST \citep{Momcheva2016}\\
~                    & 8 & \citet{Steidel2003}\\
~                    & 16 & AEGIS-X \citep{Nandra2015}\\
~                    & 32 & \citet{Huang2009}\\
~                    & 64 & C3R2 DR1 \& DR2 \citep{Masters2019}\\
\hline
~                    &  1 & \citet{Berta2007}\\
ELAIS-N1             & 2 &  SDSS DR13 \citep{York2000}\\
~                    & 4 & \citet{Trichas2010}\\
~                    & 8 & \citet{Swinbank2007}\\
~                    & 16 & \citet{Rowan-Robinson2013} (WIYN, Keck and Gemini sources)\\
~                    & 32 & \citet{Rowan-Robinson2013} sources taken from NED)\\
~                    & 64 & Updated Zwicky Catalog \citep[UZC][]{Falco1999}\\
~                    & 128 & \citet{Lacy2013}\\
\hline
~                       & 1 & SDSS DR13 \citep{York2000}\\
ELAIS-N2                & 2 & \citet{Berta2007}\\
~                       & 4 & \citet{Swinbank2005}\\
~                       & 8 & Updated Zwicky Catalog \citep[UZC][]{Falco1999}\\
~                       & 16 & IRAS Point Source Catalog Redshift Catalog \citep[IRASPSCZ,][]{Saunders2000}\\
~                       & 32 & \citet{Rowan-Robinson2013} (WIYN, Keck and Gemini sources)\\
~                       & 64 & \citet{Rowan-Robinson2013} sources taken from NED)\\
~                       & 128 & 2MASS Redshift Survey \citep[2MRS,][]{Huchra2012}\\
~                       & 256 & \citet{Lacy2013}\\
\hline
~                    & 1 & Australian Telescope Large Area Survey \citep{Mao2012}\\
ELAIS-S1             & 2 & \citet{Sacchi2009}\\
~                    & 4 & \citet{Feruglio2008}\\
~                    & 8 & PRIMUS \citep{Coil2011}\\
~                    & 16 & 2dF Galaxy Redshift Survey \citep{Colless2001}\\
~                    & 32 & 6dF Galaxy Survey \citep{Jones2004}\\
~                    & 64 & IRAS Point Source Catalog Redshift Catalog \citep[IRASPSCZ,][]{Saunders2000}\\
~                    & 128 & \citet{Lacy2013}\\
~                    & 256 &  OzDES DR1 \citep{Childress2017}\\
\hline
~                    & 1 & GAMA I - DR3 \citet{Baldry2018}\\
GAMA-09,12,15        & 2 & SDSS DR13 \citep{York2000}\\
~                    & 4 & WiggleZ DR1 \citep{Drinkwater2010}\\
~                    & 8 & 6dF Galaxy Survey \citep{Jones2004}\\
~                    & 16 & 2dF Galaxy Redshift Survey \citep{Colless2001}\\
~                    & 32 & 2SLAQ-QSO \citep{Croom2009}\\
~                    & 64 & 2SLAQ-LRG \citep{Croom2009}\\
~                    & 128 & 2MASS Redshift Survey \citep[2MRS,][]{Huchra2012}\\
~                    & 256 & 2dF QSO Redshift Survey \citep{Croom2004}\\
~                    & 512 & Millennium Galaxy Catalogue \citep[MGC,][]{Liske2003}\\
~                    & 1024 & Updated Zwicky Catalog \citep[UZC][]{Falco1999}\\
~                    & 2048 & NED sources\\
~                    & 4096 & GAMA I (Liverpool Telescope) \citet{Baldry2018}\\
~                    & 8192 & IRAS Point Source Catalog Redshift Catalog \citep[IRASPSCZ,][]{Saunders2000}\\
& & \textit{(Two proprietary sources are stored in HELP database for new releases of }\\
& & \textit{GAMA and WiggleZ for when they are made public)}\\
\hline
~                    &  1 & SDSS DR14 \citep{York2000}\\
HDF-N                & 2 & DEEP 3 \citep{Cooper2011}\\
~                    & 4 &  3DHST \citep{Momcheva2016}\\
~                    & 8 & \citet{Steidel2003}\\
~                    & 16 & \citet{Liu1999}\\
~                    & 32 & Team Keck Treasury Redshift Survey \citep{Wirth2004}\\
~                    & 64 & \citet{Reddy2006}\\
~                    & 128 & Caltech Faint Galaxy Redshift Survey \citep[][and references there in]{Cohen2000}\\
~                    & 256 & \citet{Chapman2004}\\
~                    & 512 & \citet{Chapman2005}\\
~                    & 1024 & \citet{Swinbank2004}\\
~                    & 2048 & \citet{Dawson2001}\\
~                    & 4096 & \citet{Pope2008}\\
~                    & 8192 & \citet[][their sources 1, 3, 7, 12, and 14]{Barger2008}\\
~                    & 16384 & \citet{VandenBerk2000}\\
~                    & 32768 & \citet{Cowie2004}\\
\hline
~                    & 1 & 2dF Galaxy Redshift Survey \citep{Colless2001}\\
Herschel-Stripe-82   & 2 & 2SLAQ-QSO \citep{Croom2009}\\
~                    & 4 & 2SLAQ-LRG \citep{Croom2009}\\
~                    & 8 & 6dF Galaxy Survey \citep{Jones2004}\\
~                    & 16 & PRIMUS \citep{Coil2011}\\
~                    & 32 & SDSS DR12 \citep{York2000}\\
~                    & 64 & DEEP 2 \citep{Newman2013}\\
~                    & 128 & WiggleZ DR1 \citep{Drinkwater2010}\\
~                    & 256 & 2MASS Redshift Survey \citep[2MRS,][]{Huchra2012}\\
\hline
~                    & 1 & \citet{Steffen2004}\\
Lockman-SWIRE        & 2 & \citet{Berta2007}\\
~                    & 4 & SDSS DR13 \citep{York2000}\\
~                    & 8 & \citet{Rowan-Robinson2013} sources taken from NED)\\
~                    & 16 & \citet{Rowan-Robinson2013} (WIYN, Keck and Gemini sources)\\
~                    & 32 & 2MASS Redshift Survey \citep[2MRS,][]{Huchra2012}\\
~                    & 64 & IRAS Point Source Catalog Redshift Catalog \citep[IRASPSCZ,][]{Saunders2000}\\
~                    & 128 & Updated Zwicky Catalog \citep[UZC][]{Falco1999}\\
~                    & 256 & \citet{Lacy2013}\\
\pagebreak
\multirow{2}{*}{NGP} & 1 & SDSS DR12 \citep{York2000}\\
                     & 2 & 2MASS Redshift Survey \citep[2MRS,][]{Huchra2012}\\
\hline
~                   & 1 & SDSS DR13 \citep{York2000}\\
SA13                 & 2 & \citet{Chapman2005}\\
~                    & 4 & \citet{Smail2004}\\
~                    & 8 & \citet{Cowie1996}\\
\hline
~                    & 1 & 2dF Galaxy Redshift Survey \citep{Colless2001}\\
SGP                  & 2 & 6dF Galaxy Survey \citep{Jones2004}\\
~                    & 4 & 2MASS Redshift Survey \citep[2MRS,][]{Huchra2012}\\
~                    & 8 & Southern Sky Redshift Survey \citep[SRSS,][]{daCosta1998}\\
\hline
SPIRE-NEP            & 1 & 2MASS Redshift Survey \citep[2MRS,][]{Huchra2012}\\
\hline
~                   & 1 & 2dF Galaxy Redshift Survey \citep{Colless2001}\\
SSDF                 & 2 & 6dF Galaxy Survey \citep{Jones2004}\\
~                    & 4 & IRAS Point Source Catalog Redshift Catalog \citep[IRASPSCZ,][]{Saunders2000}\\
~                    & 8 & 2MASS Redshift Survey \citep[2MRS,][]{Huchra2012}\\
\hline
~                    & 1 & SDSS \citep{York2000}\\
xFLS                 & 2 & Updated Zwicky Catalog \citep[UZC][]{Falco1999}\\
~                    & 4 & \citet{Papovich2006}\\
~                    & 8 & \citet{Marleau2007}\\
~                    & 16 & \citet{Lacy2007}\\
~                    & 32 & \citet{Lacy2013}\\
~                    & 64 & \citet{Yan2007}\\
~                    & 128 & NED sources compiled by M. Vaccari \citep{Vaccari2015}\\
\hline
~                    & 1 & SDSS DR14 \citep{York2000}\\
XMM-13hr             & 2 & \citet{Page2006}\\
~                    & 4 & \citet{Jeltema2007}\\
~                    & 8 & MCXC \citep{Piffaretti2011}\\
\hline
~                    & 1 & VIMOS VLT Deep Survey \citep{LeFevre2013}\\
XMM-LSS              & 2 & \citet{Garcet2007}\\
~                    & 4 & \citet{Lacy2007}\\
~                    & 8 & \citet{Stalin2010}\\
~                    & 16 & \citet{Lidman2013}\\
~                    & 32 & \citet{Yamada2005}\\
~                    & 64 & \citet{Ouchi2008}\\
~                    & 128 & \citet{Ono2010}\\
~                    & 256 & \citet{Sargsyan2009}\\
~                    & 512 & UDSz \citep{Bradshaw2013,McLure2013}\\
~                    & 1024 & UDS catalogue$^\dagger$ source marked `CJS-VIMOS'\\
~                    & 2048 & UDS catalogue$^\dagger$ source marked `CJS-ISIS'\\
~                    & 4096 & UDS catalogue$^\dagger$ source marked `CVB-DEIMOS'\\
~                    & 8192 & UDS catalogue$^\dagger$ source marked `JEG-LDSS2'\\
~                    & 16384 & UDS catalogue$^\dagger$ source marked `Doi-FOCASS'\\
~                    & 32768 & UDS catalogue$^\dagger$ source marked `SJC-AAOmega'\\
~                    & 65536 & UDS catalogue$^\dagger$ source marked `IRS-AAomega'\\
~                    & 131072 & UDS catalogue$^\dagger$ source marked `Aki-FOCAS'\\
~                    & 262144 & UDS catalogue$^\dagger$ source marked `Aki-2df'\\
~                    & 524288 & SDSS DR15 \citep{York2000}\\
~                    & 1048576 & 6dF Galaxy Survey \citep{Jones2004}\\
~                    & 2097152 & VIPERS PDR2 \citep{Scodeggio2018}\\
~                    & 4194304 & PRIMUS \citep{Coil2011}\\
~                    & 8388608 & Magellan/IMACS catalogue$^\star$\\
~                    & 16777216 & \citet{Lacy2013}\\
~                    & 33554432 & OzDES DR1 \citep{Childress2017}\\
~                    & 67108864 & C3R2 DR1 \& DR2 \citep{Masters2019}\\
~                    & 134217728 & VANDELS DR2 \citep{McLure2018}\\
\end{longtable}


\end{document}